\documentclass[fleqn,a4paper,11pt]{article}
\usepackage{color}

\usepackage{graphicx}
\usepackage{amssymb}
\newcommand \ee {$e^+e^-$ }
\newcommand \mumu {$\mu^+\mu^-$ }
\newcommand \kk {$K^+K^-$ }
\newcommand \mee {$m_{e^{+}e^{-}}$ }
\newcommand \mll {$m_{l^{+}l^{-}}$ }
\newcommand \mmumu {$m_{\mu^{+}\mu^{-}}$ }
\newcommand \pipi {$\pi^+\pi^-$ }
\newcommand \pt {$p_T$ }
\newcommand \mt {$m_T $ }
\newcommand \sqn {$\sqrt{s_{_{NN}}}$ }
\newcommand \sq {$\sqrt{s}$ }

\newcommand \sqnR {$\sqrt{s_{_{NN}}}$~=~200~GeV }
\newcommand \ccbar {$c\overline{c} $ }
\newcommand \ssbar {$s\overline{s} $ }
\newcommand \qqbar {$q\overline{q}$ }
\newlength{\myVSpace}
\def\JPG{{J. Phys}~{\bf G}}

\def\NIM{Nucl. Instr. and Meth.}
\def\NIMA{{Nucl. Instr. and Meth.}~{\bf A}}
\def\NPA{{Nucl. Phys.}~{\bf A}}

\def\PLB{{Phys. Lett.}~{\bf B}}

\def\PRL{Phys. Rev. Lett.\ }

\def\PRD{{Phys. Rev.}~{\bf D}}
\def\PRC{{Phys. Rev.}~{\bf C}}

\def\EPJC{{Eur.~Phys.~J.}~{\bf C}}
\def\EPJA{{Eur.~Phys.~J.}~{\bf A}}

\def\hypfig#1{\hyperref[#1]{Fig.~\ref*{#1}}}
\def\hyptab#1{\hyperref[#1]{Table~\ref*{#1}}}
\def\hypsec#1{\hyperref[#1]{Section~\ref*{#1}}}
\def\hypapp#1{\hyperref[#1]{Appendix~\ref*{#1}}}
\def\hypeq#1{\hyperref[#1]{Eq.~\ref*{#1}}}

\begin{document}


\title{Electromagnetic Probes}

\author{Itzhak Tserruya\\
   Weizmann Institute of Science, Rehovot, Israel}

\maketitle

\begin{abstract}
A review is presented of dilepton and real photon measurements in relativistic heavy ion collisions over a very broad energy range from the low energies of the BEVALAC up to the highest energies available at RHIC. The dileptons cover the invariant mass range \mll = 0 - 2.5 GeV/c$^2$, i.e.  the continuum at low and intermediate masses and the light vector mesons, $\rho, \omega, \phi$.  The review includes also measurements of the light vector mesons  in elementary reactions.
\end{abstract}

\tableofcontents
\newpage


\section{Introduction
\label{sec:intro}}

Electromagnetic probes - virtual (\ee or \mumu pairs) and real photons - are an important and powerful tool to diagnose the hot and dense matter produced in relativistic heavy-ion collisions (RHI).
They are sensitive probes of the two fundamental properties that characterize the Quark-Gluon Plasma (QGP) and are predicted by lattice QCD numerical calculations, the deconfinement of quarks and gluons and the restoration of chiral symmetry \cite{lqcd-review}.

Their importance has been emphasized time and again since the proposal made by Shuryak back in 1978 \cite{shuryak78}. Dileptons and photons interact only electromagnetically, and therefore their mean free path is large compared to the size of the system formed in nuclear collisions. They are not distorted by final state interactions and once produced can escape the interaction region, unaffected, carrying to the detectors information about the conditions and properties of the medium at the time of their production.
Electromagnetic probes are emitted over the entire space-time evolution of the collision, from the primordial parton collisions in the early stage till the hadron decays well after freeze-out. They are thus very rich in information content but this wealth is also a challenge for the measurement and subsequent analysis that need to disentangle the various sources.

The main topic of interest is the identification of radiation emitted in the form of virtual or real photons by the medium in thermal equilibrium. Such radiation is a direct fingerprint of the matter formed. Two well distinct sources are considered:

a) thermal radiation emitted by the strongly interacting QGP in the early phase of the collision. The elementary processes involved are the quark annihilation into virtual photons (${\it q}{\it \overline{q}} \rightarrow {\it l}^+ {\it l}^-$) or the annihilation (${\it q}{\it \overline{q}} \rightarrow {\it g} \gamma$) and QCD Compton (${\it q}{\it g} \rightarrow {\it q} \gamma$ or ${\it \overline{q}}{\it g} \rightarrow {\it \overline{q}} \gamma$) channels producing real photons. The identification of this signal not only serves as a proof of deconfinement but it also provides a direct measurement of one of the most basic plasma properties, the temperature, readily given by the inverse slope of its expected exponential spectral shape. Theory has singled out dileptons in the mass range \mll = 1-3 GeV/c$^2$ \cite{shuryak78,ruuskanen92} or real photons in the \pt~range \pt~= 1-3~GeV/c \cite{turbide04} as the most appropriate windows to observe the thermal radiation from the QGP phase. The thermal emission rate is a strongly increasing function of the temperature; consequently it is most abundantly produced in the early stage when the temperature and the energy density have their largest values; the sensitivity to identify it increases the higher the initial temperature is relative to the critical temperature of the phase transition. It should be easier to observe the QGP thermal radiation at LHC than at RHIC than at SPS. For almost two decades this has been a very elusive signal and it is only recently that PHENIX at RHIC \cite{phenix-photons08} and NA60 at SPS \cite{na60-imr} presented results that could signal the thermal radiation from the QGP.

b) Thermal radiation is also emitted by the high-density hadron gas (HG) in the later phase of the collision. The main elementary process here is the pion annihilation into dileptons, mediated through vector meson dominance by the $\rho$ meson ($\pi^+\pi^- \rightarrow \rho \rightarrow \gamma^* \rightarrow l^+l^-$). The HG thermal radiation, controlled by the pole at the $\rho$ mass of the pion electromagnetic form factor, contributes primarily to the low-mass region, (\mll $\leq$ 1GeV/c$^2$) around and below the $\rho$ mass \cite{kajantie86}. This component must always be present since ultimately the system ends in the hadronic phase and was readily identified in the enhancement of low-mass dileptons discovered at the SPS in the mid-nineties \cite{ceres-s,helios3}.

The enhancement of low-mass dileptons is one of the main highlights of the CERN SPS heavy-ion program. It triggered a huge theoretical activity stimulated mainly by interpretations based on in-medium modifications of the intermediate $\rho$ meson with a possible link to chiral symmetry restoration (CSR). (For theoretical reviews see \cite{rapp-wambach,brown-rho2002,gale-haglin03}). The SPS results motivated also new experiments aiming at precise spectroscopic studies of the vector meson resonances, $\rho, \omega$ and $\phi$, to explore in-medium modifications of their spectral properties (mass and width). This topic is of interest in its own right and more so because these modifications could shed light into the behavior of the resonances {\it close} to the chiral restoration boundary. By definition there are no hadrons in the QGP phase. Hadrons are formed in the mixed phase (if there is one) or in the hot and dense HG close to the phase boundary; their decay there into lepton pairs can convey valuable information about their properties close to the onset of CSR. The $\rho$ meson with its short lifetime, $\tau \sim$ 1.3 fm/c, and its strong coupling to the $\pi\pi$ channel is best suited in this context, making it the most sensitive signal for in-medium modifications and the best probe for CSR.
A basic question with no clear answer is how the restoration of chiral symmetry is realized.
The well established features are that the chiral condensate $<{\it \overline{q}}{\it q}>$ drops to zero at the phase transition and that in a chirally symmetric world the chiral doublets (e.g. $\rho$ and $a_1$) become degenerate in mass. Unfortunately, this degeneracy is practically impossible to establish experimentally.
Other implications on the hadron spectral properties (mass and width) are uncertain; there is no rigorous connection of the chiral condensate to the spectral properties of hadrons. Almost all models predict an increase of the width of the $\rho$ meson at the chiral transition \cite{pisarski82,dominguez93,chanfray-rw96,rapp-chanfray-wambach97} but for the behavior of the mass there is no clear picture; in several models the $\rho$ mass is predicted to drop to zero
\cite{pisarski82,brown-rho,hatsuda-lee} in other models it remains constant \cite{chanfray-rw96}, or even increases \cite{pisarski95}. Under these circumstances experimental insights are very valuable to guide and constrain these models. A significant breakthrough has been achieved recently in this respect with the precision measurements of NA60 \cite{na60-rho}, confirmed also by the upgraded CERES experiment \cite{ceres-pb2000}, which show a significant increase of the width, with no significant mass shift, of the $\rho$ meson at SPS energies. The experimental results seem to be telling us that the approach to CSR proceeds through broadening, and eventually melting, of the resonances rather than by dropping masses.

The interest in electromagnetic probes is strengthened by the light vector mesons (LVM) studied through their leptonic decays. The special role of the $\rho$ meson was already emphasized in the previous paragraphs; the other vector mesons, $\omega$ and $\phi$, can in principle reveal the same effects related to CSR as the $\rho$. However, the effects are relatively small. The $\omega$ and $\phi$, with their much longer lifetimes, will decay predominantly outside the medium after regaining the vacuum properties; only a small fraction of them will decay inside the medium making the effects much harder to observe through direct spectral shape analysis. A more promising way to unveil these effects is provided by the simultaneous measurement, within the same apparatus, of the $\phi$ meson yield through the \ee and \kk decay channels. With $m_{\phi} \approx 2m_{K}$ even a small in-medium effect on the $\phi$ or kaon spectral shapes may have a sizable effect on the $\phi$ yield measured through the \kk decay channel \footnote {The $\phi$ meson has attracted additional attention. The production mechanism in pp vs AA collisions is an open question. Consisting of \ssbar $\phi$ is the lightest meson with hidden strangeness and can provide information about the role of strangeness production in RHI collisions.  These are two topics of great interest in their own right but outside the scope of the present article.}.

The SPS results have also triggered a strong experimental program focused on the exploration of in-medium modifications of the LVM in elementary reactions. pA and photoabsorption reactions at relatively low energies have been mainly studied in this context. The idea is to avoid the complications due to time variations of the temperature and density inherent to RHI collisions by creating a low-momentum particle inside cold nuclear matter and observing its decay through dileptons. The disadvantage of course is that the predicted effects are much smaller than in high energy nuclear collisions. The difficulties in detecting these small effects probably explain the somewhat confusing experimental situation. Contrary to the nuclear case where all systems studied show an enhancement of dileptons, there is not yet a coherent picture that emerges from the present studies in elementary reactions.

\begin{table}[h!]
\caption{Experiments involved in dilepton, photon and light vector meson measurements in heavy-ion collisions. The energy is quoted in GeV per nucleon in the lab. system with the exception of PHENIX and STAR where the energy is in the c.m.s.}
\label{tab:list-of-exp}
\begin{center}
\leavevmode
\begin{tabular}{|c|c|c|c|c|c|}\hline

 Experiment &     Probe      &  System     & Energy  &  Mass range   & Ref                  \\
            &                &             & (GeV)   &  (GeV/c$^2$)  &                       \\ \hline
 DLS        & \ee            & C+C,~Ca+Ca  & 1.0     & 0.05 -- 1.0   & \cite{dls97}             \\ \hline
 HADES      & \ee            & C+C         & 1, 2     & 0 -- 1        & \cite{hades-prl07,hades-plb08,hades-09} \\ \hline
 HELIOS-3   & \mumu          & p+W,~S+W    & 200     & 0.3 -- 4.0    & \cite{helios3,helios3-vectormesons}\\ \hline
            & \ee            & p+Be,Au    & 450     &               & \cite{ceres-p}  \\
 CERES      & $e^+e^-, \gamma$& S+Au        & 200     & 0 -- 1.4      & \cite{ceres-s,ceres-photons}  \\
            & \ee            & Pb+Au       & 40, 158  &            & \cite{ceres-pb2000,ceres-pb158,ceres-pb40,ceres-pb95-96,ceres-phi} \\ \hline
 NA38       &                & p+W,~0,S+U  & 200     &               & \cite{na38-lvm,na38-lvm2,na38-lvm3,na38-imr-pt} \\
 NA50       & \mumu          & p+A         & 450     & 0.3 -- 7.0    &              \\
            &                & Pb+Pb       & 158     &               & \cite{na50-lm,na50-imr,na50-pt,na50-phi}\\\hline
 NA60       & \mumu          & In+In       & 158     & 0.2 -- 5.0    & \cite{na60-imr,na60-rho,na60-flow,na60-polar,na60-ff,na60-phi}\\ \hline
 WA80       & $\gamma$       & S+Au        & 200     &               & \cite{wa80} \\
 WA98       & $\gamma$       & Pb+Pb       & 158     &               & \cite{wa98} \\ \hline
 PHENIX     & $e^+e^-, \gamma$& p+p         & 200     & 0 -- 4        & \cite{phenix-pp-lmr,phenix-pp-photons}\\
            & $e^+e^-, \gamma$& Au+Au       & 200     & 0 -- 4.5      & \cite{phenix-photons08,phenix-lmr,phenix-photons05}    \\
            & $\phi \rightarrow$ \kk         & Au+Au        & 200      & $\phi$  & \cite{phenix-phi}    \\ \hline
 STAR       & $\rho \rightarrow \pi^+ \pi^-$ & Au+Au        & 200      & $\rho$  & \cite{star-rho} \\
            & $\phi \rightarrow$ \kk         & Cu+Cu, Au+Au & 62.4--200 & $\phi$  & \cite{star-phi130,star-phi200,star-phi-cu} \\ \hline
\end{tabular}
\end{center}
\end{table}

The physics potential of electromagnetic probes is confirmed by the wealth of interesting results obtained so far and by the relatively large number of experiments focusing on their study. All the experiments involved in  measurements of dileptons, photons and LVM in nuclear collisions are listed in Table~\ref{tab:list-of-exp} together with references to their results published (or to be published) in the refereed literature. The measurements cover a very broad energy range comprising three different energy scales; the DLS at the BEVALAC and HADES at the GSI studied dielectron production at low energies of 1-2~AGeV;  at CERN SPS, low-mass electron pairs were systematically studied by the CERES experiment in S, Pb+Au collisions from 40 to 200~AGeV; dimuons over a broader mass range were measured by HELIOS-3, NA38/50 and more recently by the high statistics and high mass resolution NA60 experiment. There is redundancy in these two energy regimes in the sense that at least two different experiments have performed similar measurements with reasonable agreement between them giving confidence on the robustness of the results. Given the difficulties of the measurements, this redundancy is an important element of the program. At the high energies of RHIC, PHENIX and STAR have studied LVM through hadronic decays, in particular the $\phi \rightarrow$ \kk decay.    The study of electromagnetic probes is just beginning with first results available from PHENIX on electron pairs.
Apart the energy scale, the measurements of dileptons are better divided, from the thematic point of view, into (i) the low-mass region (LMR, \mll $\leq 1~GeV/c^2$) where the main contributions are Dalitz and two-body resonance decays and where the main topics of interest are CSR and  in-medium modification of the LVM and  (ii) the intermediate mass region (IMR, \mll = 1-3 $GeV/c^2$) dominated by correlated pairs from semi-leptonic open charm decays and Drell-Yan pairs and where the main topic is thermal radiation from  the QGP.

This article is organized as follows. Section~\ref{sec:challenge} discusses the main difficulties of the measurement of dileptons and compares the merits of real vs virtual photon measurements. Section~\ref{sec:ref} presents results on p+p and p+A collisions which serve as a crucial reference for the nuclear case. Section~\ref{sec:lowmass} reviews the results on low-mass dileptons, considering first the SPS results from CERES and NA60 in Section~\ref{subsec:sps}.
The RHIC results, mainly from PHENIX are presented in Section~\ref{subsec:rhic} whereas the low-energy results on low-mass dileptons at the BEVALAC and GSI are discussed in Section~\ref{subsec:low-e}. Section~\ref{sec:lvm} presents results on the light vector mesons in nuclear collisions. Section~\ref{sec:imr} contains a discussion on IMR dileptons measured at the SPS, first by NA38/50 and more recently by NA60, and at RHIC by PHENIX. Section~\ref{sec:spectroscopy} shows results on LVM spectroscopic studies performed with elementary collisions. Thermal photons are discussed in Section~\ref{sec:photons} focusing on the attempts of the WA80/98 experiment at CERN and the recent PHENIX results at RHIC. A summary and outlook is given in the last Section~\ref{sec:summary}.

\section{Experimental challenge}
\label{sec:challenge}

The measurement of dileptons is notoriously difficult which explains why the first results at SPS and RHIC became available only several years after the beginning of the heavy ion program in these machines. There are two main difficulties. The first one is the huge {\it combinatorial background} of uncorrelated lepton pairs. It arises from the fact that, since single leptons do not preserve any information about their parent particle, all leptons are paired with all anti-leptons in the same event to form the invariant mass spectrum. This background therefore depends quadratically on the particle multiplicity and strongly increases as the coverage moves to low-\pt leptons.

In the measurement of \ee pairs, the combinatorial background originates mainly from $\pi^0$ Dalitz decays and $\gamma$ conversions. In an ideal detector with 100\% track reconstruction efficiency, the combinatorial background would not be a severe problem since the tracks from the overwhelming yield of  $\pi^0$ Dalitz decays and $\gamma$ conversions form pairs with distinctive features (very small opening angle and very small invariant mass). Therefore once such a pair is reconstructed, its two tracks can be taken out from further pairing with other tracks of the same event thereby considerably reducing the background.
However, in a real detector, there is a finite track reconstruction efficiency and in particular a minimum \pt is required for a track to be reconstructed. Consequently, very often only one of the two tracks of a real pair is reconstructed. These single tracks when paired to other tracks in the same event give rise to the combinatorial background.

The combinatorial background (B), can be determined from the yield of like-sign pairs or evaluated using an event mixing technique, and then subtracted from the inclusive yield of unlike-sign pairs (U) to determine the real signal (S). This subtraction, S~=~U~-~B, induces a large statistical and a large systematic uncertainty in S.
For example, in the measurement of low-mass electron pairs at the SPS, the S/B ratio, before (after) rejection has a typical value of  $\sim$1/200 (1/20). This implies that the statistical error in S is a factor of 20 (6) larger compared to a situation with no background and a 1\% systematic uncertainty in B results in a factor of 2 (20\%) uncertainty in S, $\Delta$S = 2S (0.2S).

The combinatorial background is the main factor affecting the quality of dilepton measurements and it is therefore imperative to reduce it as much as possible. In the double RICH spectrometer of the CERES experiment \cite{ceres-nim}, this is mainly done by the RICH-1 detector located before the magnetic field where the original direction of the electrons is preserved allowing to recognize conversions and $\pi^0$ Dalitz decays by their small opening angle. The rejection of close tracks is further supplemented by a doublet of silicon drift detectors \cite{ceres-pb95-96}. After all rejection cuts, CERES reached typical values of S/B = 1/15 -1/20 in central Pb+Au collisions at 158 AGeV.  The NA60 experiment measures dimuons where the main combinatorial background originates from charged pion and kaon decays. The main rejection tool in NA60 is provided by a precision vertex telescope located close to the target. By matching the muon tracks identified after the hadron absorber to the vertex telescope the combinatorial background is considerably reduced allowing to reach a S/B value of 1/11 in central In+In collisions at 158~AGeV \cite{na60-rho,na60-telescope}. The PHENIX detector measures electron pairs at mid-rapidity; the present set-up with a strong magnetic field starting at the vertex has no rejection resulting in a S/B $\sim$ 1/200 in minimum bias Au+Au collisions at \sqnR \cite{phenix-lmr}. The HBD upgrade providing electron identification in an almost field free region close to the vertex will considerably improve the situation \cite{hbd1,hbd2,hbd-it}.

The second difficulty is the {\it physics background}. Photons and dileptons can be emitted by a variety of sources and therefore before claiming observation of any new effect, it is mandatory to have a thorough understanding of the expected contribution from all known sources.  Drell-Yan and semi-leptonic decays of charm mesons, produced in the primary hard collisions, are the main contributions in the IMR. In the LMR, the physics background is dominated by the electromagnetic decays of hadrons (Dalitz decays: $\pi^o$,$\eta$,$\eta$' $\rightarrow$ $l^+l^-\gamma$, Dalitz-$\pi^0$ decay: $\omega$ $\rightarrow$  $l^+l^-\pi^o$, and resonance decays: $\rho,\omega,\phi$ $\rightarrow$  $l^+l^-$) which mostly take place at a late stage of the collision, after freeze-out. Most experiments have adopted a systematic approach, performing precise measurements of the dilepton spectrum in p+p collisions as the reference for the physics background and p(d)+A collisions to establish cold nuclear matter effects. Together they form the basis for identifying any possible deviation from the known physics in the hot and dense matter formed in nucleus-nucleus collisions.

The physics potential discussed in the Introduction is in principle as relevant for virtual as for real photons, since both are expected to carry the same physics information. However, their sensitivities are quite different. For real photons the physics background is larger than that of dileptons by orders of magnitude, making the measurement of photons much less sensitive to a new source \cite{it-qm95}. The real advantage of dileptons over photons comes from the fact that they are characterized by two parameters, m and \pt, whereas real photons are fully characterized by one single variable, \pt.

\section{p+p and p+A collisions: the reference measurements}
\label{sec:ref}

A precise knowledge of the conventional sources contributing to the dilepton spectrum is an essential pre-requisite to unveil effects which are specific to nucleus-nucleus collisions.

In the LMR, the main dilepton sources that compose the so-called hadronic cocktail, are the Dalitz decays, $\pi^0, \eta, \eta' \rightarrow l^+l^- \gamma $, ($\pi^0$ contributes only to di-electron measurements) the Dalitz-$\pi^0$ decay $\omega \rightarrow l^+l^- \pi^0$ and the vector meson decays $\rho, \omega, \phi \rightarrow l^+l^-$. In di-muon experiments there is also a background of correlated pairs coming from hadron decays, mainly the $\rho$ where both pions decay $\rho \rightarrow \pi^+ \pi^- \rightarrow \mu^+ \mu^- \nu_{\mu} \overline{\nu}_{\mu}$. A weaker contribution comes from  $\phi \rightarrow K^+ K^- \rightarrow \mu^+ \mu^- \nu_{\mu} \overline{\nu}_{\mu}$. The contribution of correlated pairs from the semi-leptonic decays of open charm mesons, negligible at SPS energies \cite{pbm99}, is an important source that has to be included at RHIC energies. Finally hadronic bremsstrahlung that contributes to the pair spectrum mainly at very low mass and very low \pt was found to be significant at very low energies \cite{bratkovskaya98} and negligible at SPS energies \cite{helios1-pbe}.

In the IMR, at SPS energies, the dilepton spectrum is dominated by two sources: Drell-Yan pairs and semi-leptonic decays of charmed mesons pairs. Going up to RHIC energies, the Drell-Yan contribution becomes negligible whereas the semi-leptonic decays of B mesons become significant.

This reference information is provided by measurements of p+p collisions. The measurements of p+A collisions allow the study of cold nuclear matter effects. In this section a brief review of reference measurements performed at SPS and RHIC is presented together with a description of the inclusive dilepton spectrum in terms of these sources.

\subsection{Reference measurements at SPS}
\label{subsec:ref-sps}

\begin{figure}[!tb]
 \begin{center}
     \includegraphics[height=80mm, width=72mm]{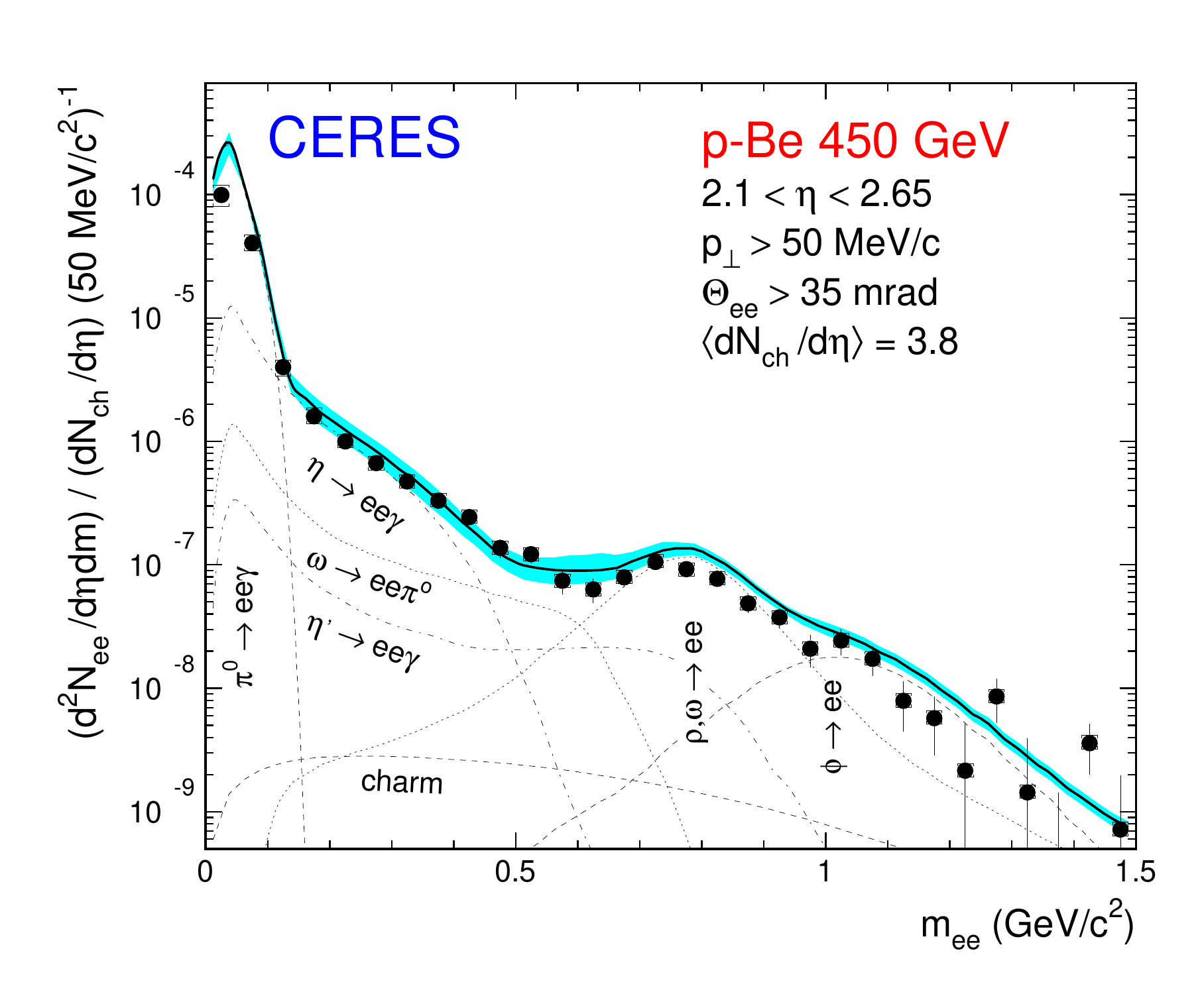}
     \includegraphics[height=80mm, width=72mm]{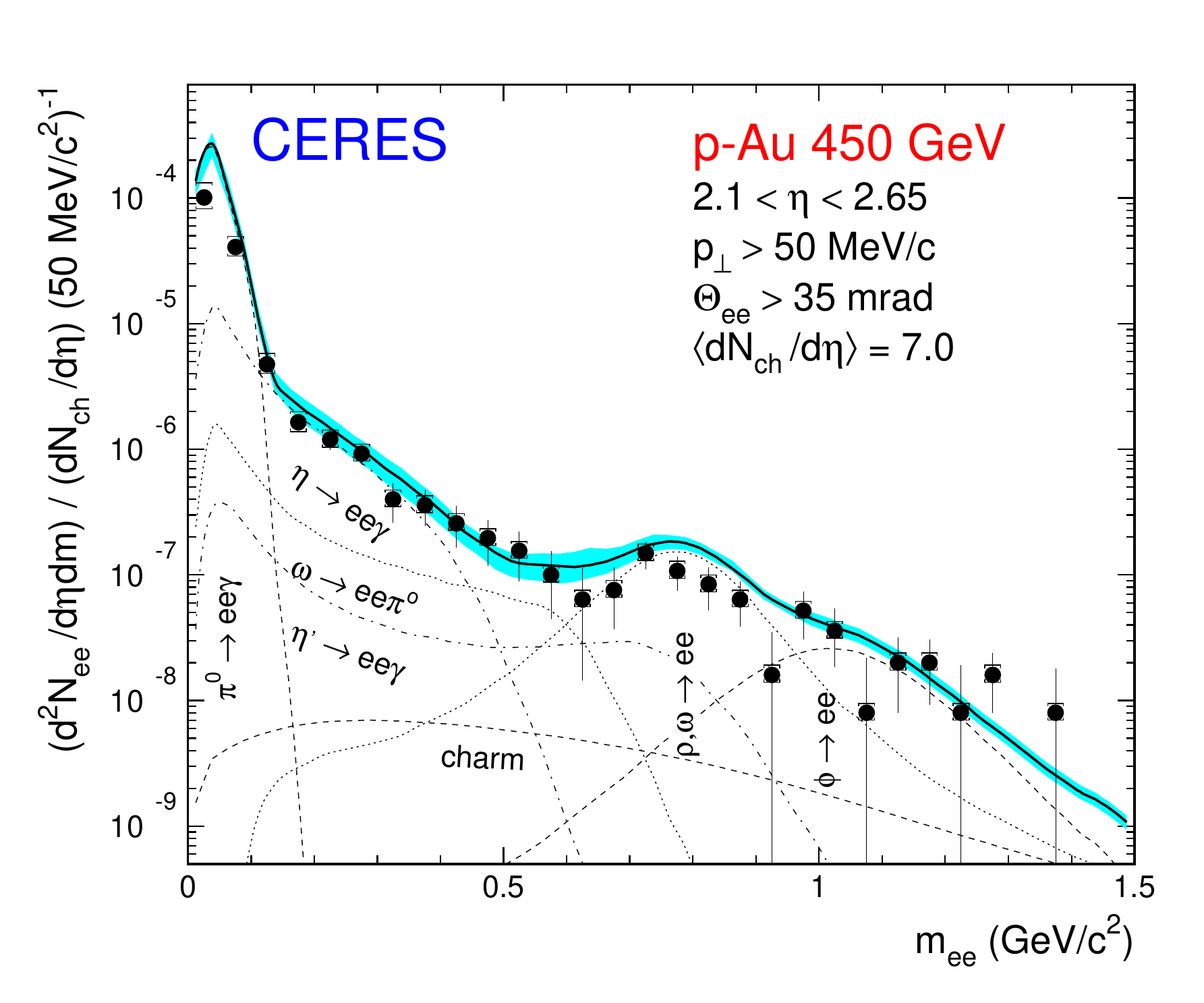}
    \caption{Invariant \ee mass spectrum measured by CERES in 450 GeV p+Be (left panel) and p+Au (right panel) collisions compared to the hadronic cocktail calculated with the GENESIS code. The lines represent the contributions from the known hadron decays and the shaded area gives the $\pm 1 \sigma$ systematic error on the summed contributions. Statistical (vertical bars) and systematic (brackets) errors are plotted independently of each other. The data are not corrected for pair acceptance and the calculated spectrum is subject to the same filter as the data, in terms of acceptance, \pt and opening angle cuts \cite{ceres-p}.}
  \label{fig:ceres-pbe}
 \end{center}
 \vspace{-6mm}
 \end{figure}

The di-electron sources have been determined at the SPS by the CERES experiment in p+Be collisions, a pretty good approximation to p+p collisions. In addition to the inclusive \ee pair yield, shown in Fig.~\ref{fig:ceres-pbe}, CERES measured production cross sections and \pt distributions of $\pi^0$ and $\eta$ via their $\gamma \gamma$ decay modes,  $\omega$ via its  $\pi^0 \gamma$ decay channel, and  fully reconstructed Dalitz decays  $\pi^0, \eta \rightarrow \gamma$ \ee \cite{ceres-p}. These data, complemented with other results from NA27 \cite{na27}, formed the basis of a Monte Carlo event generator, GENESIS, used to calculate  the inclusive \ee pair spectrum (see \cite{ceres-p} for a detailed description of GENESIS and \cite{ceres-pb95-96} for slight improvements). Mesons that decay into electron pairs are first generated according to measured production cross sections, \pt and rapidity distributions. Mesons are then allowed to decay into \ee using known branching ratios \cite{pdg} and the Kroll-Wada expression \cite{kroll-wada} with measured form factors for the Dalitz decays \cite{lepton-g,landsberg} or the Gounaris-Sakurai expression for the decays of the vector mesons $\rho, \omega$ and $\phi$ into \ee \cite{gounaris-sakurai}. The contribution of correlated pairs from the semi-leptonic decays of open charm mesons, is estimated using the PYTHIA code \cite{pythia}.
For a meaningful comparison to the data, the simulated electron data are finally convoluted with the experimental resolution and subject to the same filter as the data in terms of acceptance, \pt and opening angle cuts.

The left panel of Fig.~\ref{fig:ceres-pbe} shows a comparison of the invariant \ee mass spectrum measured by CERES in 450~GeV p+Be collisions with the calculated hadronic cocktail \cite{ceres-p}. One sees that the data are very well reproduced by electron pairs from the known hadronic sources.
Both the data and the cocktail are normalized to give the pair density per charged particle density within the acceptance of the CERES spectrometer. For the cocktail this is done by first normalizing to the yield of $\pi^0$  which is one of the ingredients of the cocktail and then to the charged particle density using the ratio $N_{\pi^0} / N_{ch}$ = 0.44 \cite{ceres-pb95-96}. This is a convenient normalization for further comparison to other collision systems. If one assumes that the relative particle abundances remain unchanged and independent of the collision system,  then the \ee yield is expected to scale with the number of charged particles, and the absolute vertical scale should remain unchanged. This is the case for p+Au collisions as demonstrated in the right panel of Fig.~\ref{fig:ceres-pbe} that compares the inclusive \ee spectrum with the same cocktail demonstrating that the cold nuclear matter effects are negligible in the dilepton low-mass spectrum at SPS energies. The bands in Fig.~\ref{fig:ceres-pbe} reflect the total uncertainty in the cocktail calculation which in the mass region m~=~0.2-0.8~GeV/c$^2$ is dominated by the uncertainties in the branching ratios and transition from factors of the $\eta$ and $\omega$. These uncertainties are now greatly reduced with the new values derived from the low-mass dimuon spectra measured by NA60 in peripheral In+In collisions \cite{na60-ff}.

NA60 performed measurements of p+Be, In and Pb at 400 GeV \cite{na60-hp04}. The inclusive dimuon mass spectrum was analyzed using a similar procedure, using in particular the same GENESIS code that was developed by CERES to generate the hadronic cocktail but adapted for the case of dimuons.

The hadronic cocktail for nucleus-nucleus collisions contains the same components mentioned above but some of the ingredients need to be adjusted to reflect known changes of the relative particle production cross sections, rapidity and \pt distributions and their dependence with centrality, of the various sources. This is best done by performing independent measurements of the hadrons that contribute to the dilepton spectrum. PHENIX does so for the main sources, for the $\pi^0 \rightarrow \gamma\gamma$ and similarly for the $\eta \rightarrow \gamma\gamma$. For the weaker sources where direct experimental information is not available, one can exploit the systematic behavior of the data, e.g. using the statistical model that so successfully predicts the particle ratios or the systematic behavior of the transverse momentum spectra with mass \cite{ceres-pb95-96}. NA60 fixes the main sources of the cocktail, for each centrality bin, using the measured inclusive dimuon mass distribution itself \cite{na60-rho,na60-flow}.

NA38/50 performed systematic studies of the IMR dimuon yield in 450 GeV p+A collisions where A stands for Al, Cu, Ag and W \cite{na50-imr}. A very good description of the IMR data in all four cases could be achieved by a superposition of Drell-Yan  and open charm pairs. The event generator PYTHIA was used to calculate the shape of these two contributions \cite{pythia}. The simultaneous fit of the four data sets, assuming that both cross sections scale as A$^\alpha$ with $\alpha$ = 1, and imposing the same ratio of the two cross sections for the four targets, provided the absolute normalization of the two components with a total $c \overline{c}$ production cross section in p+p collisions at 450 GeV of $\sigma_{c \overline{c}}^{pp} = 36.2 \pm 9.1 \mu$b.

\subsection{Reference measurements at RHIC}
\label{subsec:ref-rhic}

\begin{figure}[!h]
  \begin{center}
    \includegraphics [width=120mm]{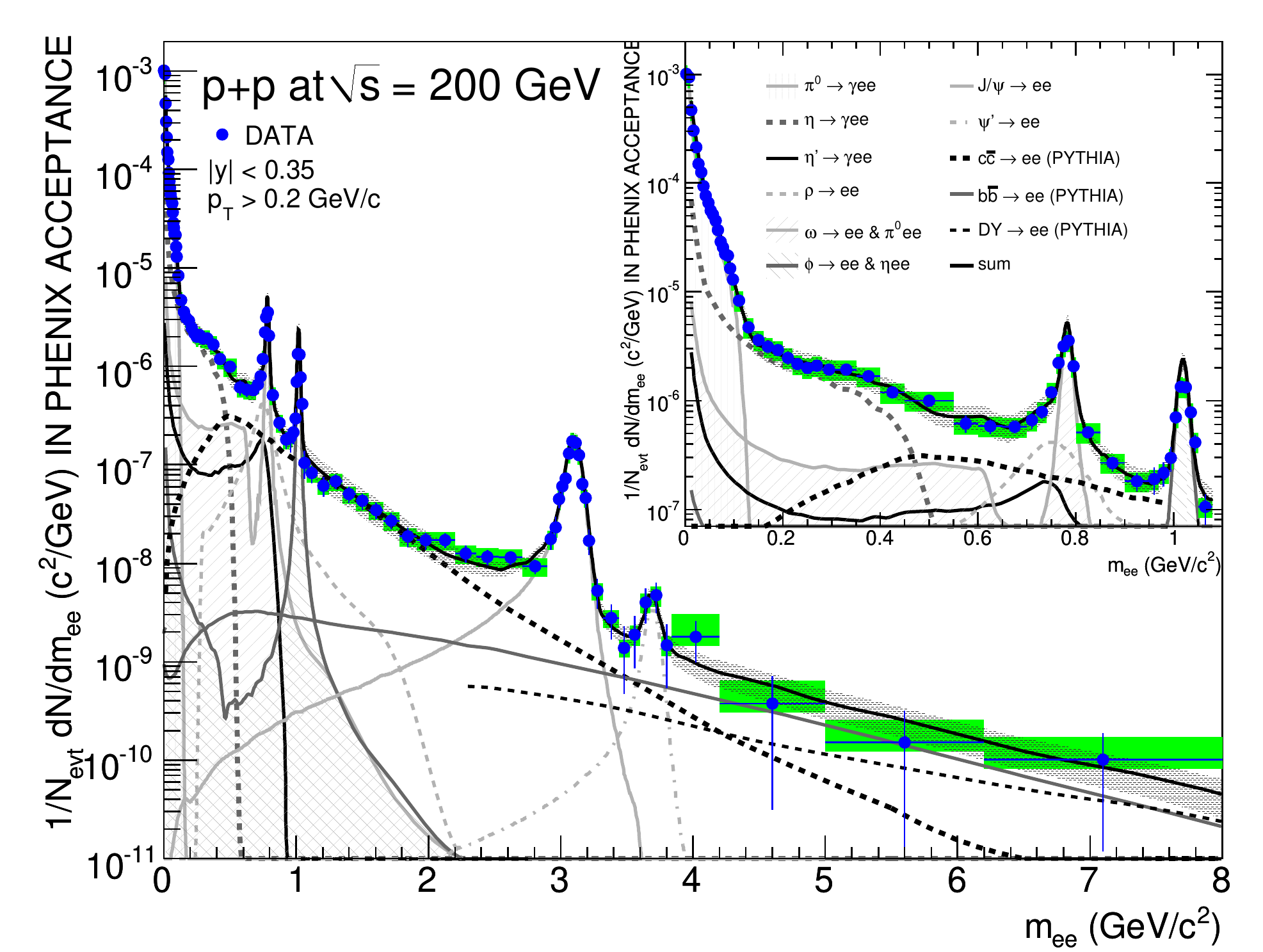}
  \vspace{-5mm}
  \caption{Inclusive mass spectrum of \ee pairs measured by PHENIX in p+p collisions at \sq = 200~GeV compared to the cocktail of hadronic sources. The insert is a zoom on the low-mass region. Statistical (vertical bars) and systematic (shades) are plotted separately. See text for more details \cite{phenix-pp-lmr}.}
  \label{fig:phenix-pp}
  \vspace{-2mm}
\end{center}
\end{figure}

PHENIX has studied in great detail the dielectron spectrum in p+p collisions at \sq = 200 GeV. The inclusive spectrum is shown in Fig.~\ref{fig:phenix-pp} covering the mass range from 0 up to 8~GeV/c$^2$ \cite{phenix-pp-lmr}. Following the same rational as CERES, a Monte Carlo simulation code, EXODUS, was developed for the calculation of the \ee pair spectrum. The contributing sources are the same as those outlined above and the ingredients are similar with abundances, \pt and rapidity distributions of the various components, taken from other PHENIX measurements or from systematics. The main change is that this time the contribution from heavy flavor (charm and bottom) is substantial. Fitting the data, after subtracting the cocktail of the light meson decays, with the charm and bottom spectral shapes calculated with PYTHIA, yields a total cross section of $\sigma_{c \overline{c}}^{pp}~=~544\pm39(stat)\pm142(syst)\pm 200(model)~\mu$b, in reasonable agreement with an independent determination of the charm cross section derived from single electron measurements \cite{phenix-single-e}. The comparison of the measured \ee pair spectrum with the various sources over the entire mass range is shown in Fig.~\ref{fig:phenix-pp}.

\section{Low-mass continuum in nuclear collisions
\label{sec:lowmass}}
Low-mass dileptons in nuclear collisions have been measured at three different energy ranges: 1-2~AGeV at the BEVALAC and GSI, 40-200~AGeV at the CERN SPS and  \sqnR at RHIC. The three experiments that measured lepton pairs at the CERN SPS in the early nineties, CERES, HELIOS-3 and NA38, reported almost simultaneously an enhanced production of dileptons (either \ee or \mumu) in S induced reactions over a very broad invariant mass range from $m_{ll} \sim 200~MeV/c^2$ up to the J/$\psi$ \cite{it-qm95}. Since then an enhancement of low-mass dileptons has been observed at all energies and all systems studied  although it is not clear that the enhancement reflects the same physics in all cases. At the two lower energies, measurements were done by two different experiments at least, with consistent results. At RHIC energies, PHENIX is presently the only experiment capable of measuring low-mass dileptons.

The excess of low-mass dileptons discovered at the SPS provided a tremendous stimulus. It triggered a huge theoretical activity mainly motivated by a possible link with CSR. It also prompted additional measurements by CERES and motivated a new generation experiment NA60 at the SPS. This section  reviews the low-mass dilepton results focussing first on the SPS results of CERES and NA60, then on the RHIC results of PHENIX and finally on the low energy results of DLS and HADES.

\subsection{Low-mass dileptons at the SPS}
\label{subsec:sps}

\subsubsection{CERES results}
\begin{figure}[!tb]
\begin{center}
     \includegraphics[width=70mm, height=70mm]{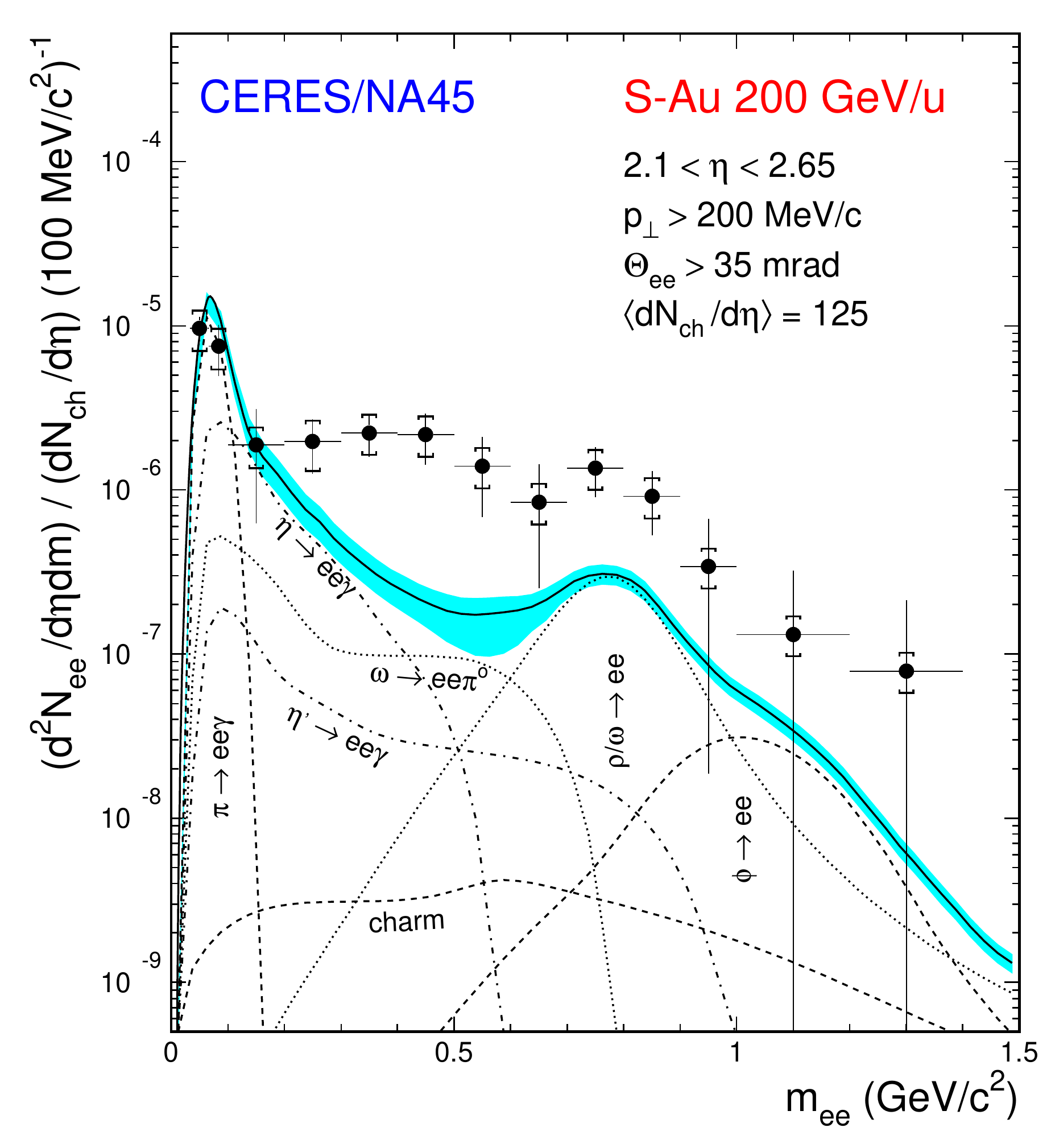}
      \vspace{0.0cm}
      \includegraphics[width=70mm, height=70mm]{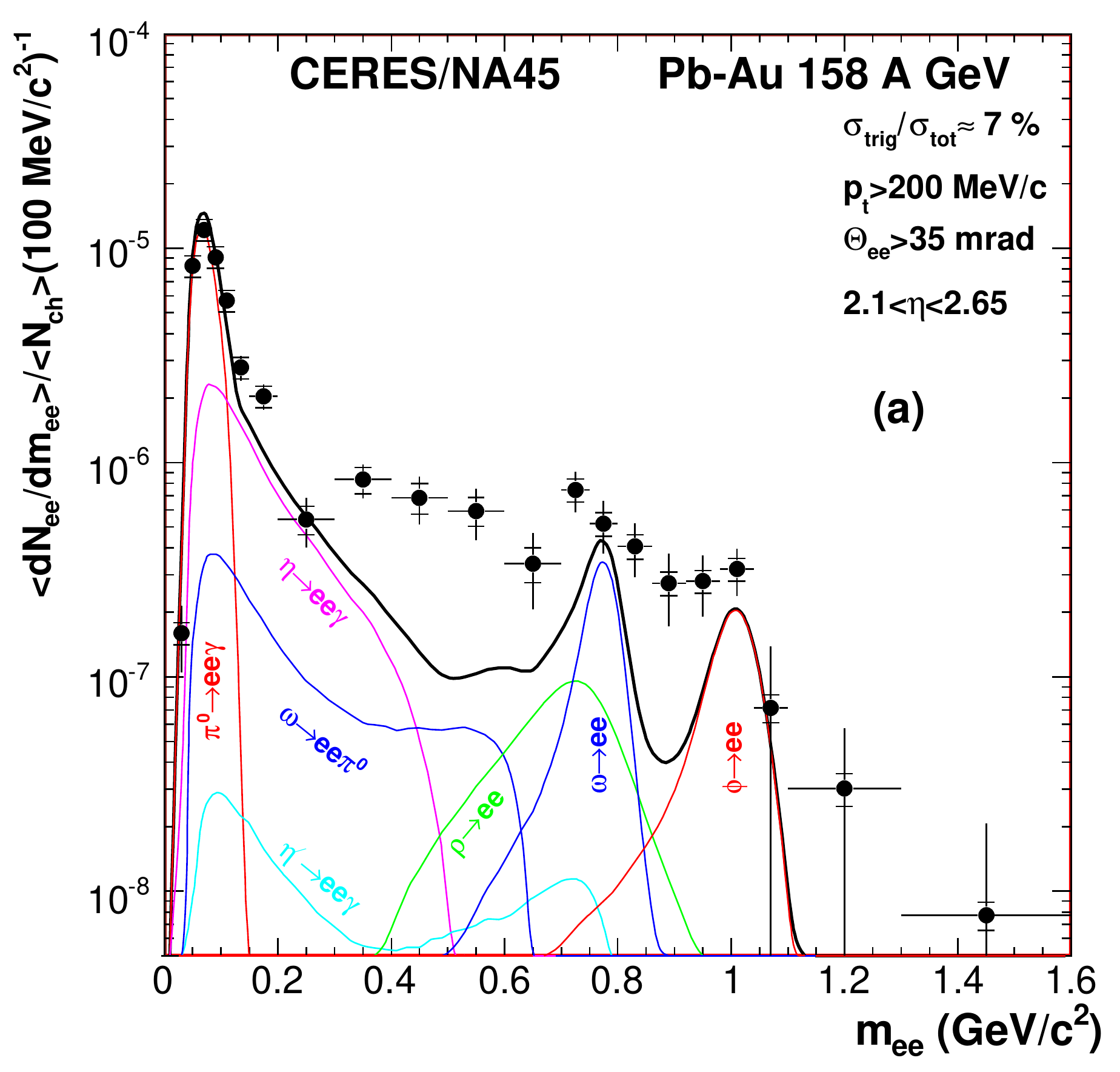}
  \caption{The enhancement of low-mass \ee pairs observed by CERES for the first time in S+Au collisions at 200~AGeV (left panel) \cite{ceres-s} and for the last time in Pb+Au collisions at 158~AGeV (right panel) \cite{ceres-pb2000} compared to the cocktail of known hadronic sources (solid black lines), showing also the individual contributions.}
  \label{fig:ceres}
\end{center}
\vspace{-0.5cm}
\end{figure}

The low-mass pair continuum has been systematically studied by the CERES experiment at CERN. In addition to the measurements in p+Be and p+Au at 450 GeV discussed in the previous section, CERES measured low-mass electron pairs in S+Au  at 200~AGeV \cite{ceres-s} and Pb+Au collisions at 158~AGeV \cite{ceres-pb2000,ceres-pb158,ceres-pb95-96} and at 40~AGeV \cite{ceres-pb40}. The most prominent feature observed in all heavy-ion collisions studied is the enhancement of electron pairs in the mass region m~=~0.2-0.6~GeV/c$^2$ whereas in p+Be and p+Au collisions the spectrum is well described by the known hadron decays. Fig.~\ref{fig:ceres} shows the effect as observed by CERES, for the first time in S+Au collisions (left panel) and in the last data set in Pb+Au collisions at 158~AGeV (right panel).
The data and cocktail in Pb+Au collisions reflect the much better mass resolution ($\Delta$m/m = 3.8\%) that resulted from the improved tracking with a doublet of silicon drift chambers close to the vertex \cite{ceres-si} and the upgrade of the CERES detector with the addition of a radial TPC downstream of the double RICH spectrometer \cite{ceres-tpc}. With the improved mass resolution a hint of the $\omega$ and $\phi$ meson peaks is seen for the first time in the CERES Pb+Au data. The results are presented in the form of electron pair production per event and per charged particle within the spectrometer acceptance and are compared to the cocktail of known hadronic sources. The yield is clearly enhanced with respect to the cocktail. The enhancement factor, defined as the ratio of the measured over the calculated yield in the mass range m~=~0.2~-~1.1~GeV/c$^2$, is similar in both cases and for the Pb case amounts to 2.45$\pm$0.21(stat)$\pm$0.35(syst)$\pm$0.58(decays) (the last error represents the uncertainties in the ingredients of the cocktail calculation). CERES has also shown that this excess is mainly due to soft p$_T$ pairs and that it increases faster than linearly with the event multiplicity \cite{ceres-pb95-96}.

\begin{figure}[!h]
\begin{center}
  \vspace{2mm}
    \includegraphics[width=72mm]{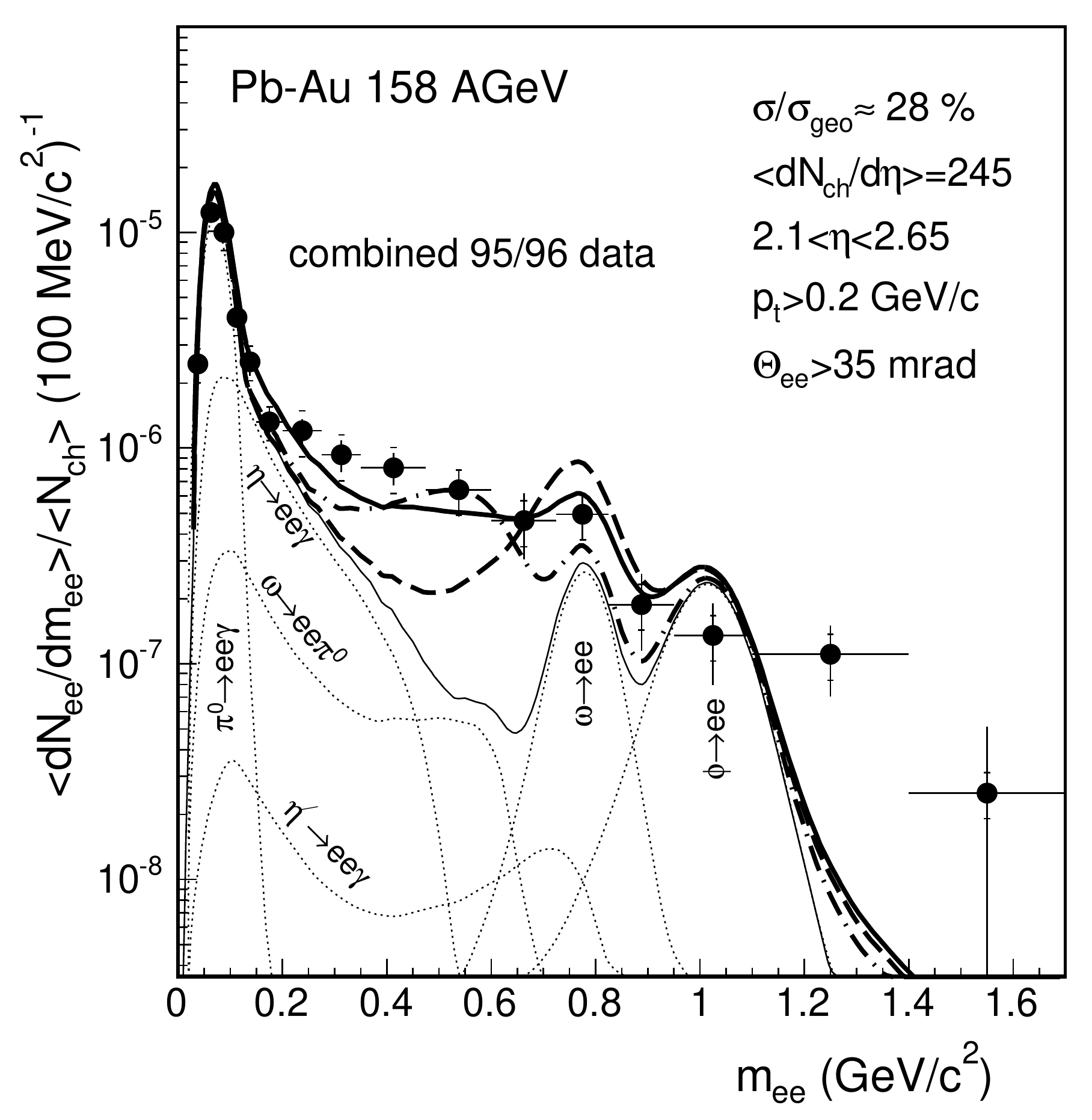}
      \vspace{0.0cm}
     \includegraphics[width=72mm, height=72mm]{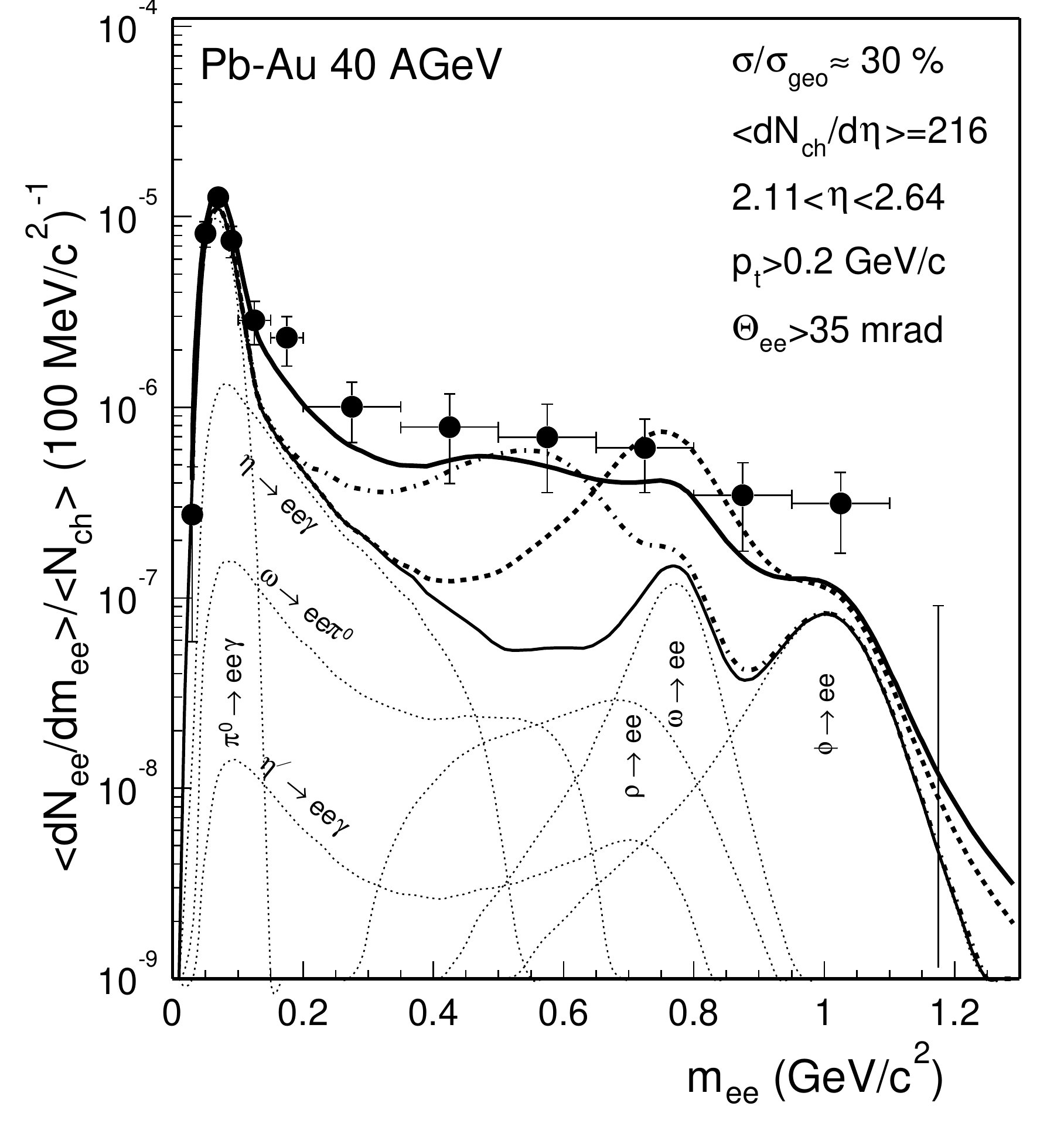}
  \caption{Inclusive \ee mass spectrum measured by CERES in central Pb+Au collisions at 158~AGeV (left panel) \cite{ceres-pb95-96} and 40~AGeV (right panel) \cite{ceres-pb40} compared to the hadronic cocktail (thin solid line). The various components of the cocktail are represented by the dotted lines. The figures show also model calculations including the \pipi annihilation assuming the vacuum $\rho$ spectral function (dashed), dropping $\rho$ mass (dash-dotted) and in-medium $\rho$ broadening (solid). }
\label{fig:ceres-95-96}
\end{center}
\end{figure}

This low-mass dilepton enhancement is one of the highlights of the heavy-ion program at CERN SPS. Its possible connection to CSR triggered a wealth of theoretical activity (for recent reviews see \cite{rapp-wambach,brown-rho2002,gale-haglin03}). The prime candidate to explain the excess was the thermal radiation from the hadronic phase, dominated by the two-pion annihilation ($\pi^+\pi^-\rightarrow \rho \rightarrow e^+e^-$). This channel contributes a substantial yield of \ee pairs below the $\rho$ mass. However all attempts using the vacuum $\rho$ properties failed to quantitatively reproduce the data. To do so, it was necessary to introduce in-medium modifications of the intermediate $\rho$ meson. Two main venues were used: (i) a decrease of the $\rho$-meson mass in the dense fireball \cite{li-ko-brown95} as a precursor of CSR, following the original Brown-Rho scaling \cite{brown-rho}. In this scenario, the $\rho$-meson mass scales with the quark condensate $<\overline{q}q>$ and the latter drops due to the high baryon density (rather than high temperature) of the medium and (ii) a broadening of the $\rho$-meson spectral function resulting from the scattering of the $\rho$ meson mainly off the baryons in the dense hadronic medium \cite{rapp-wambach,chanfray-rw96,rapp-chanfray-wambach97}. Both approaches rely on the high baryon density at mid-rapidity which, at CERN energies, mainly originates from baryon stopping, and both achieved good agreement with the CERES data in the mass region m~=~0.2-0.6~GeV/c$^2$ as illustrated in the left panel of Fig.~\ref{fig:ceres-95-96} \cite{ceres-pb95-96}.

The importance of baryon density prompted the CERES collaboration to measure low-mass electron pairs in Pb+Au collisions at the lower energy of 40~AGeV where a higher baryon density and a lower temperature are expected at mid-rapidity compared to full SPS energy. The results are shown in the right panel of Fig.~\ref{fig:ceres-95-96}. In line with the theoretical expectations, an even stronger, or at least similar, enhancement is observed.  The enhancement factor  with respect to  the hadronic cocktail, is 5.1$\pm$1.3(stat)$\pm$1.0(syst)$\pm$1.5(decays) in the mass range m~=~0.2~-~1~GeV/c$^2$. The enhancement is equally well reproduced by the two models, dropping mass and broadening of the $\rho$, mentioned above \cite{ceres-pb40}. However, the errors bars are too large and preclude a more critical assessment.

The success of these two different approaches, one relying on quark degrees of freedom, with a direct link to CSR, and the other one based on a many-body hadronic model, attracted much debate. The observation that at high temperatures close to T$_C$, the dilepton production rates calculated within the hadronic approach become very similar to the \qqbar annihilation rate computed within perturbative QCD  raised the interesting hypothesis of quark-hadron duality down to relatively low-masses \cite{rapp-wambach,rapp-duality} and provided a more direct connection of the widening of the $\rho$ meson spectral function to CSR. This duality will be further discussed below and in the context of the IMR results in Section~\ref{sec:imr}.

\begin{figure}[!h]
\begin{center}
      \includegraphics[width=73mm]{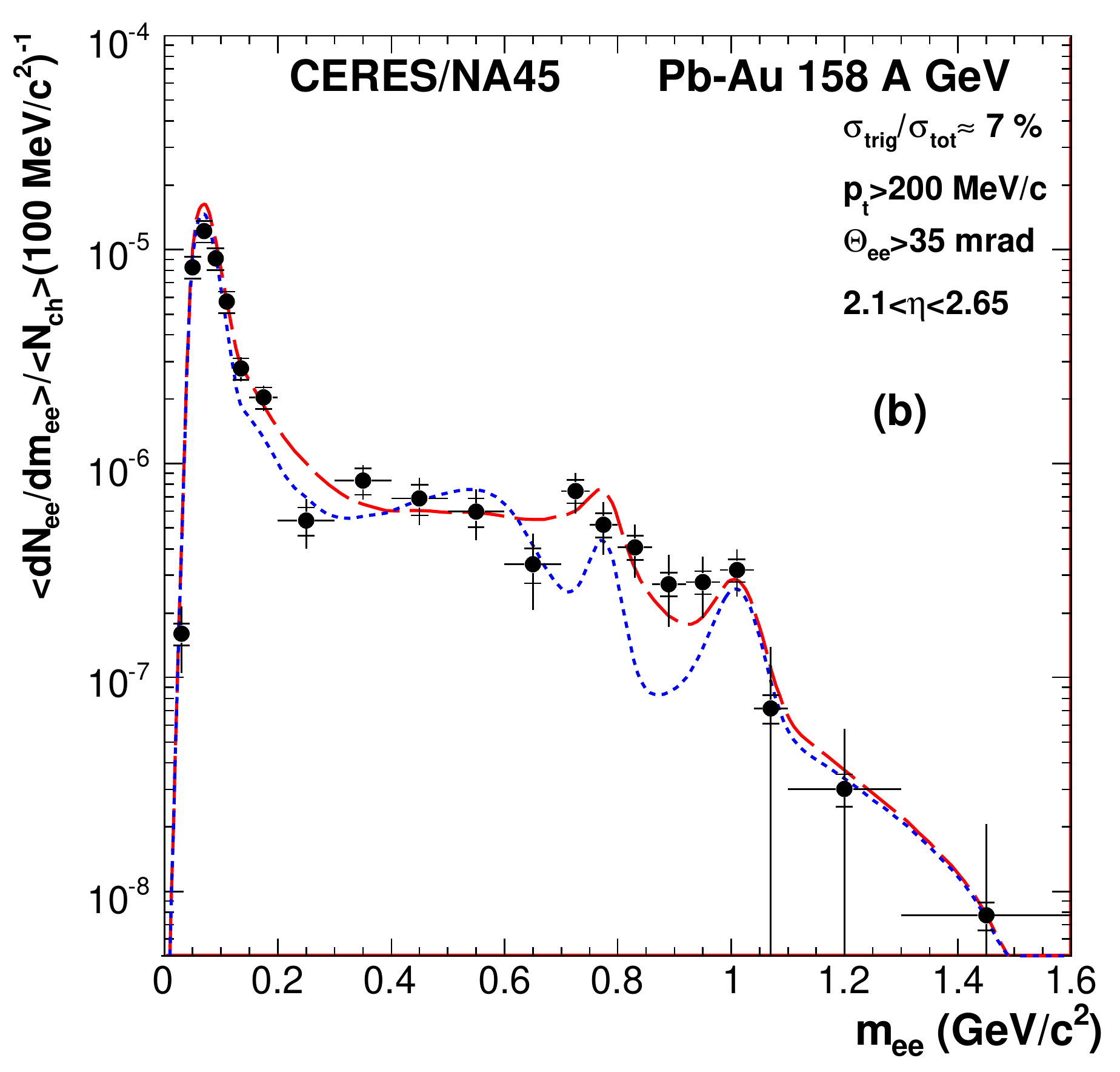}
      \vspace{0.0cm}
       \includegraphics[width=73mm]{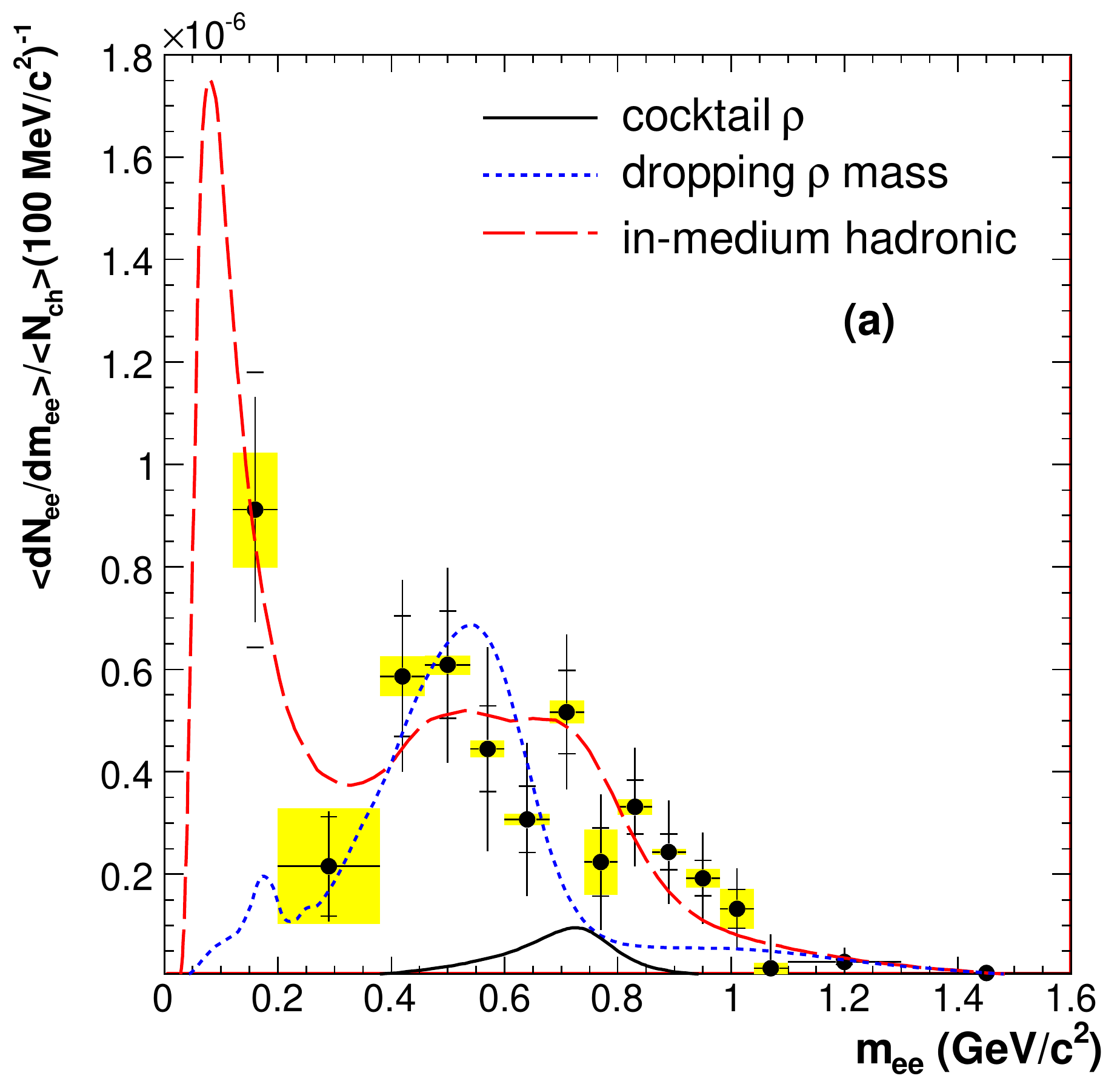}
  \caption{CERES invariant mass spectrum of \ee pairs in Pb+Au collisions at 158~AGeV (left panel) compared to calculations based on a dropping $\rho$ mass model (dashed line) and on in-medium $\rho$ spectral function (long-dashed line). The right panel shows the same data and the same calculations after subtracting the hadronic cocktail (excluding the $\rho$ meson) \cite{ceres-pb2000}.}
  \label{fig:ceres-2000}
\end{center}
\vspace{-0.5cm}
\end{figure}

The quality of the CERES data, prior of the upgrade, was insufficient to discriminate among the two approaches. A breakthrough was recently achieved primarily with the high quality data of NA60 (discussed in the next section) \cite{na60-rho} and confirmed by the last CERES results with the upgraded spectrometer shown in Fig.~\ref{fig:ceres-2000} \cite{ceres-pb2000}.
The results are compared to dropping mass and broadening calculations \cite{ceres-rapp}. Whereas the two models give very similar results for masses m $<$ 0.8 GeV/c$^2$ where the precision of the data is still insufficient to discriminate between them, the data between the $\omega$ and the $\phi$ favor the broadening scenario. This is better seen in the right panel, where the excess mass spectrum is plotted {\it a-la} NA60, after subtracting from the data the hadronic cocktail excluding the $\rho$ meson, and compared to the same model calculations.
It is interesting to note that a calculation by Kaempfer based on a parametrization of the dilepton yield in terms of \qqbar annihilation inspired by quark-hadron duality reproduces the data equally well \cite{kaempfer01}.

\subsubsection{NA60 results}
NA60 measured dimuons from threshold up to $\sim$5 GeV/c$^2$ in In+In collisions at 158~AGeV. The  primary motivation of the experiment was to elucidate the origin of the dilepton excess in the LMR discussed in the previous section and in the IMR discussed in Section~\ref{sec:imr}. The detector, based on the NA50 spectrometer, was supplemented by a silicon pixel vertex tracker \cite{na60-telescope} that improved the mass resolution to 2.2\% at the $\phi$ mass and  significantly reduced the combinatorial background produced by muons from $\pi$ and K decays. The vertex tracker also allowed distinguishing prompt dimuons from displaced vertex dimuons, a feature essential to clarify the origin of the dimuon excess at intermediate masses (see Section \ref{sec:imr-sps}).

\begin{figure}[!h]
\begin{center}
  \vspace{0.3cm}
     \includegraphics[width=72mm, height=72mm]{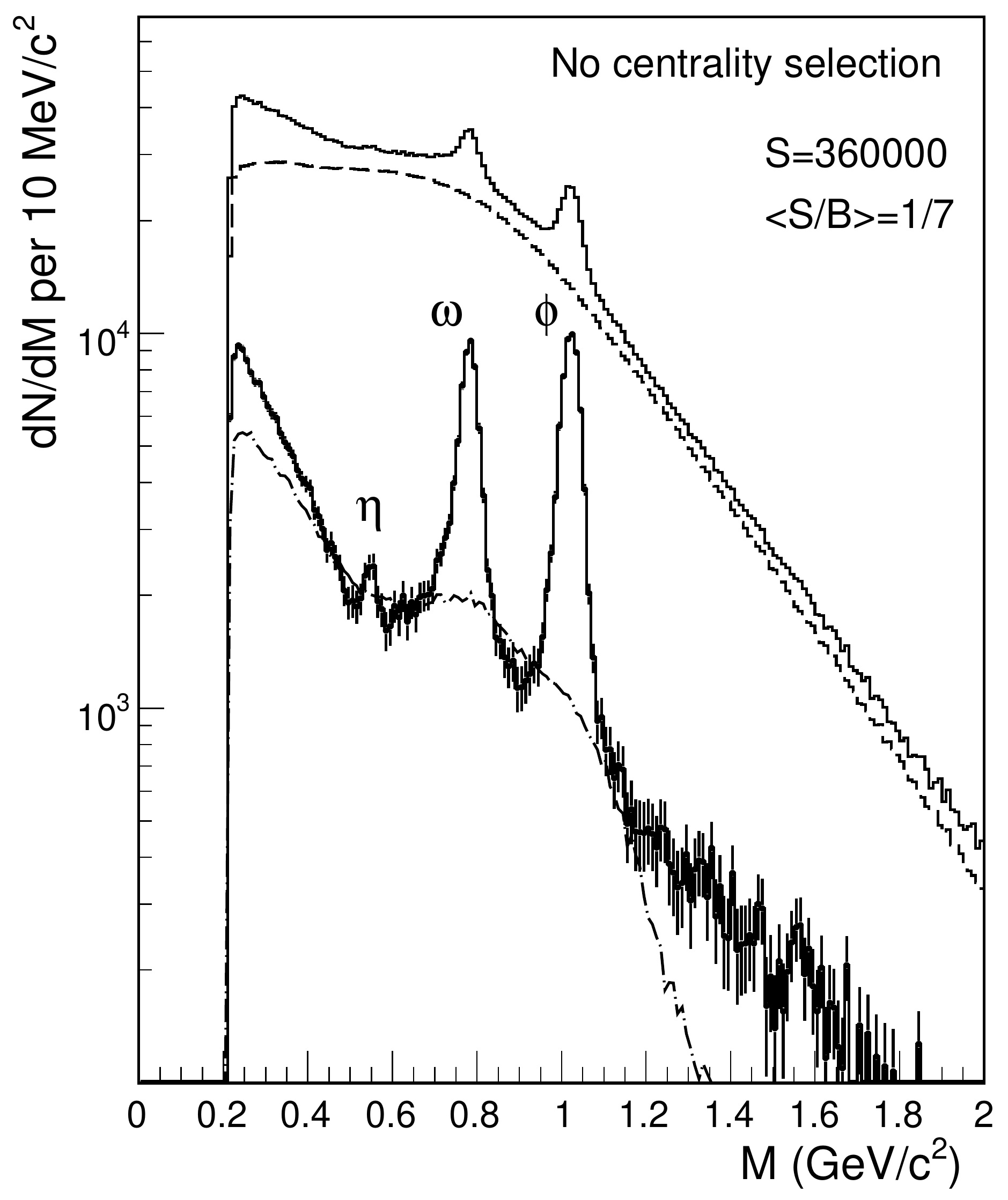}
      \vspace{0.0cm}
     \includegraphics[width=74mm, height=72mm]{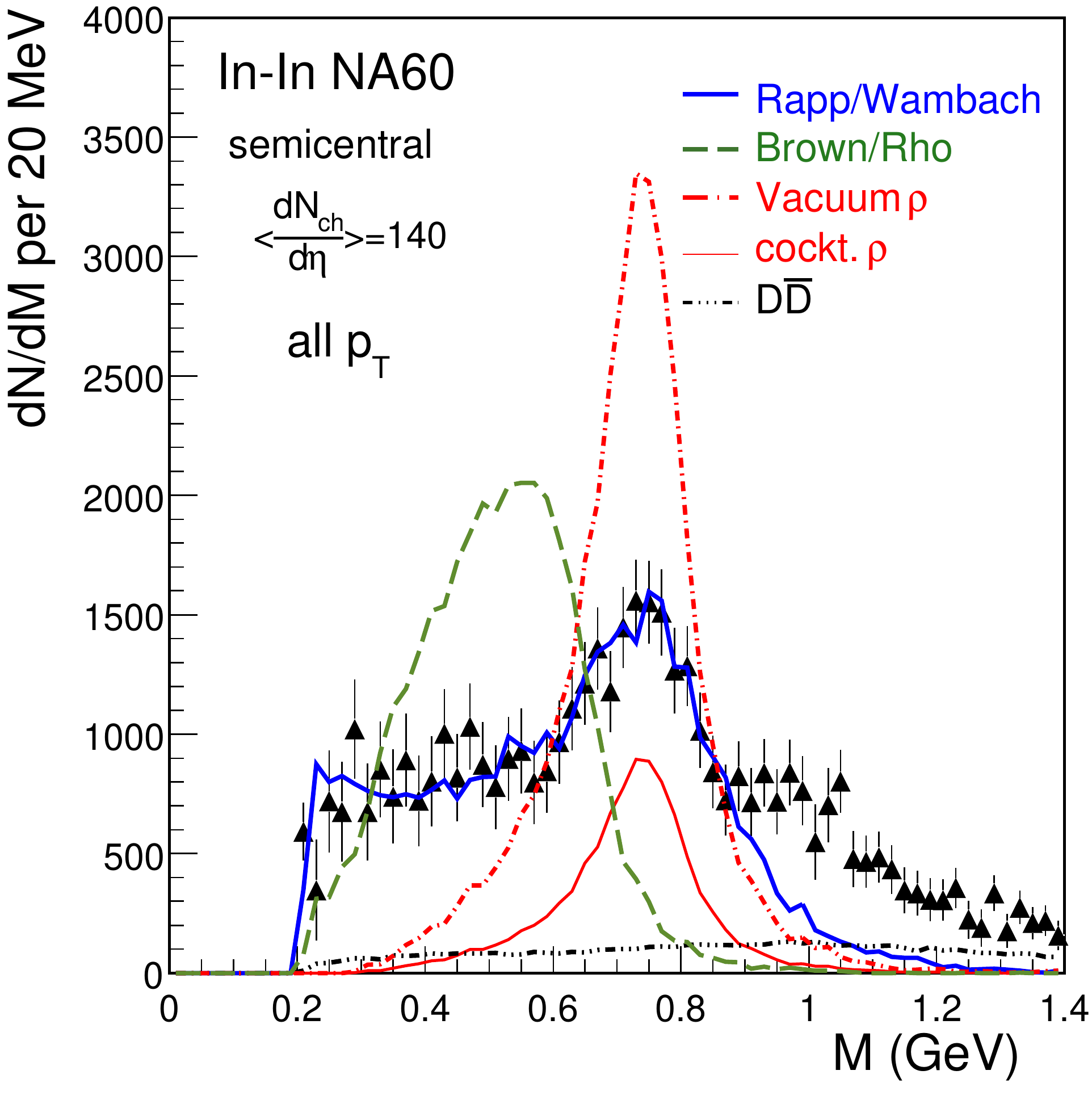}
  \caption{NA60 data in 158~AGeV In+In collisions. Left panel: invariant mass spectra of unlike sign dimuons (upper histogram), combinatorial background (dashed line), fake signal (dashed-dotted line) and the resulting signal (lower histogram) \cite{na60-rho}. Right panel: excess dimuons after subtracting the hadronic cocktail, excluding the $\rho$, compared to cocktail $\rho$ (thin solid line), $\pi^+\pi^-$ annihilation with an unmodified $\rho$ (dash-dotted line), dropping $\rho$ mass (dashed line) and in-medium $\rho$ broadening (thick solid line) \cite{na60-hp08}.}
  \label{fig:na60}
\end{center}
\end{figure}

\begin{figure}[!h]
  \vspace{-8mm}
  \begin{center}
     \includegraphics[width=100mm]{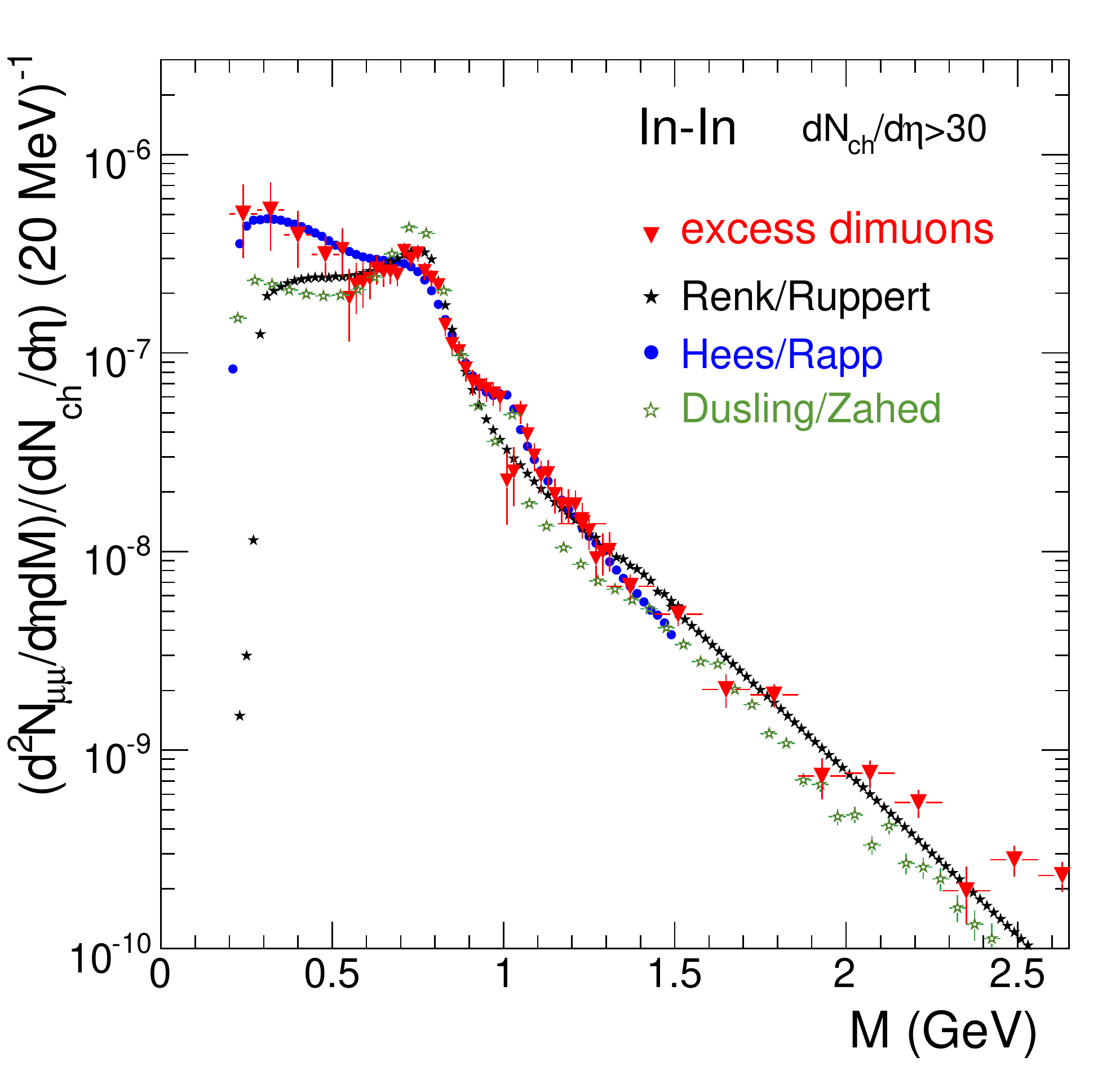}
  \end{center}
  \vspace{-8mm}
  \caption{Absolutely normalized excess dimuon mass spectrum corrected for acceptance and reconstruction efficiency, measured by NA60 in In+In collisions at 158 AGeV and compared to theoretical calculations of Renk/Ruppert \cite{ruppert-prl}, Hees/Rapp \cite{hees-rapp08,hees-rapp06} and Dusling/Zahed \cite{zahed07}. Both the data and the calculations are subject to a \pt cut of 200 MeV/c on the single muon tracks \cite{na60-imr,na60-erice}.}
  \label{fig:na60-abs-norm}
\end{figure}

The left panel of Fig.~\ref{fig:na60} exhibits the superb quality of the raw dimuon data in terms of resolution and statistics \cite{na60-rho}.
After subtracting the combinatorial background the spectrum contains a total of 360 000 dimuons and the resonances $\omega$ and $\phi$ are clearly visible and resolved; even the $\eta$ decay into \mumu is seen. Whereas the spectrum in peripheral collisions is well reproduced by the cocktail of expected hadronic sources -resonance decays of the $\eta, \rho, \omega, \phi$ and Dalitz decays of the $\eta, \eta^{'}, \omega$-, this is not so any longer for more central collisions. The data show a clear excess of dimuons which increases with centrality and is more pronounced at low pair \pt. The results confirm, and are consistent with, the CERES results described previously. The NA60 data quality allows to go one step further. The excess mass spectrum, obtained after subtracting from the data the hadronic cocktail without the $\rho$ is shown in the right panel of Fig.~\ref{fig:na60} (see Refs. \cite{na60-rho, na60-flow, na60-hp08} for details on the subtraction procedure). The excess represents the space-time average of the $\rho$ spectral function convoluted with the photon propagator and the Boltzmann factor and filtered through the acceptance of the NA60 apparatus.
It exhibits a peak at the nominal position of the $\rho$ meson mass, seating on top of a broad structure whose width and yield increase with centrality. The figure shows also a comparison of the excess with the two main models discussed above in the context of the CERES data: in-medium $\rho$ broadening (thick solid line) according to the Rapp-Wambach model \cite{rapp-wambach,chanfray-rw96,rapp-chanfray-wambach97} and dropping $\rho$ mass (dashed line) according to the Brown-Rho scaling \cite{brown-rho2002,brown-rho}. For reference, the figure also shows the calculated contributions of the cocktail $\rho$ , the vacuum $\rho$ (dash-dotted line), and the charm decays (dash-double dotted line). All the curves were obtained after applying the NA60 acceptance filter to the theoretical calculations. The comparison clearly rules out not only the $\pi^+\pi^-$ annihilation with a vacuum $\rho$ which was already ruled out by the CERES data, but also the dropping $\rho$ mass scenario. The predictions based on the many-body approach of Rapp and Wambach reproduce very well the data for \mmumu $\leq$ 0.9GeV/c$^2$. These early calculations were focussed on the low masses and were not intended to reproduce the yield at masses above the $\rho$. Recent calculations of the same group, that include contributions from multi-pion states, mainly 4$\pi$ states, reproduce the entire mass spectrum as shown in Fig.~\ref{fig:na60-abs-norm} \cite{hees-rapp08,hees-rapp06}.

\begin{figure}[!h]
  \begin{center}
     \includegraphics[width=100mm]{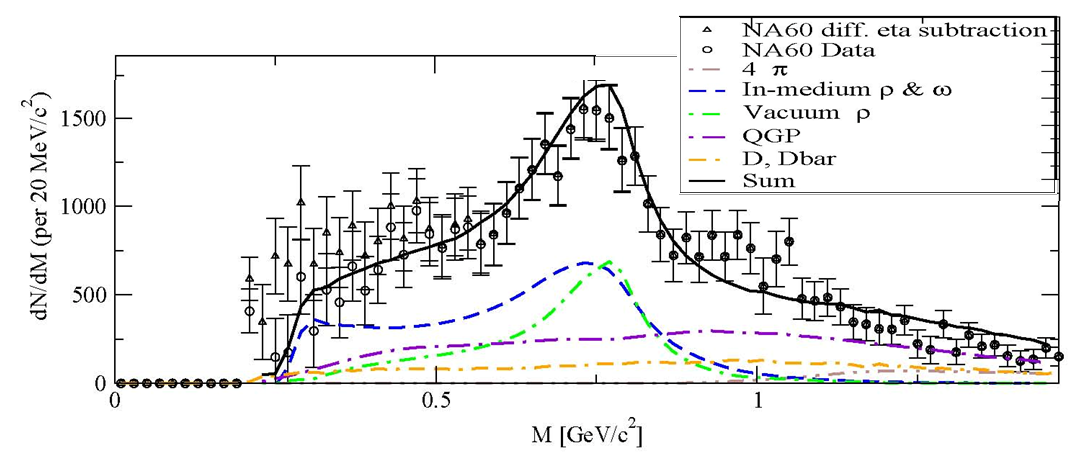}
  \end{center}
  \vspace{-5mm}
  \caption{The NA60 dimuon excess in semi-central In+In collisions at 158~AGeV, integrated over all \pt compared to calculations of \cite{ruppert-prl}. The curves represent partial contributions and their sum as indicated in the figure.}
  \label{fig:na60-ruppert}
\end{figure}

Other recent calculations reproduce quite well the NA60 excess \cite{ruppert-prl,zahed07}. In Ref. \cite{ruppert-prl}, excellent agreement is found over the entire mass range as demonstrated in Fig.~\ref{fig:na60-ruppert}. The low-mass excess is described also here by the in-medium broadening of the $\rho$ spectral function but at masses above
1~GeV/c$^2$ the thermal radiation from the QGP, and not the 4$\pi$ annihilation, is found to dominate. Ref. \cite{zahed07} provides a very good description of the peripheral and semiperipheral mass excess with hadronic rates constrained by chiral symmetry arguments and experimental data.

The excess dimuon mass spectrum, fully corrected for acceptance and reconstruction efficiency, with a \pt cut of 200 MeV/c on the single muon tracks is shown in Fig.~\ref{fig:na60-abs-norm}. The figure displays the excess in the LMR discussed here as well as  the excess observed in the IMR discussed in Section \ref{sec:imr-sps}. The figure also shows the calculations from the three groups mentioned above \cite{na60-imr, na60-erice}.

The NA60 data pose a new constraint on the theoretical models which have to simultaneously account for the CERES and NA60 results. But the deeper impact of these results is their possible relevance to the broader context of chiral symmetry restoration. If the system reaches, or is near to, chiral symmetry restoration then the dilepton results could be telling us that the approach to such a state proceeds through broadening and eventually subsequent melting of the resonances rather than by dropping masses.

\subsection{Low-mass dileptons at RHIC}
\label{subsec:rhic}

RHIC allows the study of Au+Au collisions up to an energy of \sqnR. This energy, more than one order of magnitude higher than the SPS top energy of \sqn = 17.6 GeV, offers much better conditions for studying the QGP. Higher collision energies result in higher initial temperatures or equivalently higher initial energy densities that lead to longer lifetimes and larger volumes allowing for a better study of the system before it freezes out. Based on the insight gained at the SPS with CERES and NA60, the study of low-mass dileptons at RHIC was anticipated to be very interesting. The key factor responsible for in-medium modifications of the $\rho$ meson at SPS energies, as discussed in Section \ref{subsec:sps}, is the baryon density or to be more precise the {\it total} baryon density. The latter is almost as high at RHIC as at SPS. At SPS, most of the baryons at mid-rapidity are participant nucleons. At RHIC there is a strong decrease in nuclear stopping but this is compensated by a copious production of baryon-antibaryon pairs such that the total baryon density is practically the same in both cases \cite{it-pramana03}. Consequently, the same model calculations that successfully reproduce the SPS results predict that the enhancement of low-mass electron pairs should persist at RHIC with at least comparable strength \cite{rapp-rhic}.

PHENIX is the only experiment at RHIC that can measure low-mass electron pairs. The mid-rapidity spectrometers
have very good electron identification capabilities by combining a RICH detector with an electromagnetic calorimeter \cite{phenix-nim}. The experiment has also an excellent mass resolution of $\sim$ 1\% at the $\phi$ mass, an essential requirement for precise spectroscopy studies of the light vector mesons. However, the experimental set-up is limited by a large combinatorial background. The strong magnetic field, starting at the collision vertex, causes a limited acceptance of low-momentum tracks in the central-arm detectors (which cover each 90$^o$ in the azimuthal direction), very often resulting in the detection of only one of the two tracks from $\pi^0$ Dalitz decays and $\gamma$ conversions which leads to an overwhelming yield of combinatorial background pairs. With a \pt~cut of 200~MeV/c on single electron tracks, the signal to background ratio is of the order of S/B $\sim$1/200 at \mee~=~400~MeV/c$^2$ in minimum bias Au+Au collisions at \sqnR.

 \begin{figure}[!h]
  \begin{center}
    \includegraphics[width=100mm]{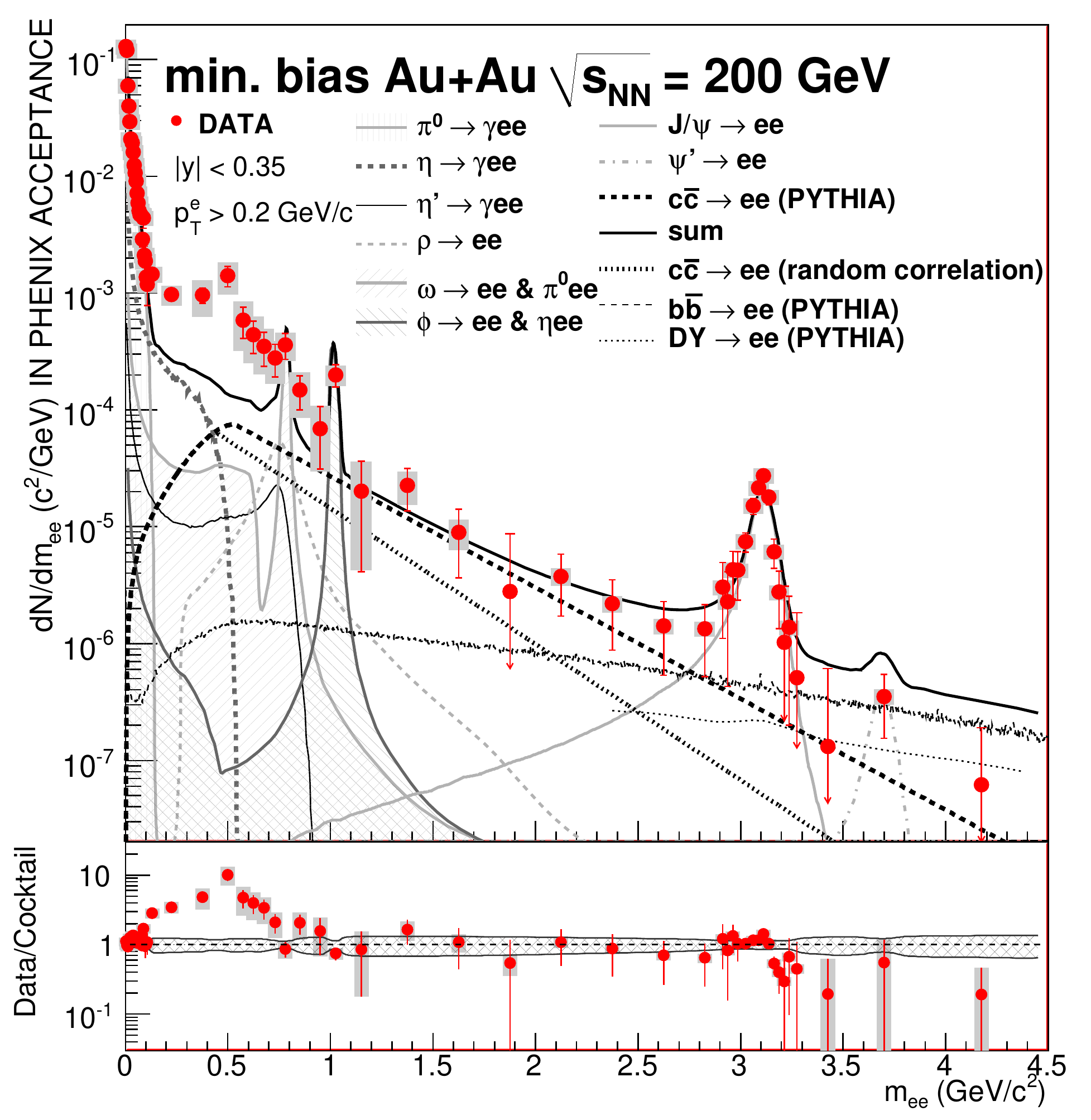}
  \end{center}
  \vspace{-5mm}
  \caption{Invariant mass \ee spectrum measured by PHENIX in \sqnR minimum bias Au+Au collisions at mid-rapidity. The data are compared to the cocktail of expected yields from light mesons and semi-leptonic open charm decays. Statistical (bars) and systematic (boxes) errors are plotted separately. The bottom panel shows the data to cocktail ratio with the band around 1 representing the systematic uncertainty in the cocktail  \cite{phenix-lmr}. }
  \label{fig:phenix-auau-mb}
\end{figure}

In spite of that, in a real tour de force, PHENIX mastered the mixed event technique for the evaluation of the combinatorial background to an unprecedented precision of 0.25\%. Fig.~\ref{fig:phenix-auau-mb} shows the first results of PHENIX on  \ee pairs measured in Au+Au collisions at \sqnR  after subtracting the combinatorial background \cite{phenix-lmr}. The expected yields from the light mesons, calculated using the EXODUS event generator, and from the correlated semi-leptonic decays of open charm, calculated using PYTHIA and constrained by the measured \ccbar cross section, are also indicated in the figure.
At very low masses, up to $\sim$100~MeV/c$^2$, the agreement is excellent; at high masses, above the $\phi$ meson, it is pretty good also; in particular the continuum in the IMR is fully accounted by the contribution from charm decays. In the low-mass continuum, from 150 to 750~MeV/c$^2$ there is a considerable excess of \ee pairs, with an enhancement factor of 3.4$\pm$0.2(stat.)$\pm$1.3(syst.)$\pm$0.7(model) where the last error reflects the systematic uncertainty in the total calculated yield.

\begin{figure}[!h]
  \begin{center}
    \includegraphics[width=100mm]{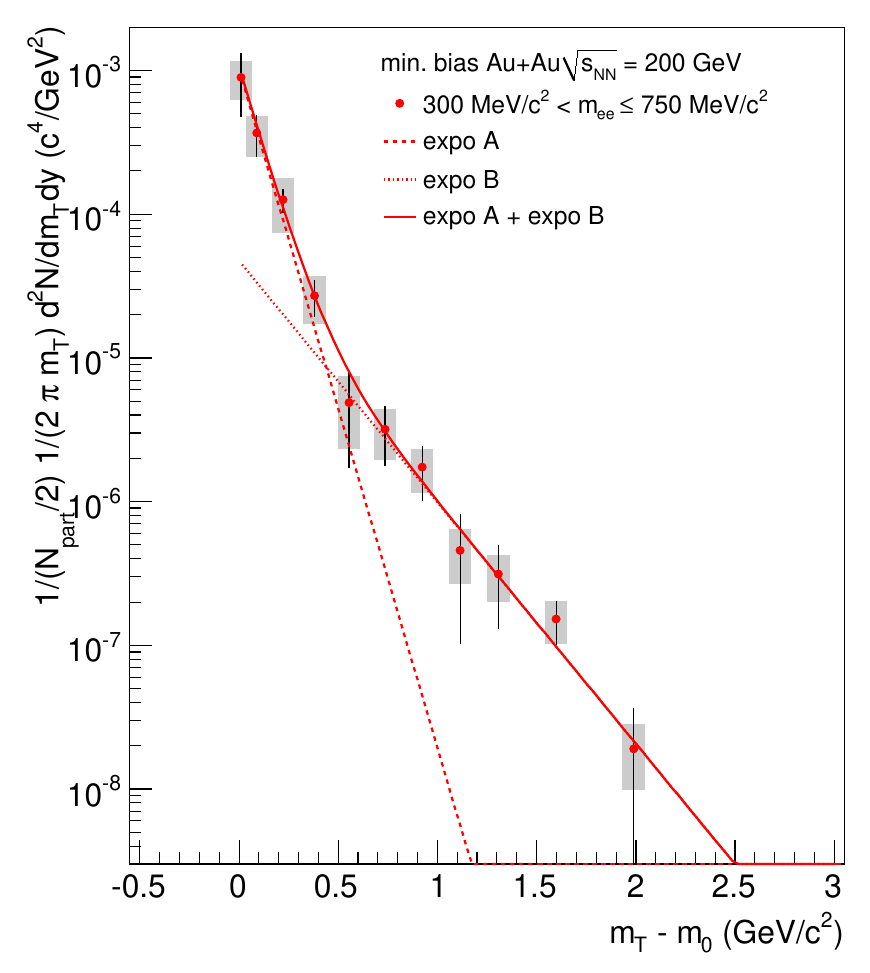}
  \end{center}
  \vspace{-5mm}
  \caption{Invariant $m_T$ spectrum of the \ee pair excess (after subtracting the cocktail and the charm contribution) measured in the mass range 0.3 $< m_{ee} < 0.75$ GeV/$c^2$ by PHENIX in \sqnR minimum bias Au+Au collisions at mid-rapidity. The solid line represents the fit to the sum of two exponential functions shown separately by the dashed and dotted lines. Statistical (bars) and systematic (boxes) errors are plotted separately \cite{phenix-lmr}. }
  \label{fig:phenix-pt}
\end{figure}

The excess is present at all pair \pt but it is more pronounced at low pair \pt (\pt~$<$~0.7~GeV/c) \cite{phenix-lmr}. Fig.~\ref{fig:phenix-pt} shows the \pt dependence of the excess, after subtracting the hadronic cocktail and the charm contribution, plotted for the pair mass range $0.3 < m_{ee} < 0.75$ GeV/$c^2$ in the form of invariant yield vs $m_T-m_0$ ($m_T$ is the transverse mass $m_T=\sqrt{p_T^2+m_0^2}$), $m_0$ is the mean value of $dN/dm_{ee}$ in this mass range and the lower bound of 300 MeV/$c^2$ is chosen to avoid the region of no acceptance due to the \pt cut, \pt $>$ 200 MeV/c, on the single tracks). The distribution exhibits a clear change of slope at $m_T - m_0 \simeq$ 0.4 GeV/c and can be well described by the sum of two exponential distributions with inverse slope parameters T1 = 92 $\pm 11.4^{stat} \pm8.4^{syst}$ MeV and T2 = 258.3 $\pm 37.3^{stat} \pm 9.6^{syst}$ MeV.

\begin{figure}[!h]
  \begin{center}
    \includegraphics[width=110mm, height=100mm]{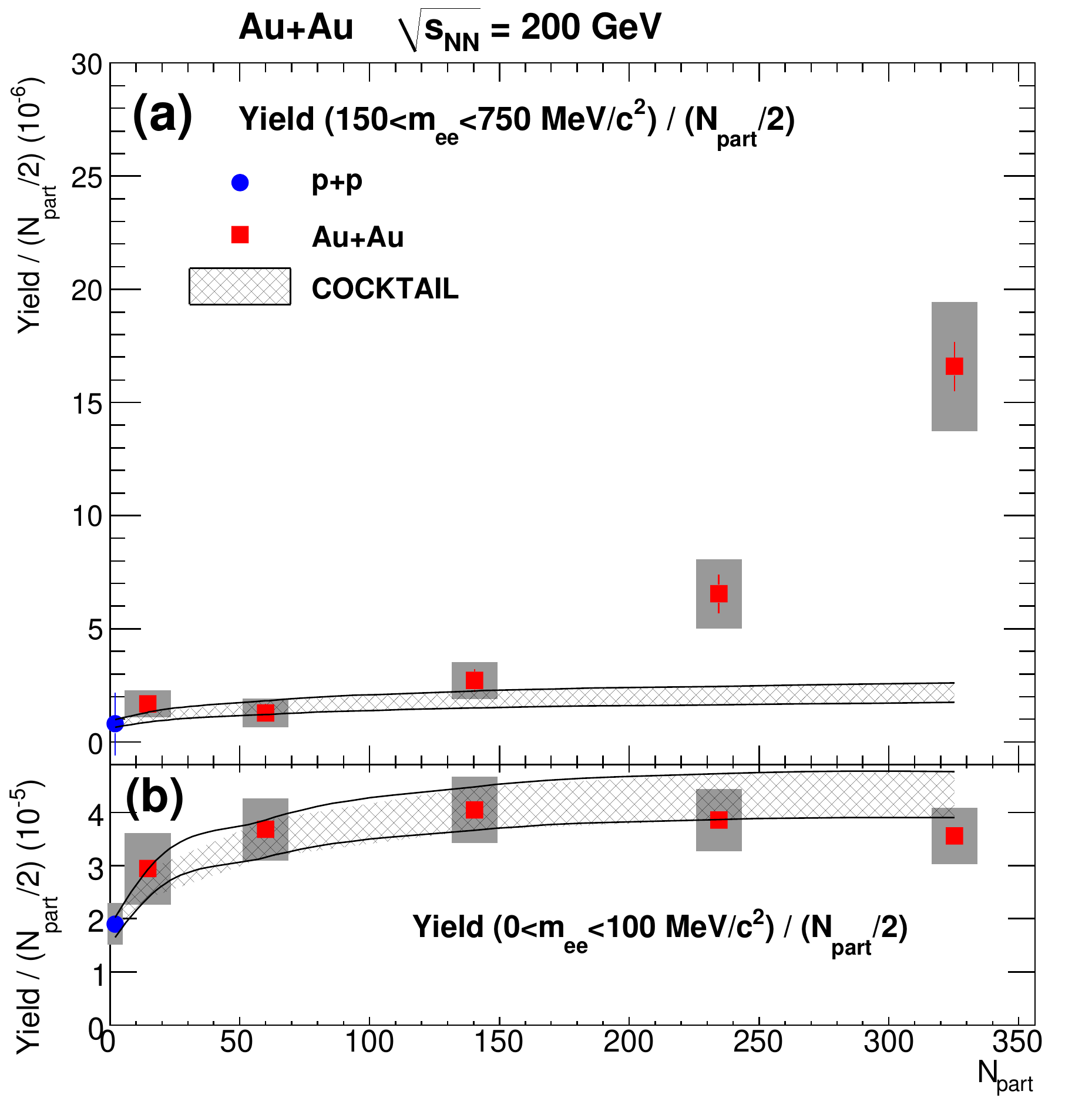}
  \end{center}
  \vspace{-5mm}
\caption{Integrated yield per pair of participating nucleons as function of $N_{part}$ compared to the calculated yield from the cocktail of light mesons and open charm decays, for two different mass windows, 0 $<$ \mee $<$ 100 MeV/c$^2$ (bottom panel) and  150 $<$ \mee $<$ 750 MeV/c$^2$ (top panel) from  PHENIX \cite{phenix-lmr}.}
  \label{fig:phenix-auau-cent}
  \vspace{-2mm}
\end{figure}

To further characterize this enhancement PHENIX has studied its centrality dependence. The integrated yield, divided by the number of participating nucleon pairs ($N_{part}$/2), is shown in Fig.~\ref{fig:phenix-auau-cent} as a function of $N_{part}$ for two mass windows: 0~$<$~\mee~$<$~100~MeV/c$^2$ dominated by the $\pi^0$ Dalitz decay and 150~$<$~\mee~$<$~750~MeV/c$^2$ where most of the enhancement is observed. The figure also shows the expected integrated yield in the same mass windows. Whereas the low-mass yield (bottom panel) is in very good agreement with the cocktail, reflecting the expected increase with the pion yield, a very strong centrality dependence is observed for the low-mass continuum (top panel). The enhancement appears concentrated in the two most central bins only, 0-10\%  where the enhancement reaches a factor of almost 8, and 10-20\%.

The PHENIX results in the LMR are intriguing. They appear different from those observed at the SPS. The excess is more spread towards lower masses compared to the SPS where the excess is closer to the $\rho$ meson (see right panels of Figs.~\ref{fig:ceres-2000} and \ref{fig:na60}). The centrality dependence is also different. Models that successfully reproduce the LMR results of CERES and NA60, based on a broadening of the $\rho$ meson spectral function fail to reproduce the PHENIX results \cite{phenix-lmr,cassing08,zahed08}.
The results are also intriguing in the IMR but these will be discussed later in Section~\ref{sec:imr}.

A new qualitative level in the PHENIX \ee data is expected with the recently installed Hadron Blind Detector (HBD) \cite{hbd1,hbd2,hbd-it}.
The PHENIX detector was designed anticipating that the measurement of low-mass \ee pairs would be feasible with an appropriate upgrade.
In particular, provision was made for the installation of an inner coil. When operated in the  $+-$ configuration, the inner coil counteracts the main field of the outer coil creating an almost field-free region close to the vertex, extending out to $\sim$50-60 cm in the radial direction. The HBD is located in this field-free region where the pair opening angle is preserved and its main task is to recognize and reject electron tracks originating from $\pi^0$ Dalitz decays and $\gamma$ conversions by exploiting their small opening angle. The HBD is a windowless Cherenkov detector operated with pure CF$_4$, in a proximity focus configuration. The detector consists of a 50~cm long radiator directly coupled to a triple GEM detector which has a CsI photocathode evaporated on the top face of the upper-most GEM foil and pad readout at the bottom of the GEM stack.
A large suppression of the combinatorial background, well above a factor of 20 depending on the cuts applied, was demonstrated with p+p data taken during the 2009 RHIC run. A similar improvement is expected in the 2010 RHIC run which is devoted to Au+Au collisions.

\subsection{Low-mass dileptons at low energies}
\label{subsec:low-e}

\begin{figure}[!h]
\begin{center}
       \includegraphics[width=70mm, height=70mm]{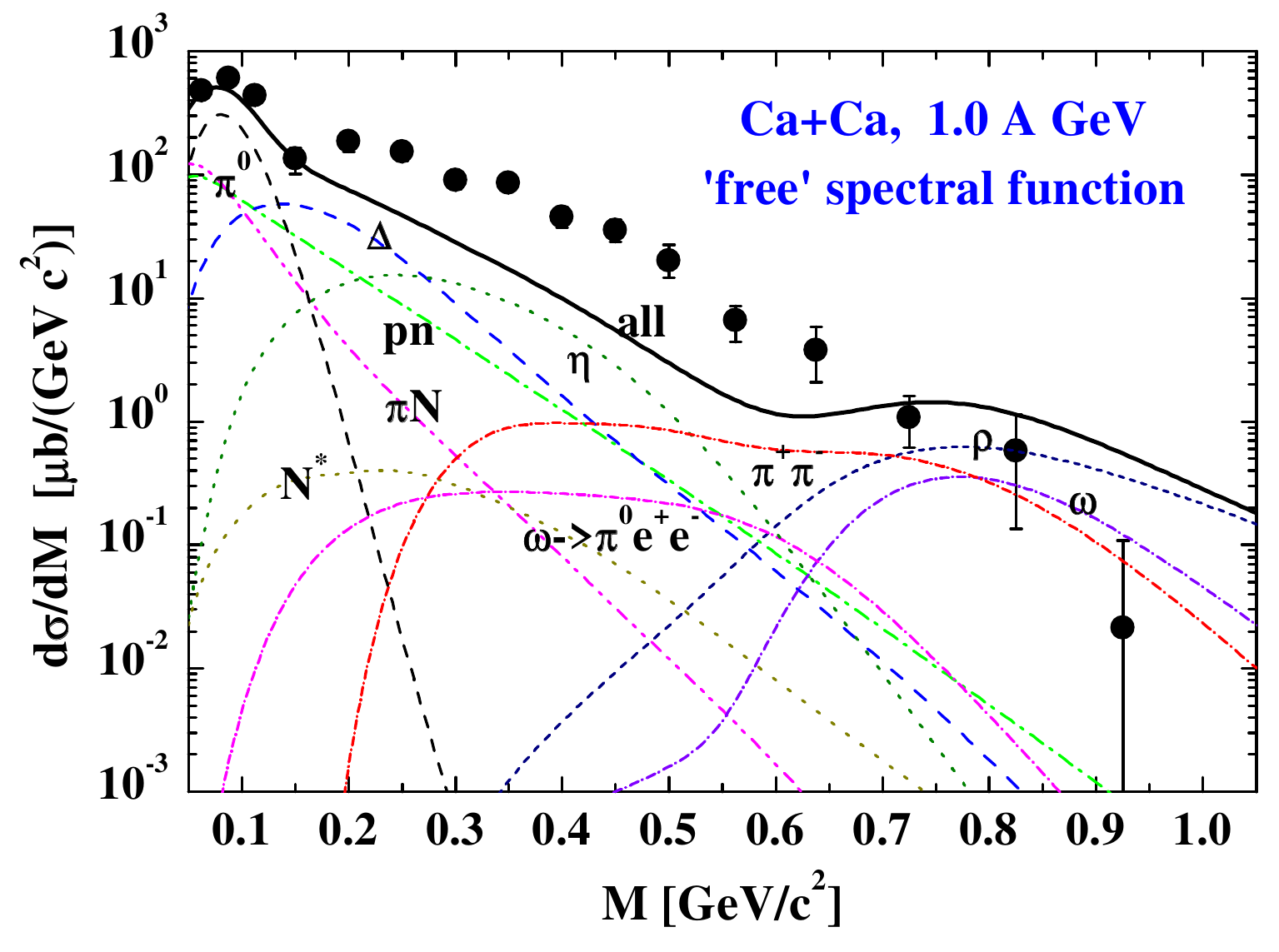}
  \hspace{0.3cm}
      \vspace{-0.4cm}
    \includegraphics[width=70mm, height=70mm]{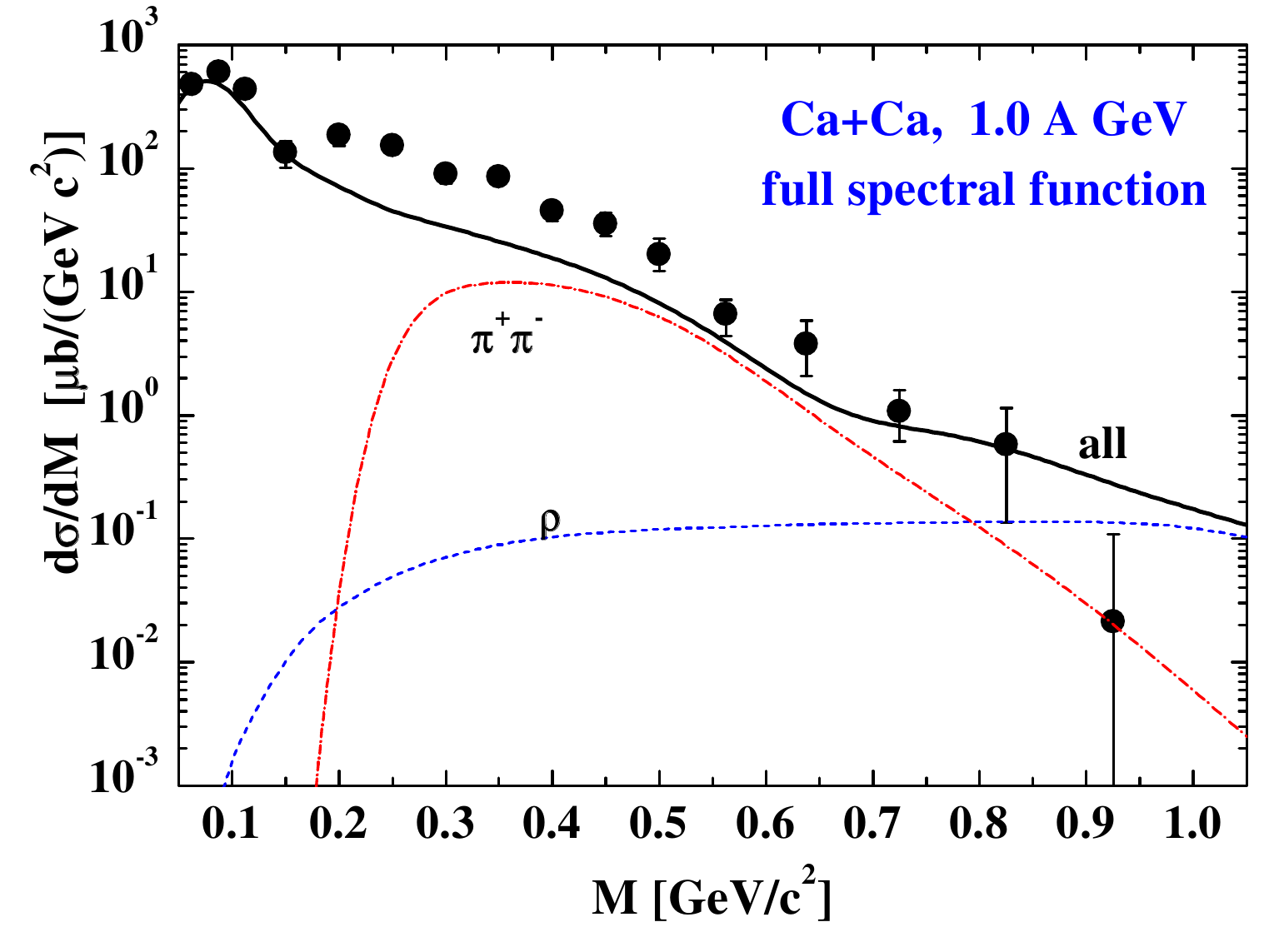}
\caption{DLS results in Ca+Ca collisions at 1~AGeV \cite{dls97} compared to HSD calculations \cite{bratkovskaya98}. The left panel uses the free $\rho$ meson spectral function whereas the right panel shows the calculations with the full $\rho$ spectral function. }
  \label{fig:dls}
\end{center}
\end{figure}

Low-mass electron pairs have also been measured at much lower energies. The DLS Collaboration measured low-mass \ee pairs in 1~AGeV C+C and Ca+Ca collisions at the BEVALAC \cite{dls97}. The Ca results shown in the left panel of Fig.~\ref{fig:dls} exhibit an enhancement of low-mass pairs in the mass range m = 0.2 - 0.6~GeV/c$^2$, in comparison to transport model calculations based on the HSD (Hadron String Dynamics) approach that include all the expected sources: hadron decays, proton-neutron (pn) and pion-nucleon ($\pi$N) bremsstrahlung and the pion annihilation ($\pi^+\pi^- \rightarrow \rho \rightarrow e^+e^-$) using the vacuum or free $\rho$ spectral function \cite{bratkovskaya98}. The enhancement looks qualitatively similar to the one observed at the SPS. But contrary to the situation at SPS energies, all attempts to reproduce it failed. The right panel of Fig.~\ref{fig:dls} compares the data with the same cocktail using this time the $\rho$ meson spectral function modified in the medium. Part of the enhancement is explained but the calculations are still a factor of 2-3 below the data \cite{bratkovskaya98}. Other attempts failed also \cite{ernst98,cassing99,shekhter03}. This intriguing situation led to the denomination "DLS puzzle" that persists for almost a decade and motivated a new experiment, HADES, at the GSI. HADES, was designed and built for systematic studies of low-mass dileptons in proton (p+p,d,A), pion ($\pi$+p,A) and heavy-ion (C+C, Ni+Ni, Au+Au) induced reactions at energies up to 8~AGeV.

\begin{figure}[!h]
\begin{center}
  \includegraphics[width=70mm, height=70mm]{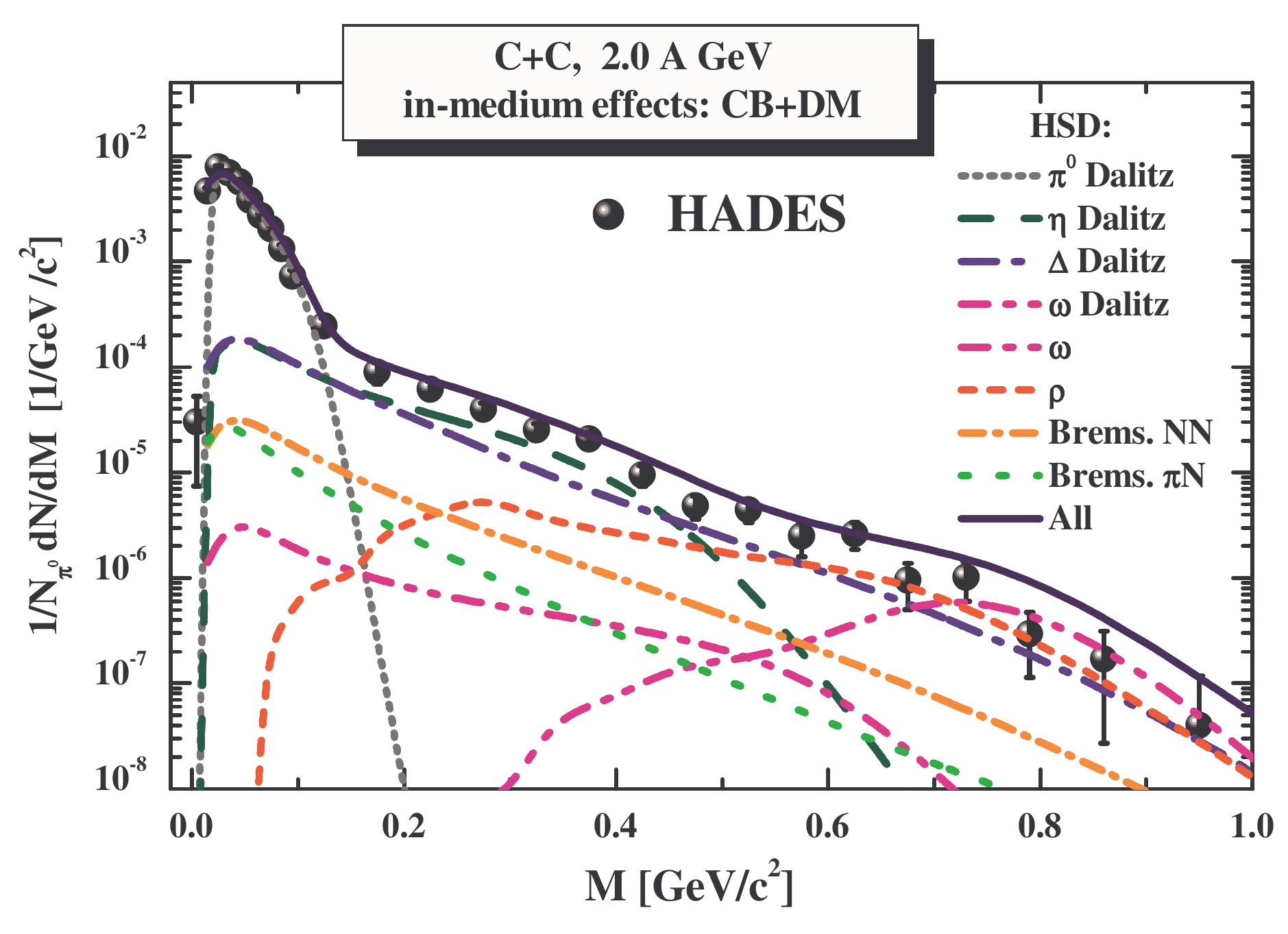}
      \vspace{0.5cm}
   \includegraphics[width=70mm]{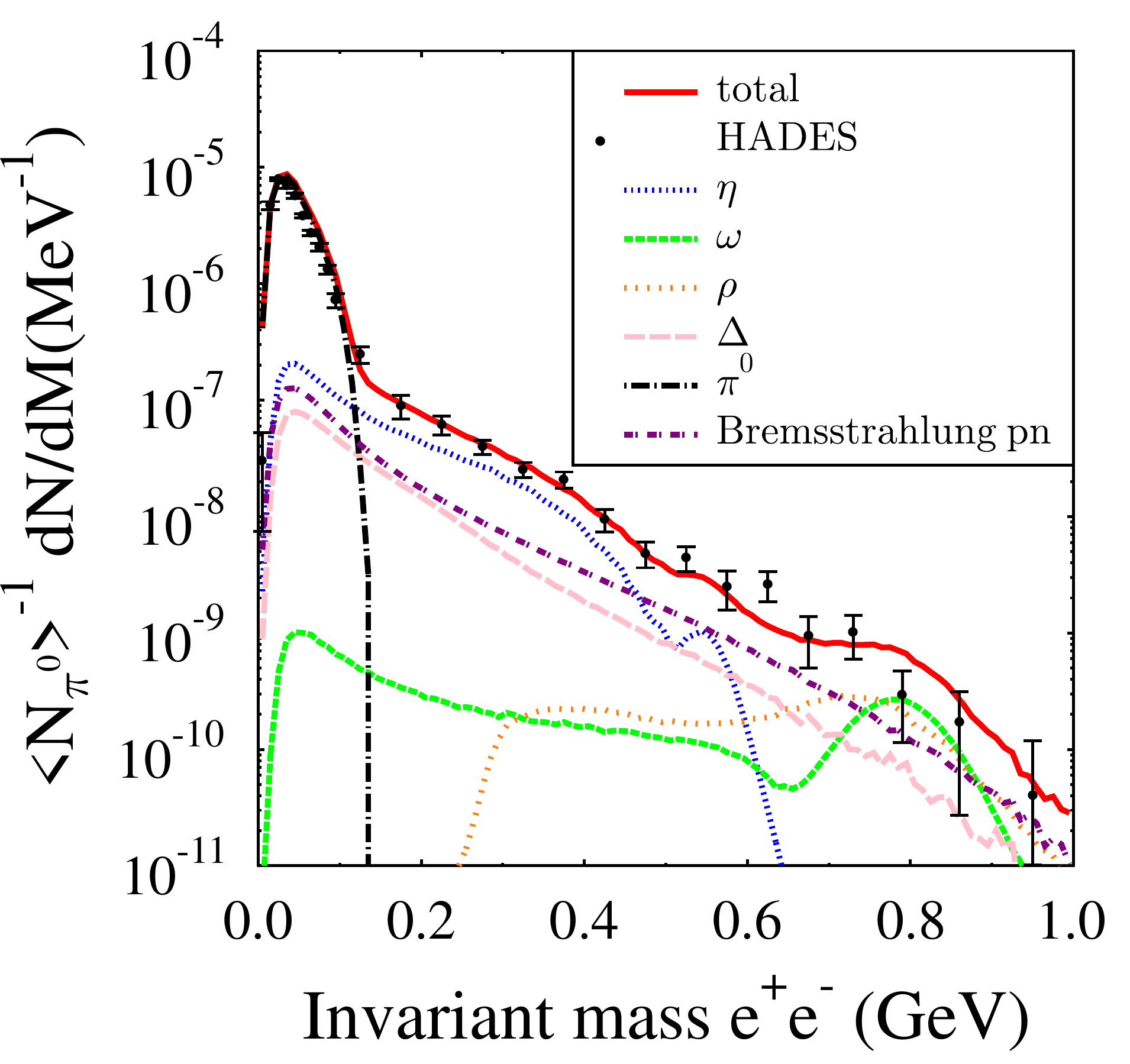}
  \vspace{-0.5cm}
\caption{Invariant mass \ee pair spectrum measured by HADES in C+C collisions at 2~AGeV \cite{hades-prl07} compared to model calculations. Left panel: new HSD calculations with an enhanced bremsstrahlung contribution and including collision broadening and dropping mass effects \cite{bratkovskaya08}. Right panel: IQMD simulations with a reduced $pn \rightarrow pn\omega$ cross section using vacuum spectral shapes for the $\rho$ and $\omega$ \cite{thomere07}.}
  \label{fig:brat}
\end{center}
\end{figure}

Several interesting developments have occurred over the last couple of years. From the experimental side, first results from HADES are now available \cite{hades-prl07,hades-plb08,hades-09}. They are in very good agreement with the DLS results and show an enhancement of low-mass dileptons in C+C collisions at 1 and 2~AGeV with respect to model calculations. The experimental results are therefore on a very robust ground. On the theoretical side, new transport calculations based on the HSD approach seem to get close to explaining the excess. Using an enhanced bremsstrahlung contribution, in line with recent OBE (one boson exchange) calculations \cite{kampfer06}, Bratkovskaya and Cassing achieved very good agreement with the published DLS and HADES data on C+C collisions, as illustrated in the left panel of Fig.~\ref{fig:brat} \cite{bratkovskaya08}. The same calculations applied to the Ca+Ca system are closer to the data (compared to the status of the previous HSD calculations shown on the right panel of Fig.~\ref{fig:dls}) but they still underestimate the measured yields.

Recently, Ref. \cite{thomere07} demonstrated again how crucial it is to have a complete and precise knowledge of the elementary reactions for a reliable interpretation of the more complex nuclear collisions. In particular, they showed that IQMD calculations are able to reproduce reasonably well the HADES C+C data at 2~AGeV by lowering the unknown cross section $pn \rightarrow pn\omega$ (cf. right panel of Fig.~\ref{fig:brat}) or/and invoking an in-medium mass decrease of the $\omega$ according to $m_{\omega} = m_{\omega}^0 (1 - 0.13\rho/\rho_0)$.

The measurement of elementary reactions is part of the HADES program  and the first dilepton measurements on p+p and d+p collisions at 1.25~AGeV were recently performed. The preliminary results are very interesting as they show that the 1/2(pp + pn) spectrum is in good agreement with the C+C spectrum at 1 AGeV, when both spectra are normalized to $\pi^0$. The C+C data can thus be understood as a superposition of NN collisions and there is no compelling evidence to invoke any new source of dilepton or any in-medium modification of mesons in this light system and at this low energy \cite{hades-09, witold-panic08}. This solves the DLS puzzle in the sense that there is no excess of dileptons at 1-2 AGeV; the problem is reduced to an understanding of the elementary reactions.

\section{Light vector mesons in nuclear collisions}
\label{sec:lvm}

The previous section demonstrated the particular role of the $\rho$ in the low-mass dilepton spectrum and provided evidence for in-medium effects on its spectral function. Similar effects are also predicted for the $\omega$ and $\phi$ mesons.  Their observation in nuclear collisions is however a formidable experimental challenge.  First, the predicted effects are smaller compared to the $\rho$ due to the weaker interaction of the $\omega$ and $\phi$ with the medium. Furthermore, with their relatively long lifetimes (assuming that they are unmodified in the medium, $\tau_{\phi}$ = 46 fm/c and $\tau_{\omega}$ = 23 fm/c), the $\omega$ and $\phi$ predominantly decay outside the medium after regaining the vacuum properties, with only a small fraction decaying inside the medium. Since the measurement integrates over the entire collision's history, this will result in a small modification of their spectral shape. Most of the yield will exhibit the vacuum spectral shape with only a small fraction showing up as a tail in the low-mass side (in the case of dropping masses) or both in the low and high mass sides (in the broadening scenario). The identification of these small effects sitting on top of a physics background and on top of a usually very large combinatorial background is not an easy task. NA60 with a mass resolution of 2.2\% and a S/B ratio of 1/3 at the $\phi$ mass is best suited for these studies at the SPS. First hints of modifications of the $\omega$ and $\phi$ have recently been reported \cite{na60-imr,na60-hp08}.
At RHIC, PHENIX with its excellent mass resolution of the order of 1\% at the $\phi$ pole should be able to perform spectroscopic studies of the $\omega$ and $\phi$ mesons once the combinatorial background is reduced with the HBD upgrade. In the meantime, PHENIX has attempted a direct spectral shape analysis of the $\phi$ meson measured through the \kk decay channel in \sqnR Au+Au collisions \cite{phenix-phi}. The $\phi$ yield fitted with a relativistic Breit-Wigner function convoluted with a Gaussian to account for the experimental mass resolution, revealed
no significant change in the centroid and width values of the $\phi$ meson from the PDG accepted values.

\begin{figure}[!h]
  \begin{center}
     \includegraphics[width=80mm]{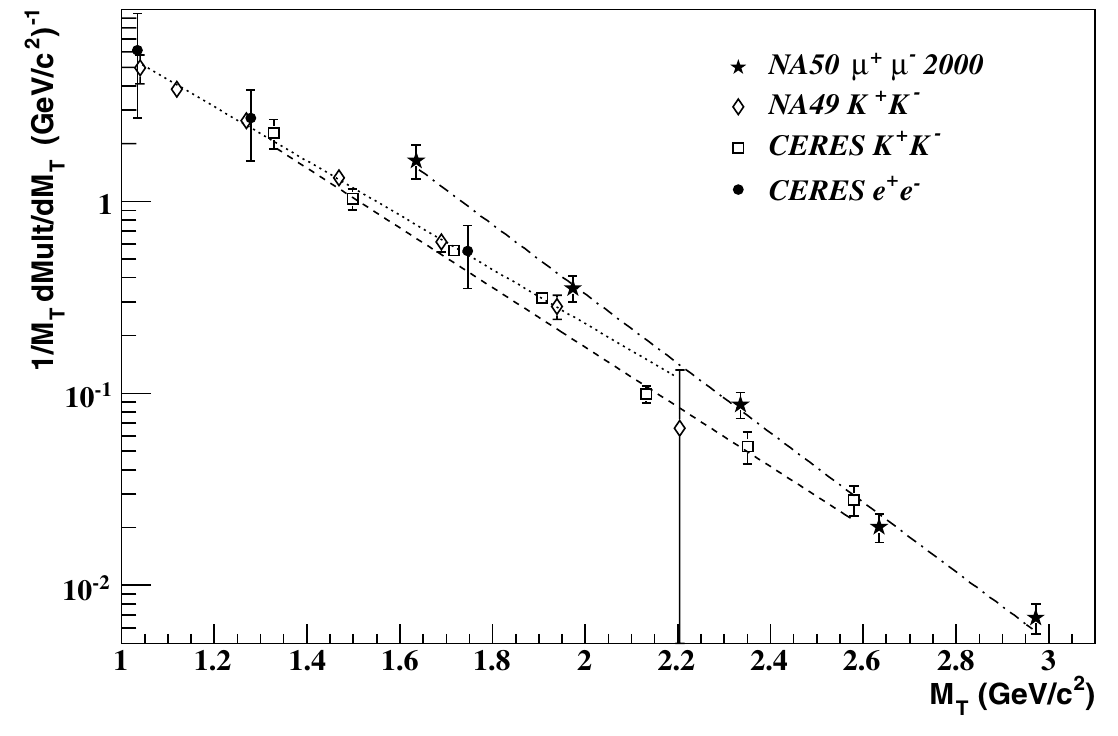}
  \end{center}
     \vspace{-5mm}
\caption{Transverse mass distributions of $\phi$ mesons measured at the SPS through the \kk (NA49 and CERES),  \ee (CERES) and \mumu (NA50) decay channels in central Pb+Pb (NA49, NA50) and Pb+Au collisions (CERES) \cite{na50-qm08}.}
  \label{fig:phi1}
\end{figure}

The comparison of the $\phi$ meson decay into \kk and \ee pairs provides an alternative and powerful tool to search for in-medium modifications. Since the $\phi$ mass is close to twice the kaon mass, any in-medium modifications of the $\phi$ or the kaons may induce a large change in the $\phi$ yield derived from the $\phi \rightarrow$ \kk decay mode. This type of studies has been carried out at the SPS by NA49, NA50, CERES and more recently NA60. The $\phi$ meson was studied via the \kk decay channel by NA49 \cite{na49-phi} and via the \mumu decay channel by NA50 \cite{na50-phi} in 158~AGeV Pb+Pb collisions. Differences in the $\phi$ yield by factors of 2 to 4 were found between NA49 and NA50 in the common \mt range covered by the two experiments, leading to what became to be known as the SPS $\phi$ puzzle. Recent reanalysis of the NA50 data \cite{na50-qm08} shows a smaller difference but the disagreement still persists, now at a factor of $\sim$2 (see Fig.~\ref{fig:phi1}). There is also significant disagreement in the inverse slope parameters derived from the two experiments. The differences are too large to be explained by the different rapidity coverage or slightly different centrality selection of the two experiments.

CERES is the first experiment that measured simulataneously the $\phi$ meson through both the \ee and \kk decay channels within the same apparatus \cite{ceres-phi}. The measurements were done in central 158~AGeV Pb+Au collisions. Consistent results were found in the two channels within the experimental uncertainties. The \kk yield was found in good agreement with the NA49 results whereas the  \ee data were incompatible with the original NA50 results. However, this is not anymore the case with the reanalyzed \mumu data of NA50. Within 1-2 $\sigma$ the CERES \ee data are compatible with the new NA50 results or in other words, the precision of the CERES \ee data is insufficient to rule out the present level of difference between NA49 and NA50. The $\phi$ puzzle in Pb+Pb is therefore not yet fully resolved. There is room for a substantial difference between the hadronic and leptonic decays of the $\phi$ meson, that could result from in-medium modifications of the $\phi$ meson \cite{hees-rapp08} or kaon absorption or rescattering that deplete the $\phi$ meson reconstruction in the \kk channel \cite{ko02}.

\begin{figure}[!h]
\begin{center}
      \includegraphics[width=80mm]{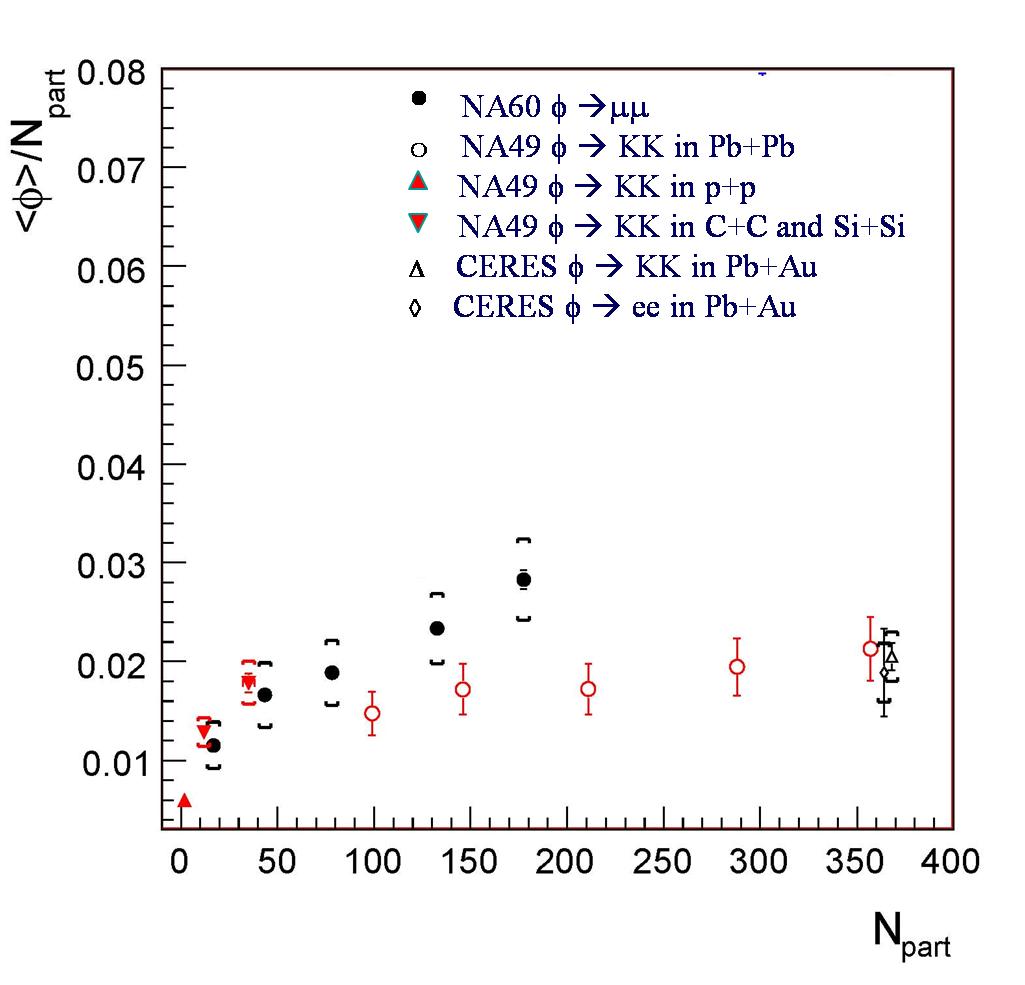}
\caption{Centrality dependence of the $\phi$ meson yields measured through dileptons and kaons at the SPS  by NA49, CERES and NA60. The data represent integrated yield over the full phase space per participant. Brackets show the systematic errors, for Pb+Pb the error bars include also the systematic error \cite{na60-phi}.}
  \label{fig:phi2}
\end{center}
\vspace{-0.6cm}
\end{figure}

The recent results from NA60 on $\phi$ production via muon and kaon decays in 158 AGeV In+In collisions \cite{na60-phi, na60-qm09} provide additional insight but no coherent picture emerges yet from all SPS measurements. In NA60, the yields and inverse slopes from the two decay channels are in agreement within errors \cite{na60-qm09}, i.e. the discrepancy seen in Pb+Pb between the \mumu (NA50) and \kk (NA49) decay channels is not seen in In+In. On the other hand, the multiplicity dependence of the $\phi$ meson yield per participant, integrated over full phase space, $<\phi>/N_{part}$, is different in the In+In muon channel and the Pb+Pb kaon channel. As shown in Fig.~\ref{fig:phi2}, $<\phi>/N_{part}$ increases monotonically with centrality in In+In whereas it seems to saturate for $N_{part} > 50$ in Pb+Pb. Furthermore, for $N_{part} > 100$ the $\phi$ meson yield per participant is higher in the muon channel of In+In collisions than in the kaon channel of Pb+Pb collisions when compared at the same number of participants.

The story of the $\phi$ meson at RHIC is similarly unsettled. PHENIX has the capability of measuring the $\phi$ meson through the \ee and the \kk decay channels at mid-rapidity. Preliminary results \cite{phenix-kozlov} of the rapidity density per pair of participants in \sqnR Au+Au collisions may indicate a possible larger yield in the dilepton channel compared to the kaon one. However, within the large  statistical and systematic uncertainties of the electron data the two channels yield consistent results. A more definite statement will have to await for the improvement in data quality expected with the HBD upgrade of the PHENIX experiment.

We conclude this section by mentioning the STAR results on the $\rho$ meson reconstruction via its  $\pi^+ \pi^-$ decay. STAR reported a downward shift of $\sim$40~MeV in the $\rho$ meson peak position in p+p collisions and a stronger shift of $\sim$70~MeV in peripheral Au+Au collisions at \sqnR \cite{star-rho}.  The observation has been attributed to in-medium modification of the $\rho$ meson in the dilute hadronic matter of the late stages of Au+Au collisions and has been addressed within the two scenarios, discussed previously in the context of the CERES data. While the collision broadening qualitatively reproduces the trend but fails to reproduce the magnitude of the shift \cite{rapp03}, the Brown-Rho scaling succeeds in reproducing the magnitude of the downward mass shift \cite{shuryak-brown03}.

\section{Intermediate mass region}
\label{sec:imr}
Dileptons in the intermediate mass region are particularly interesting. As discussed in the Introduction, several early calculations have singled out this region as the most appropriate window to observe the thermal radiation from the QGP \cite{shuryak78,ruuskanen92}. This section reviews dilepton  measurements in the IMR at SPS and RHIC energies.

\subsection{IMR dimuons at SPS}
\label{sec:imr-sps}

\begin{figure}[h!]
   \vspace{7mm}
  \begin{center}
      \includegraphics[width=90mm, height=70mm]{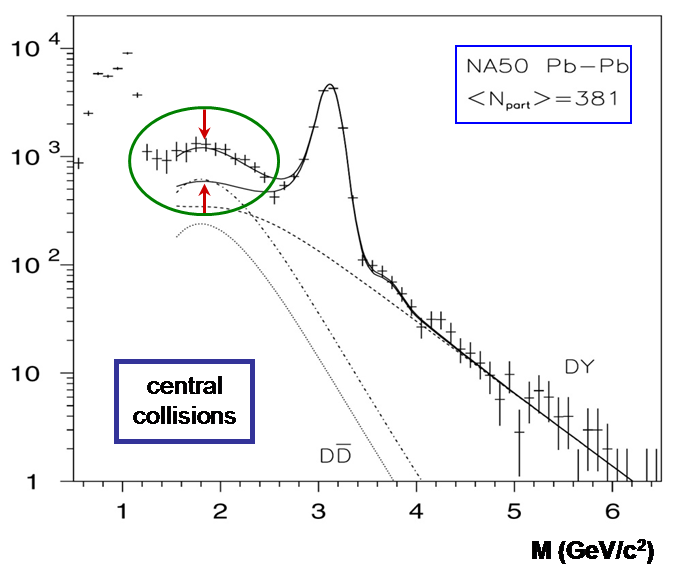}
  \end{center}
  \vspace{-5mm}
\caption{Di-muon invariant mass spectrum measured by NA50 in central Pb+Pb collisions at 158~AGeV together with the total expected yield (lower solid line) obtained from the sum of the Drell-Yan (dashed line) and open charm (dotted line) contributions. Adding the latter multiplied by a factor of 2.6 (dash-dotted line) leads to a total yield (upper solid line) that closely follows the data \cite{na50-imr}.}
  \label{fig:na50-imr}
\end{figure}

Dileptons in the IMR have been measured at SPS energies by HELIOS-3 (p,S+W at 200~AGeV)
\cite{helios3}, NA38 (p+W, S+U at 200~AGeV) and NA50 (p+Al, Cu, Ag, W at 450 GeV, Pb+Pb at 158~AGeV) \cite{na38-imr-pt,na50-imr,na50-pt} and more recently by NA60 in In+In at 158~AGeV \cite{na60-imr}. All experiments reported an excess with respect to the expected yield from the two main contributions in this mass region, Drell-Yan and semi-leptonic charm decay. An example of the enhancement seen by NA50 is shown in Fig.~\ref{fig:na50-imr}. The shape of the excess is very similar to the open charm contribution and in fact enhancing the latter by a factor of 3.6 nicely accounts for the inclusive spectrum. This is the basis for the hypothesis of enhanced charm production put forward by NA38/50 to explain the excess \cite{na50-imr}. This line of thought was pursued in a more formal way in Ref.~\cite{wong-wang96} suggesting gluons produced in soft baryon-baryon collisions as an additional source of \ccbar~pairs.

\begin{figure}[!h]
\begin{center}
   \includegraphics[width=70mm, height=58mm]{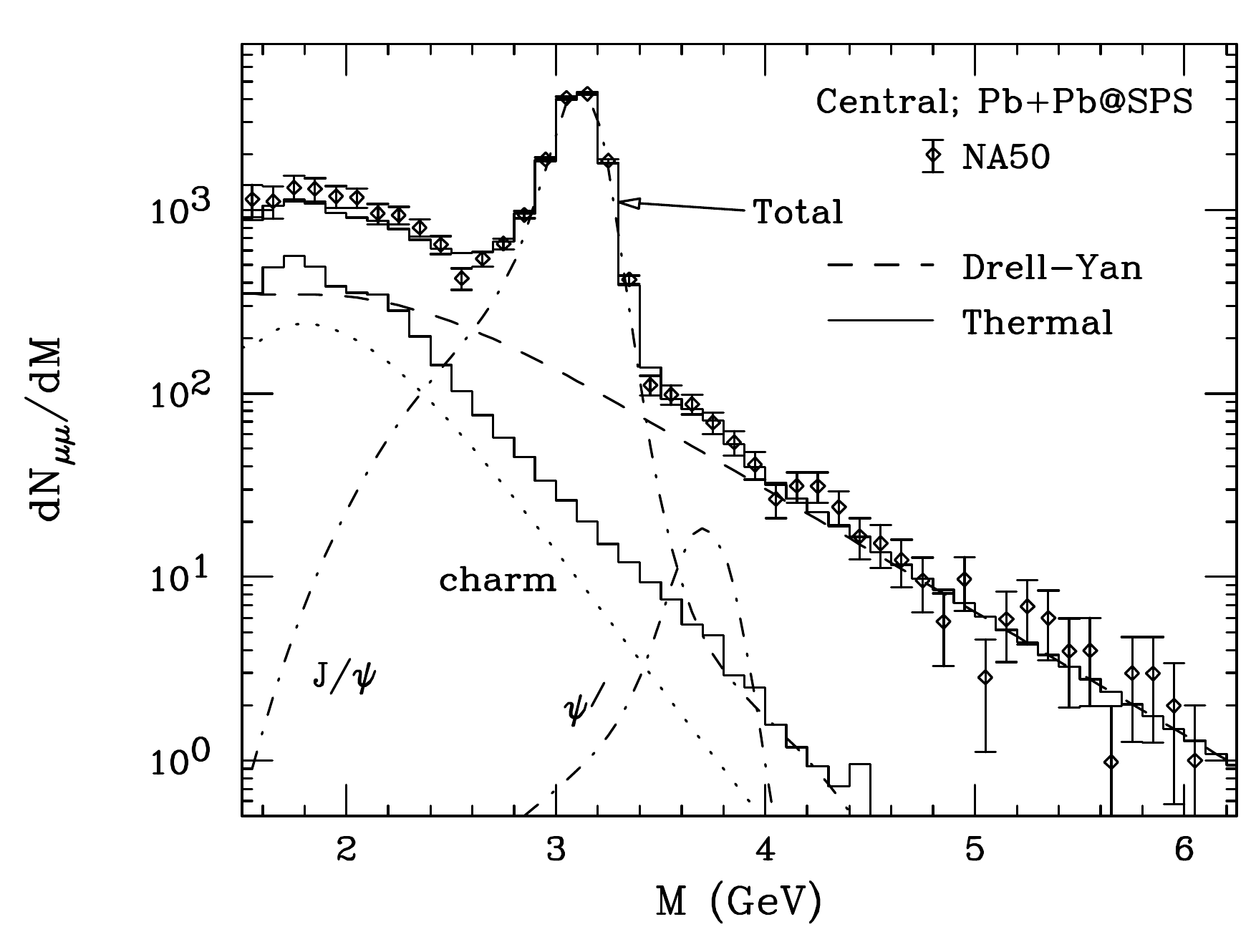}
    \includegraphics[width=70mm]{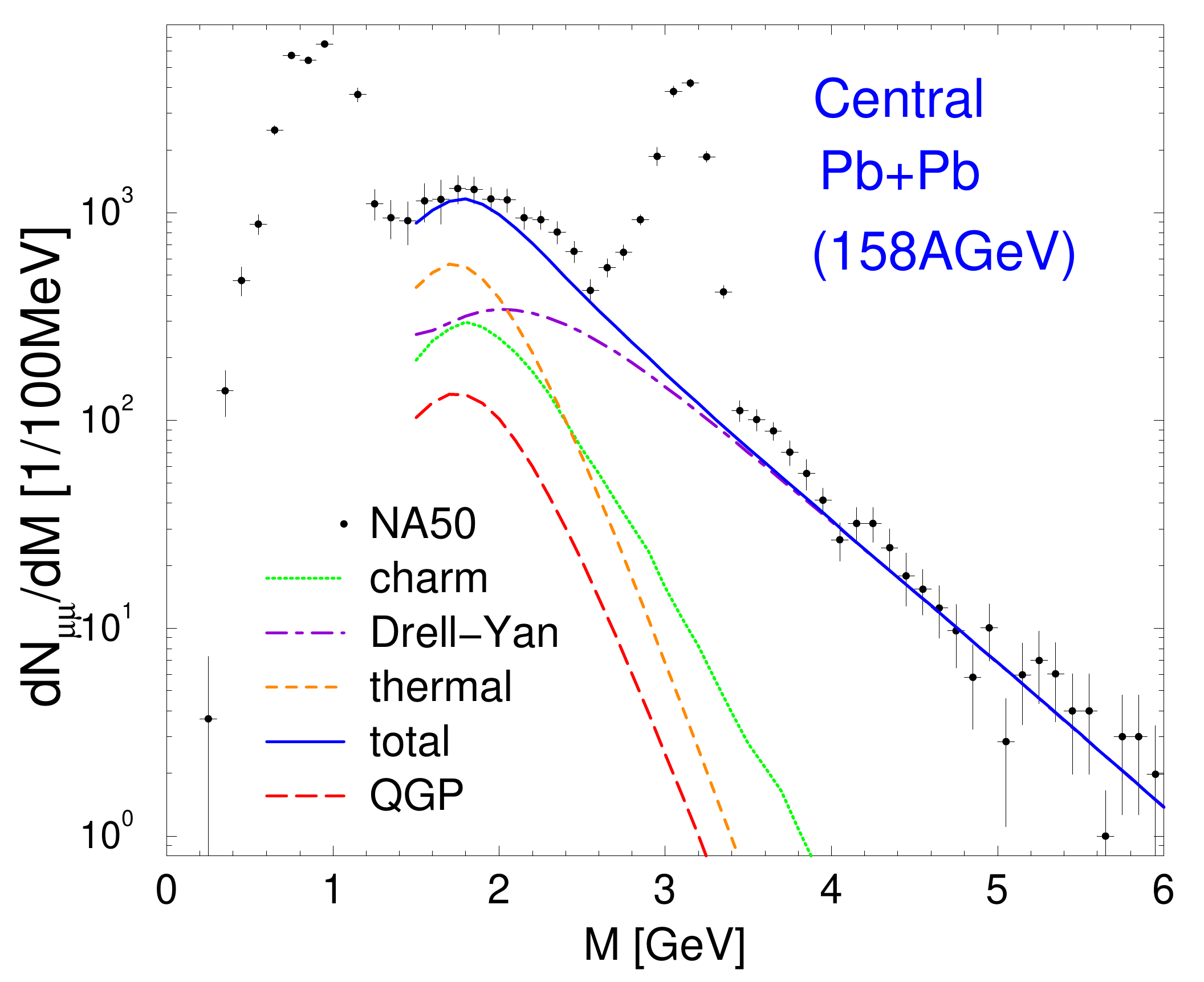}
  \caption{NA50 dimuon data compared to calculations of Ref.~\cite{gale02} (left panel) and Ref.~\cite{rapp-shuryak} (right panel). The various contributions (open charm, Drell Yan and thermal) and their sum are indicated in the figures. The thermal contribution is calculated using hadronic (partonic) rates in the left (right) panel.}
  \label{fig:na50-gale-shuryak}
\end{center}
\vspace{-3mm}
\end{figure}

Other possibilities to explain the IMR excess were also explored, like secondary Drell-Yan emission \cite{spieles98} and final-state rescatterings that broaden the \mt~distribution of charmed mesons enhancing the yield in the phase space covered by the NA50 experiment \cite{lin-wang98}. Most of the attention focussed on the possibility of explaining the excess as thermal radiation. Gale and co-workers followed a similar line to the one that successfully reproduced the low-mass dileptons discussed in Section~\ref{subsec:sps} in terms of thermal radiation from the hadronic gas. On top of the {\it physics} background of Drell-Yan and open charm pairs, they systematically considered all sources of muon pairs from secondary meson interactions that could contribute to the IMR: $\pi\pi,~\pi\rho,~\pi\omega, {\pi}a_1$ and $K\overline{K} \rightarrow l\overline{l}$ and were able to reproduce the mass and \pt~distributions of HELIOS-3 in S+W \cite{li-gale98,li-gale98-2} and NA50 in Pb+Pb collisions (see Fig.~\ref{fig:na50-gale-shuryak}, left panel) \cite{gale02}.
On the other hand Rapp and Shuryak \cite{rapp-shuryak} exploiting the quark-hadron duality, used the pQCD \qqbar annihilation rates to calculate the dilepton yield throughout the entire space-time evolution of the collision which they described with a simple fireball model. This thermal radiation, added to the background (Drell Yand and open charm) contributions, is also able to explain the NA50 enhancement in the IMR  as shown in Fig.~\ref{fig:na50-gale-shuryak} (right panel). Only a small fraction of this radiation is emitted at the early stages and is associated with the QGP phase. These calculations also reproduce the transverse momentum dependence of the muon pairs in the IMR. Similar results using a similar approach were obtained in Ref.~\cite{kampfer00}. This ambiguity between the hadronic and partonic description of the IMR excess might be resolved with the new NA60 results that will be discussed now.

\begin{figure}[!tb]
\begin{center}
      \includegraphics[width=70mm, height=60mm]{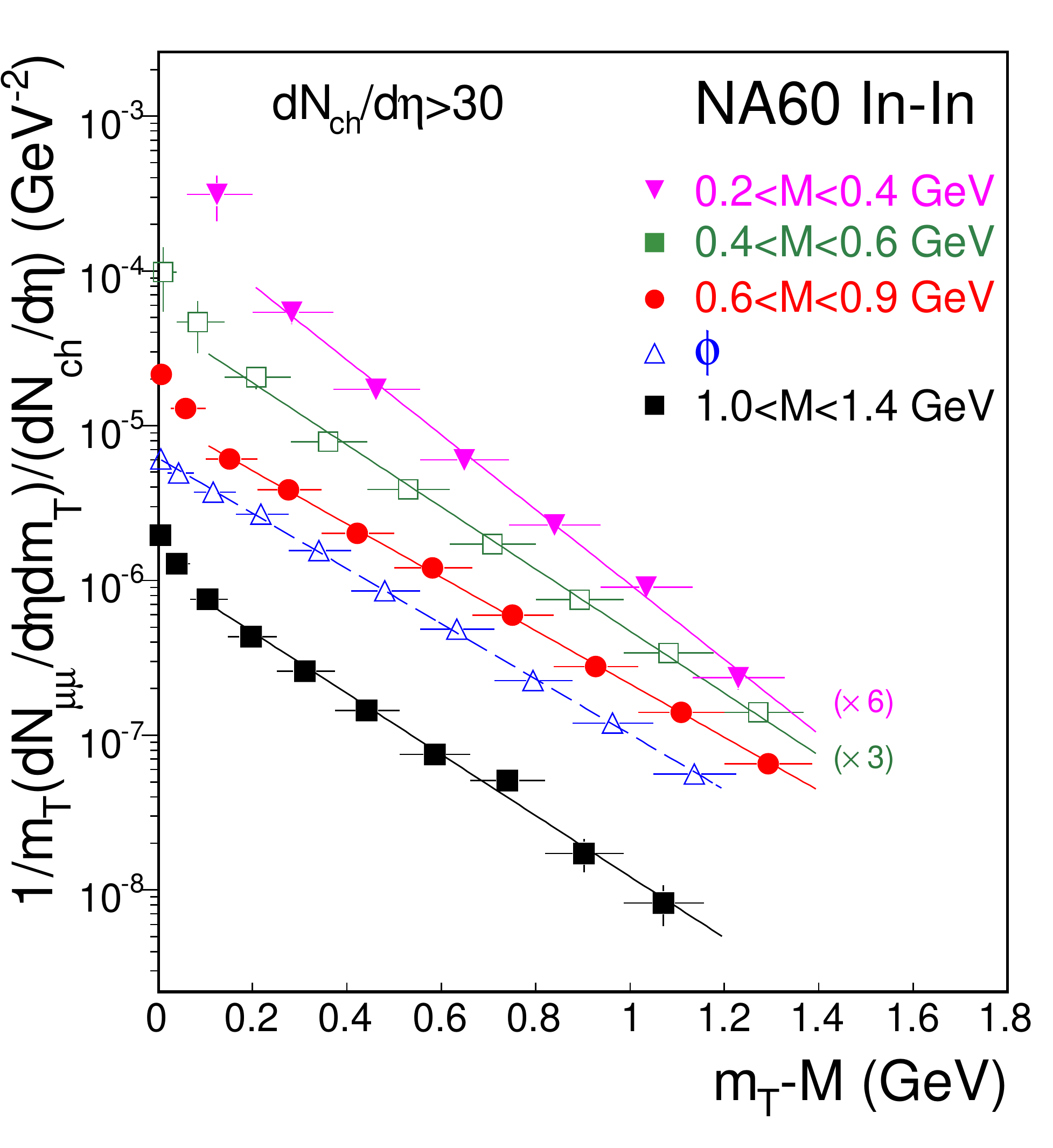}
      \includegraphics[width=70mm, height=60mm]{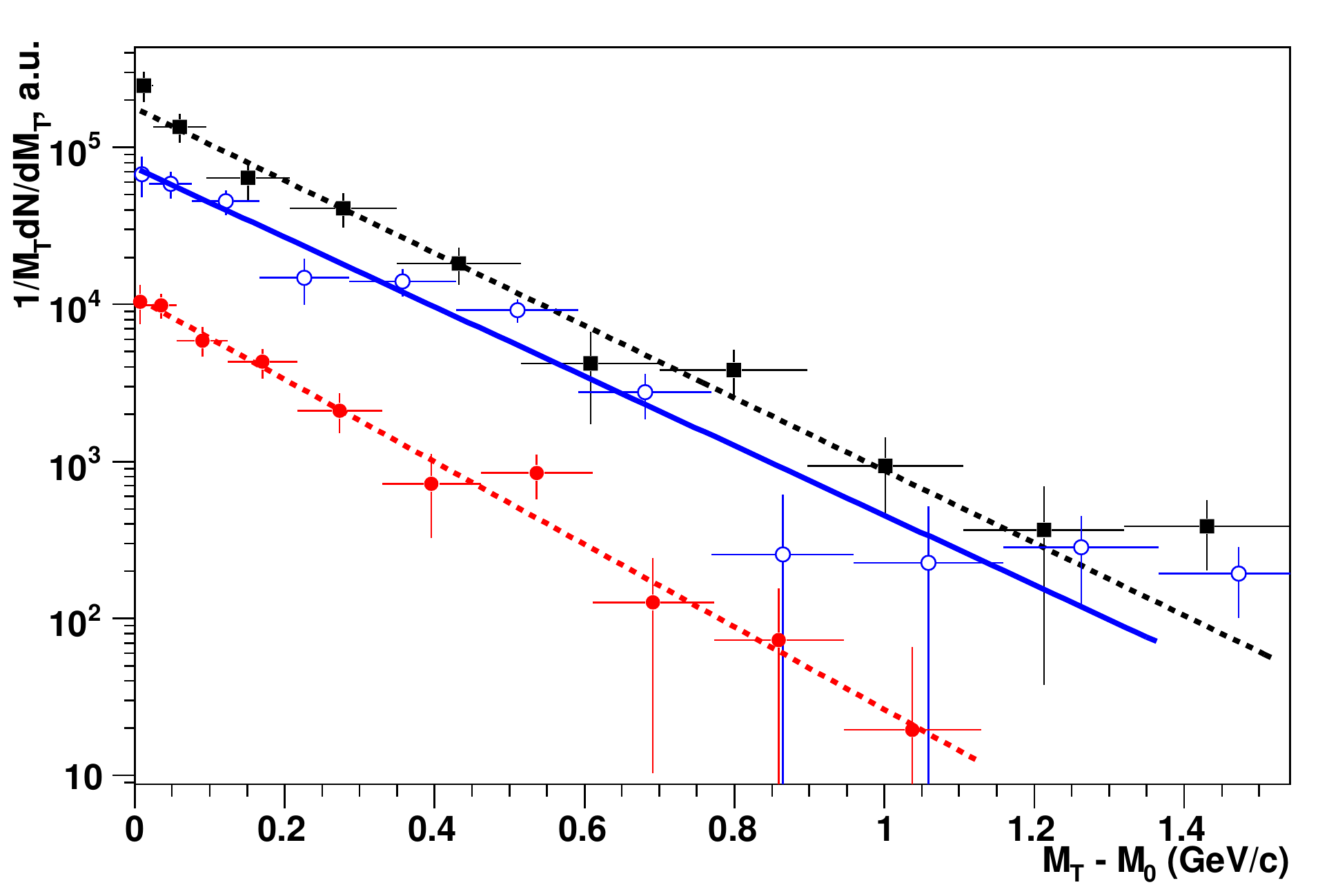}
\caption{Transverse mass distributions of the excess of muon pairs measured by NA60 in 158~AGeV In+In collisions for four mass windows in the LMR  and the $\phi$ (left panel) \cite{na60-flow, na60-hp08} and for three adjacent mass windows from top to bottom, 1.16-1.4, 1.4-2.0 and 2.0-2.56 GeV/c$^2$ in the IMR (right panel) \cite{na60-imr}.}
  \label{fig:na60-mt}
\end{center}
\vspace{-5mm}
\end{figure}
NA60 measured the dimuon spectrum in 158~AGeV In+In collisions up to 5~GeV/c$^2$ and finds also an excess in the IMR \cite{na60-imr}. The excess has the same shape as the open charm contribution calculated with PYTHIA, the same observation already made by NA50 in Pb+Pb collisions and mentioned previously. Using the vertex telescope information, NA60 is able to go one step further. With a vertex resolution of the order of 10-15$\mu$ NA60 can distinguish dimuons from a prompt source originating at the vertex and dimuons from the semi-leptonic decays of D mesons originating from a displaced vertex. This allowed NA60 to establish that the IMR excess is from a prompt source, thereby ruling out charm enhancement as explanation of the excess. The yield of this excess increases with centrality with a stronger than linear, but weaker than quadratic, dependence on N$_{part}$.

\begin{figure}[!h]
  \begin{center}
      \includegraphics[width=95mm]{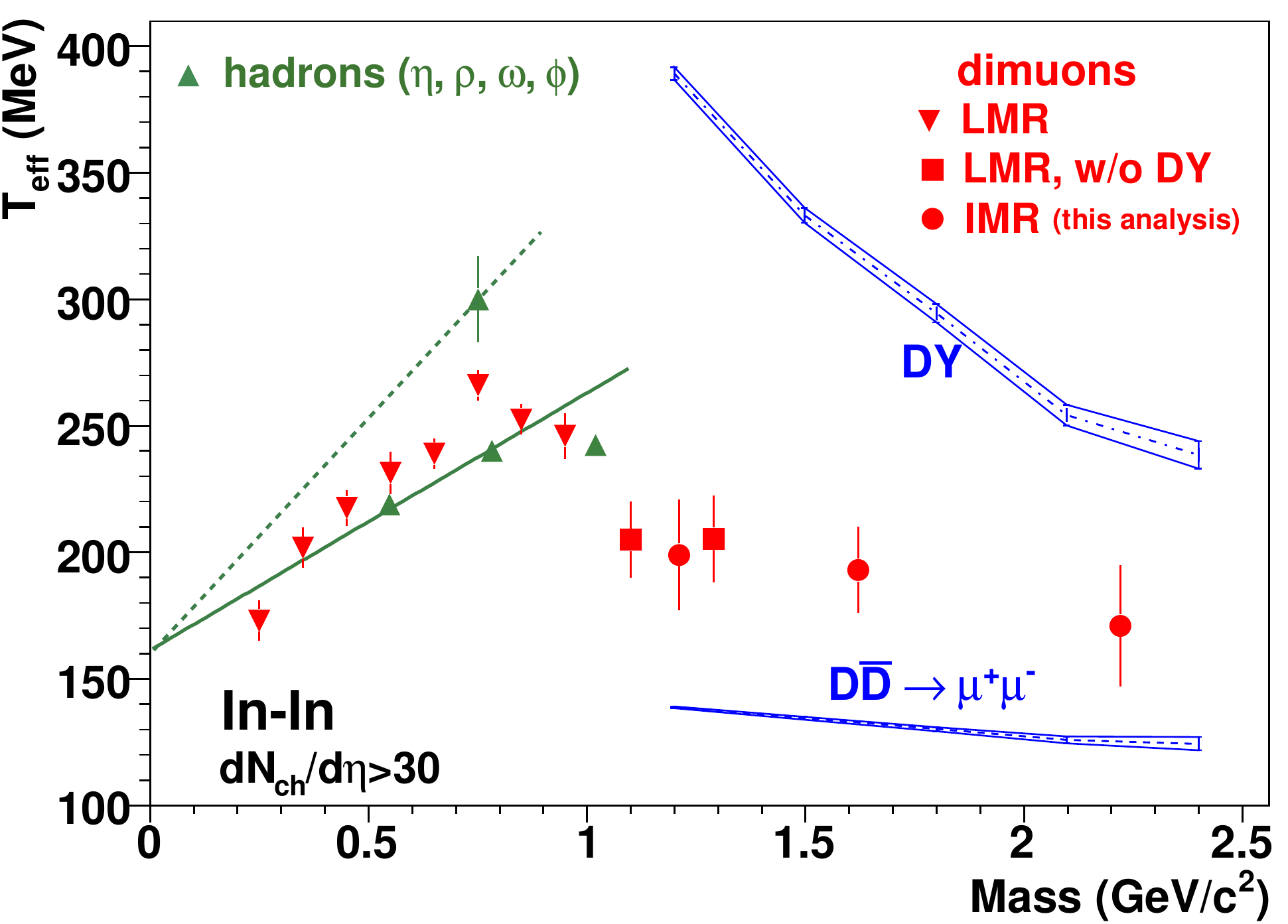}
  \end{center}
  \vspace{-5mm}
\caption{Inverse slope parameters vs. mass derived from \mt distributions of the dimuon excess and hadrons ($\eta, \rho, \omega, \phi$) measured by NA60 in 158~AGeV In+In collisions. The bands show the inverse slope parameters for Drell-Yan and open charm calculated with PYTHIA \cite{na60-imr}.}
  \label{fig:na60-teff}
\end{figure}

The centrality-integrated \mt distributions in the IMR combined with those from the LMR turned out to be very illuminating. The results are shown in Fig.~\ref{fig:na60-mt}. Ignoring the steepening of the distributions at low \mt for the low mass windows, whose origin is not understood, all distributions exhibit a pure exponential shape and can be fit by $exp(-m_{T}/T)$. Whereas in the LMR, the inverse slope T monotonically rises with mass, reaching values of $\sim$250~MeV at m~=~1~GeV/c$^2$, as expected from a hadronic source subjected to radial flow, the inverse slope in the IMR is practically independent of mass with a relatively low value of $\sim$190~MeV. The inverse slope parameters derived from the NA60 data vs. mass are displayed in Fig.~\ref{fig:na60-teff}.

To reconcile these two apparently contradictory findings (T increasing with mass in the LMR and T independent of mass in the IMR) together with the sharp drop of T in the IMR, NA60 concludes that the excess of dimuons in the IMR  is the thermal radiation emitted early in the collision, before radial flow develops and thus is most likely of partonic origin thereby providing an experimental argument that breaks the quark-hadron duality. This interpretation is in qualitative agreement with recent calculations using an improved space-time description of the fireball that report a substantial or even dominant contribution of thermal radiation from the QGP in the IMR \cite{hees-rapp08,zahed07,renk08}. It is also consistent with recent NA60 measurements showing that the excess dimuons are not polarized as expected for the thermal radiation from a randomized source \cite{na60-polar}.

\subsection{IMR at RHIC}
\begin{figure}[!h]
  \begin{center}
     \includegraphics[width=110mm]{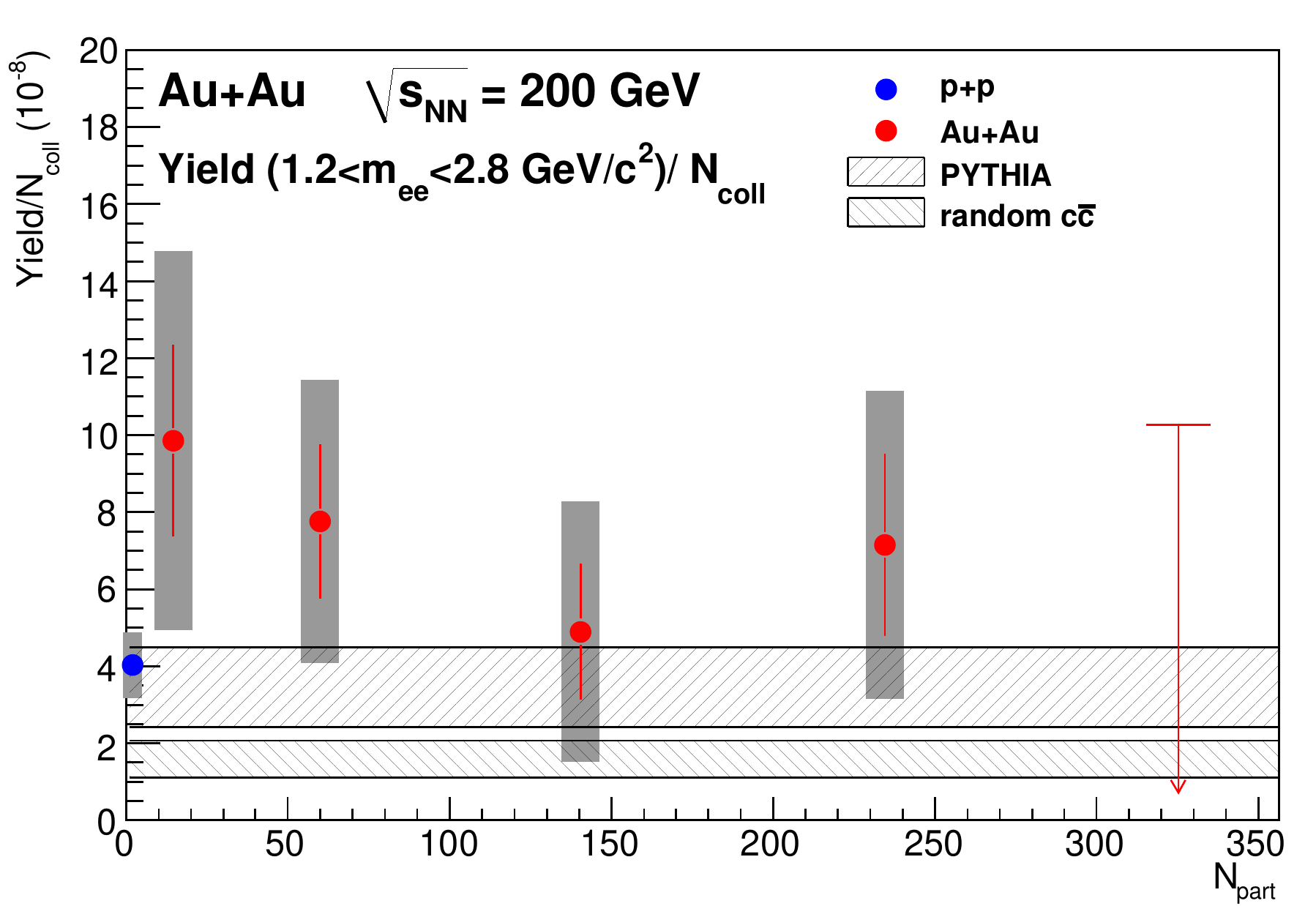}
  \end{center}
  \vspace{-5mm}
\caption{Dielectron yield in the IMR scaled with the number of binary collisions N$_{coll}$ vs. N$_{part}$ measured by PHENIX in Au+Au collisions at \sqnR \cite{phenix-lmr}.}
  \label{fig:imr-phenix}
\end{figure}

The dilepton yield in the IMR looks different at RHIC \cite{phenix-lmr}. Within the experimental uncertainties, the IMR yield scales as expected with the number of binary collisions as illustrated in Fig.~\ref{fig:imr-phenix}. But contrary to the SPS where the IMR yield shows a clear excess beyond the expected yield from Drell-Yan and open charm, that is now attributed to thermal radiation of partonic origin, at RHIC the inclusive yield seems to be exhausted by the open charm contribution as estimated from PYTHIA (see Fig.~\ref{fig:phenix-auau-mb}). There is only room for an additional source within the experimental uncertainties. PHENIX also points out that the charm contribution calculated with PYTHIA represents an upper limit, the yield could be smaller due to in-medium suppression of high \pt  charm quarks \cite{phenix-single-e-auau} that smear the correlation between the correlated \ccbar pair. The lines labeled \ccbar $\rightarrow$ \ee (random correlation) in Figs.~\ref{fig:phenix-auau-mb} and \ref{fig:imr-phenix} represent an estimate of the minimum open charm contribution under the extreme assumption of complete randomness between the \ccbar pairs that leaves some more room for additional sources. This is an interesting hypothesis that can be addressed experimentally once PHENIX is able to disentangle prompt from off-vertex electron pairs with the vertex detector under development \cite{phenix-vertex}.

\section{Light vector meson spectroscopy in elementary reactions}
\label{sec:spectroscopy}

The discovery at the SPS of in-medium modifications of the $\rho$ meson in nuclear collisions and their possible link to CSR, motivated a vigorous program of precision spectroscopic studies of the light vector mesons (LVM), $\rho, \omega$ and $\phi$ in elementary collisions, proton induced and photoproduction reactions \cite{hayano-hatsuda}. The most prominent results are reviewed in this section.

A list (probably incomplete) of recent and current experiments is given in Table~\ref{tab:elementary}. It includes measurements of LVM through their \ee decay channel which allows clean spectroscopic studies not distorted by final state interactions, but also a few measurements where the LVM are identified through hadronic decay modes. The latter benefit from considerably larger branching ratios but could be affected by rescattering effects of the hadrons in the nuclear medium. Elementary reactions are simpler systems to study. They have well controlled conditions as the nucleus is always at T=0 and at constant baryon density $\rho_0$ as opposed to nucleus-nucleus collisions where the temperature is different from zero and where both the temperature and the baryon density vary during the collision. The predicted modifications at normal nuclear matter density $\rho_0$ are expected to be much smaller than in nuclear collisions but they are still large enough to be observable. For example, models based on the Brown-Rho scaling \cite{brown-rho} and QCD sum rules \cite{hatsuda-lee} predict a decrease of the vector meson mass of $\sim$15\% at $\rho = \rho_0$.
The rho meson is advantageous because with its short lifetime it has a large probability to decay inside the nucleus. The disadvantage is that one has to disentangle the $\rho$ and $\omega$ yields and possible $\rho - \omega$ interference effects. The $\phi$ meson offers the opposite properties, a narrow and isolated resonance with a low probability to decay inside the nucleus. For the $\omega$ and $\phi$, the observation of in-medium effects requires measurements down to low momentum values.

\begin{table}[h!]
\caption{Elementary reactions used to study light vector mesons modifications in normal nuclear matter.}
\label{tab:elementary}
\vspace{0.4cm}
\begin{center}
\leavevmode
\begin{tabular}{|c|c|c|c|}\hline

 Experiment   & Reactions                              & Probe                               & Ref.  \\ \hline
 KEK-PS E325  & p+A$\rightarrow \rho,\omega$+X         & $\rho,\omega \rightarrow$ \ee       & \cite{KEK-2001,KEK-2006}\\
              & p+A$\rightarrow \phi$+X                & $\phi\rightarrow$ \ee               & \cite{KEK-2007}\\ \hline
 CLAS         & $\gamma$+A $\rightarrow$VM+A$^*$       & $\rho,\omega,\phi\rightarrow$ \ee   & \cite{CLAS-07,CLAS-08}\\ \hline
 CBELSA/TAPS  & $\gamma$+A $\rightarrow\omega$+A$^*$   & $\omega \rightarrow \pi^0 \gamma$   & \cite{CBELSA-2005,CBELSA-2008,nanova-hadron09}\\ \hline
 TAGX         & $\gamma + A \rightarrow \rho+A^*$      & $\rho \rightarrow \pi^+\pi^-$       & \cite{TAGX98-1,TAGX98-2,TAGX03} \\ \hline
\end{tabular}
\end{center}
\end{table}

\begin{figure}[!h]
\vspace{-6mm}
\begin{center}
     \includegraphics[width=70mm, height=75mm]{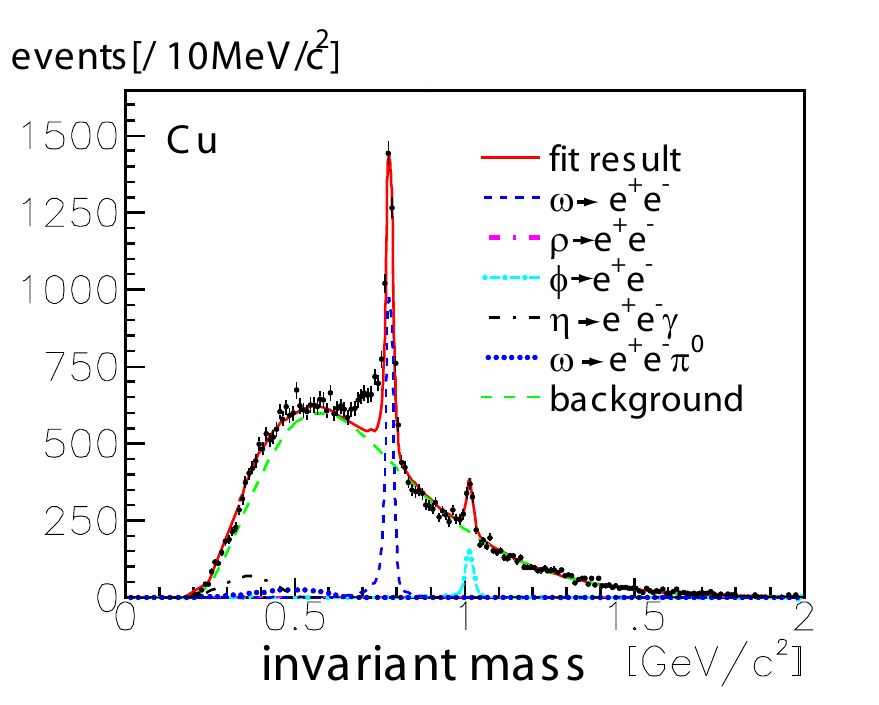}
      \vspace{0.0cm}
     \includegraphics[width=70mm]{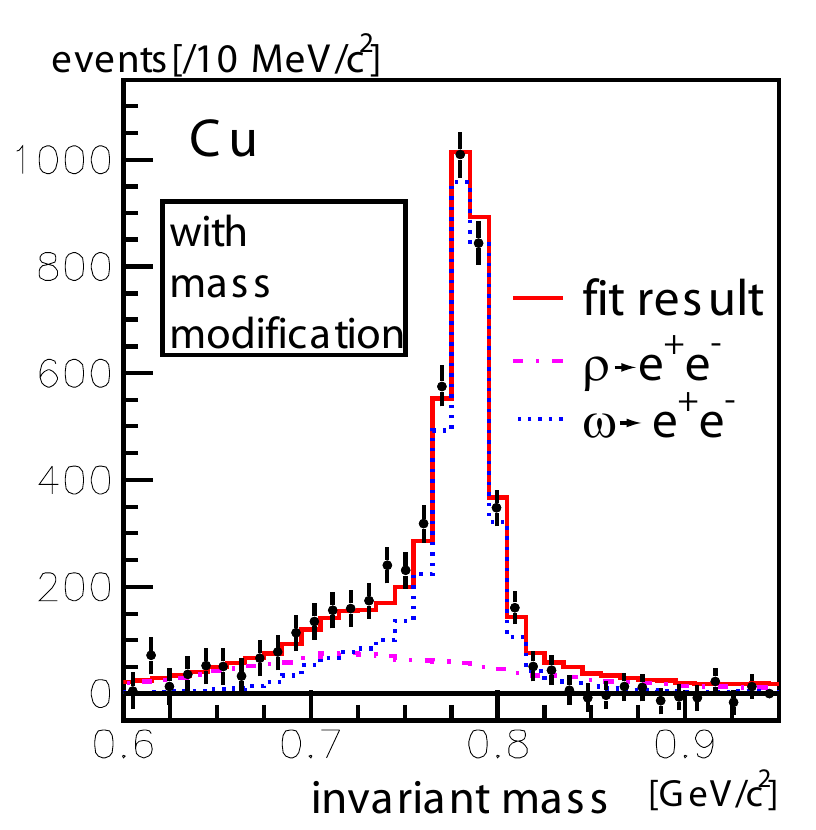}
\caption{Invariant mass spectra, measured in 12~GeV p+Cu collisions by the E325 experiment at KEK-PS \cite{KEK-2006}. Left panel: raw spectrum with a best fit including the individual contributions from hadron decays and the combinatorial background. Right panel: spectrum after subtracting the background and the contributions from $\eta \rightarrow$ \ee$\gamma$ and $\omega \rightarrow$ \ee$\pi^0$ together with results of a model assuming that the $\rho$ and $\omega$ masses decrease by 9.2\% with respect to their vacuum values at normal nuclear density.}
  \label{fig:KEK}
\end{center}
\vspace{-3mm}
\end{figure}

The experimental results reported so far are very interesting but still insufficient for a consistent picture to emerge. The first positive result came from the KEK-PS E325 experiment that measured the LVM in 12 GeV p+C and p+Cu collisions with an excellent mass resolution,  $\sim$1\% at the $\phi$ mass. The measurements were performed in the target rapidity region to enhance the decay probability of the virtual photon inside nuclear matter. The experiment reported a significant excess of \ee in the low-mass side of the $\omega$ \cite{KEK-2001, KEK-2006} and $\phi$ \cite{KEK-2007} mesons. The excess is clearly visible in Fig.~\ref{fig:KEK} (left panel) \cite{KEK-2006} which shows the raw mass spectrum obtained with the Cu target. The combinatorial background shape was determined using an event mixing technique. However, since the like-sign spectrum was not recorded, the quality of the mixed event technique could not be demonstrated and the background could not be absolutely normalized and subtracted from the raw data as usually done. Instead, the normalization was obtained by fitting the mixed event spectrum, together with the relevant hadron decays, to the raw data (see lines in the left panel of Fig.~\ref{fig:KEK}).

The right panel of Fig.~\ref{fig:KEK} shows the same data after subtracting the contributions from $\eta \rightarrow$ \ee$\gamma$ and $\omega \rightarrow$ \ee$\pi^0$ and the combinatorial background. The resulting spectrum is very well reproduced by a simple model assuming a linear decrease of the vector meson masses $m_V$ with the nuclear density $\rho$ according to $m_V(\rho)/m_V(0)$ = 1 - k($\rho$/$\rho_0$), with no in-medium broadening. The best fit to the data, indicated by the solid line in the figure, yields a value of k = 9.2\% meaning that both the $\rho$ and $\omega$ masses drop by 9.2\% at normal nuclear density.

More recently, the same experiment reported the observation of a similar excess in the low-mass side of the $\phi$ meson \cite{KEK-2007}. The excess was only observed in the lowest \pt bin covered by the measurement ($\beta \gamma < $1.25). The data required k = 3.4\% and also an increase of the $\phi$ width by almost a factor of 4. The latter was necessary to generate enough decay inside the target nucleus.

A similar effect was originally reported by the CBELSA/TAPS experiment in the photoproduction of $\omega$ mesons identified through the $\pi^0\gamma$ decay channel, on Nb and LH$_2$ targets \cite{CBELSA-2005}. At low $\omega$-momenta, a clear excess was shown in the low-mass side of the $\omega$ meson for the Nb case whereas no excess was observed with the LH$_2$ target or at high momenta. The data analysis was  consistent with a dropping mass of 13\%, in agreement with the expected predictions of Brown and Rho \cite{brown-rho} and Hatsuda and Lee \cite{hatsuda-lee}. Subsequent transparency ratio measurements on C, Ca, Nb and Pb targets yielded, through comparison to model predictions, a strong broadening of the $\omega$ meson of 130-150 MeV/c$^2$ \cite{CBELSA-2008} . This result together with criticism expressed in the literature \cite{CLAS-08,kaskulov-07} on the way the combinatorial background was handled in the data of Ref. \cite{CBELSA-2005}, prompted a reanalysis of the same data \cite{nanova-hadron09}. This reanalysis does not confirm the earlier claim of mass shift. On the contrary, the comparison of the data to BUU simulations near the $\omega$ production threshold (E$_\gamma$ = 900-1400 MeV) favors the scenario of $\omega$ broadening without mass shift. New measurements of much higher statistics have been performed to confirm this new finding.

\begin{figure}[!tb]
\begin{center}
     \includegraphics[width=70mm, height=60mm]{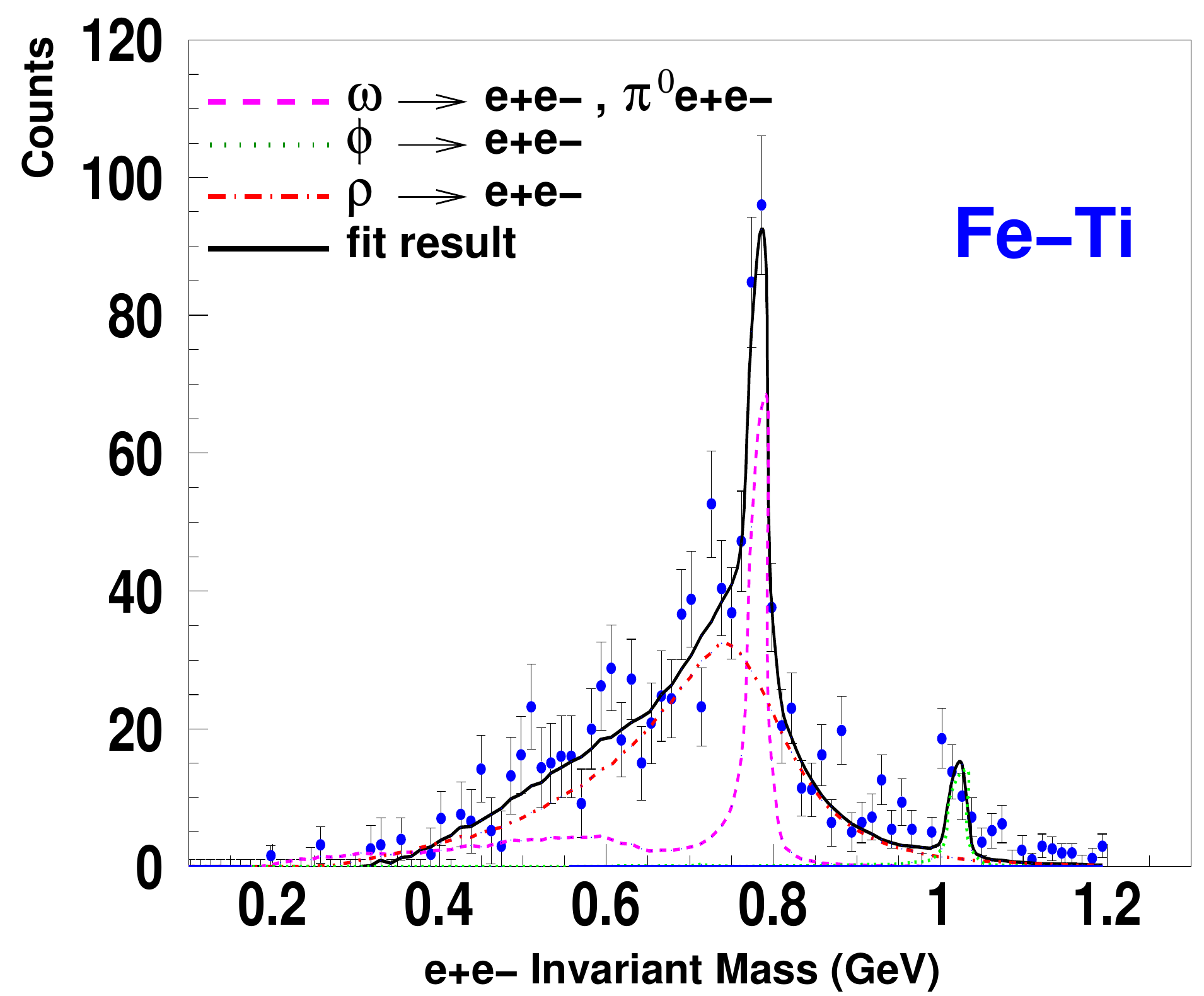}
     \includegraphics[width=70mm]{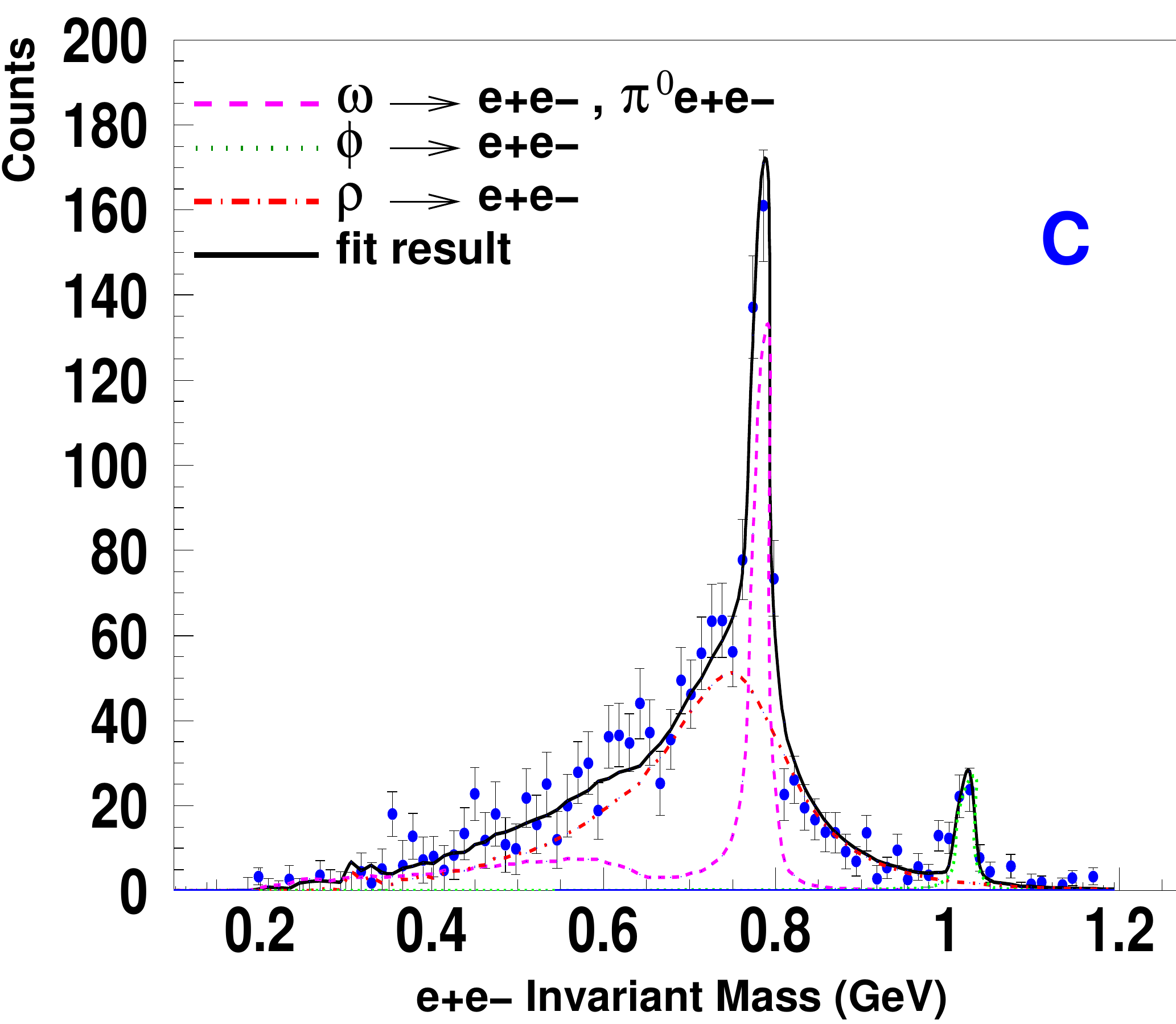}
\caption{Invariant mass \ee spectrum after background subtraction measured by the CLAS experiment in the LVM region together with a fit including the resonance decays $\rho, \omega, \phi \rightarrow$ \ee and the $\omega$ Dalitz decay, using the vacuum mass values \cite{CLAS-07}. }
  \label{fig:CLAS}
\end{center}
\vspace{-5mm}
\end{figure}

The CLAS experiment reports no effect in the search for in-medium modifications of vector mesons in photo-induced reactions on various targets $^2$H, C, Fe and Ti over an energy range of 0.6 - 3.8 GeV \cite{CLAS-07}. The vector mesons were identified through their decay into $e^+e^-$, with pair momenta of 0.8 - 2 GeV/c, similar to the momentum range covered by the KEK experiment. The spectra for C and Ti targets, after background subtraction, are shown in Fig.~\ref{fig:CLAS}. They look very similar to those measured by the KEK group (compare to Fig.~\ref{fig:KEK}). However, in contrast with the KEK experiment, the CLAS data are very well reproduced by calculations based on a transport model using the vacuum mass values of the $\rho, \omega$ and $\phi$ mesons \cite{buu}. The data analysis yields a very small mass shift of the $\rho$ meson mass, k = 2 $\pm$ 2\%, consistent with zero, and some collisional broadening of the $\rho$ meson width ($\Gamma$ = 218 $\pm$15 MeV for the Fe-Ti target, compared to the natural width of 150.7 $\pm$ 1.1 MeV).  The CLAS results are clearly inconsistent with the KEK results. They rule out the dropping mass scenarios of Refs. \cite{brown-rho,hatsuda-lee} and are consistent with other models that predict a broadening of the spectral shape without or with very small mass shift \cite{rapp-wambach,mosel,oset}.

The TAGX Collaboration measured the $\rho$ meson, identified via its \pipi decay channel, in photoproduction reactions on $^2$H,  $^3$He and  $^{12}$C at $E_\gamma$ = 0.6 - 1.12 GeV \cite{TAGX98-1,TAGX98-2,TAGX03}.
In this subthreshold energy region, the $\rho$ production occurs with the help of Fermi momentum. The $\rho$'s are produced with very low momentum such that they have a large probability to decay even inside a small nucleus like $^3$He.
The last analysis of the TAGX data shows a mass decrease of the $\rho$ in  $^3$He, of the order of 45-65 MeV/c$^2$ \cite{TAGX03}, in contrast to the CLAS results previously mentioned. On the other hand, the $^{12}$C data show mainly a broadening of the $\rho$, with no mass shift, in reasonable agreement with the many body effective Lagrangian approach of Rapp and Wambach \cite{rapp-chanfray-wambach97} or the similar picture of Ref. \cite{lpm}.

\section{Thermal photons}
\label{sec:photons}

Thermal photons are treated in depth in a dedicated article in this volume \cite{gale}. Here, only a brief account is given since real photons are expected  to carry the same physics information as virtual photons and are therefore linked to the essence of this article.

The interest in thermal photons was articulated in some detail in the Introduction. The unambiguous identification of thermal photons from the QGP is a very strong signal of deconfinement and provides a direct measurement of the plasma temperature. The absolute yields, obtained by integrating the emission rate over the space-time evolution of the collision, have been calculated by several authors and considerable progress has been achieved \cite{turbide04,arnold-moore-yaffe}. The calculations point to rather well established features. At temperatures close to the phase transition, thermal photons from partonic and hadronic processes have similar production rates in close analogy to the behavior of dilepton production rates, perhaps another manifestation of quark-hadron duality. In an elaborate calculation, the transverse momentum range \pt~=~1-3~GeV/c appears as the most promising window where the QGP radiation could shine over other contributions in central Au+Au collisions at \sqnR \cite{turbide04}.

The thermal photon yield is relatively small compared to the inclusive photon yield. As discussed in Section~\ref{sec:challenge}, the physics background for real photons (mainly from $\pi^0$ and $\eta$ decays) is larger by orders of magnitude compared to dileptons (at m $>$ 200 MeV/c$^2$), making the measurement of real photons much less sensitive to a new source than dileptons. Indeed there is no conclusive evidence for QGP radiation from real photon measurement at the SPS. The initial measurements of WA80 \cite{wa80} and CERES \cite{ceres-photons} with a S beam at 200~AGeV yielded only upper limits of $\sim$15\% of the total photon yield for any source beyond the known hadron decays. In the two experiments the sensitivity was limited not by the statistics but rather by the systematic errors, too large to identify a thermal photon source which could only be of the order of a few percent of the inclusive photon yield. In later measurements, the WA98 experiment observed a photon excess in Pb+Pb collisions at 158~AGeV \cite{wa98}. The excess is small, only a 1-2$\sigma$ effect, and was observed only in central collisions (in peripheral events the yield is consistent with the expectations from hadron decays). The WA98 results attracted considerable theoretical interest \cite{turbide04,gale-haglin03,wong98,dumitru01,srivastava01,alam01,huovinen02}.
\begin{figure}[!h]
  \begin{center}
     \includegraphics[width=120mm]{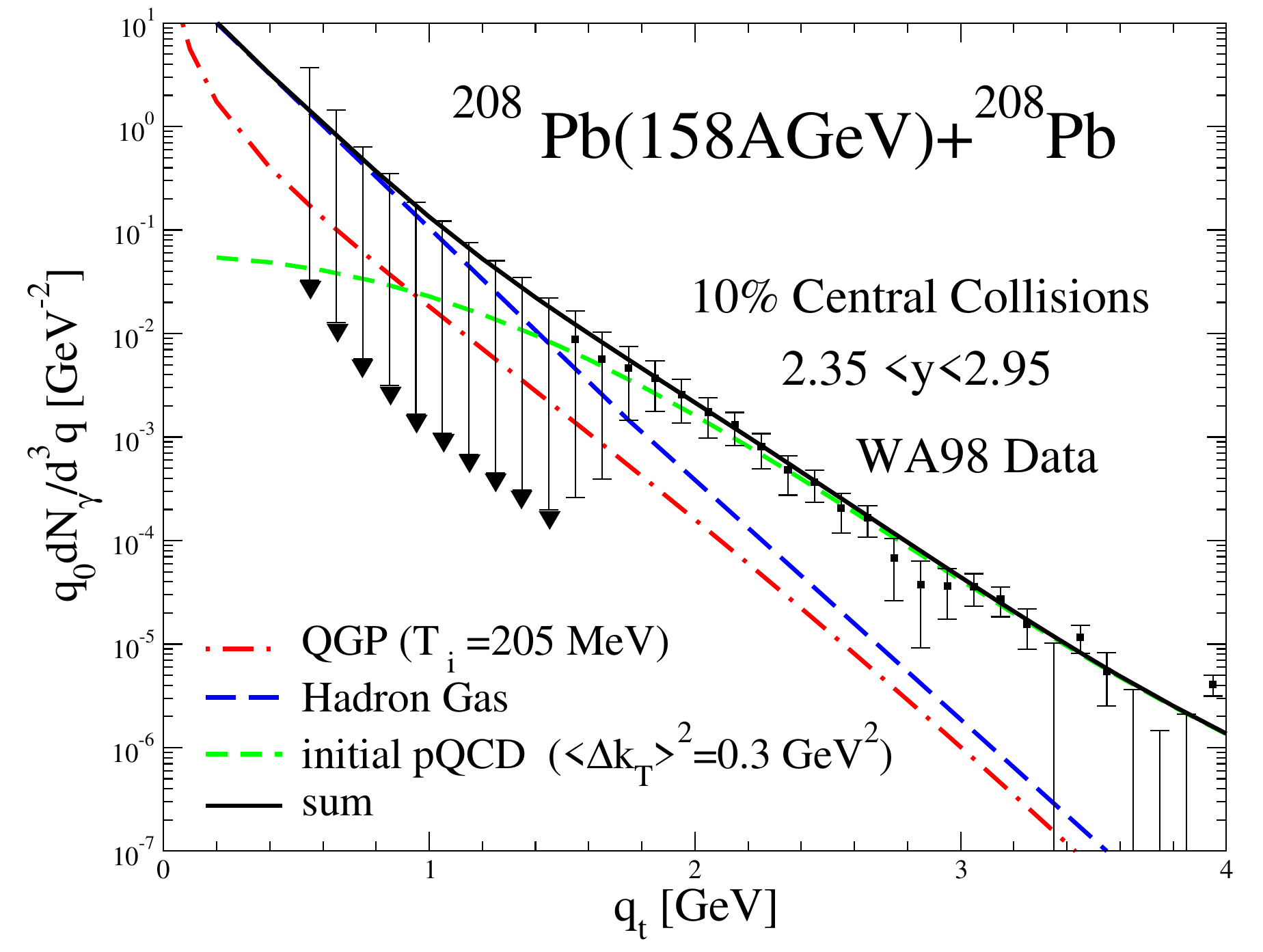}
  \end{center}
     \vspace{-5mm}
\caption{Photon excess in central Pb+Pb collisions at 158~AGeV from WA98 \cite{wa98} compared to a calculation including prompt (pQCD) and thermal (hadron gas and QGP emission) photons \cite{turbide04}.}
  \label{fig:wa98-turbide}
\end{figure}
Fig.~\ref{fig:wa98-turbide} shows the  \pt~distribution of the excess photons from WA98 together with  a state-of-the-art calculation that provides a rather good description of the data \cite{turbide04}. The calculation uses a simple fireball evolution of the collisions with a reasonable initial temperature of $T_i$ = 205 MeV and includes two main components: (i) a contribution of direct photons from primordial hard scattering during the first instants of the collision. This contribution saturates the yield at high \pt when supplemented with a transverse momentum k$_T$ broadening (Cronin effect) and (ii) thermal photons that dominate the low \pt part of the spectrum, predominantly emitted from the hadron gas ($\pi \rho a_1$ gas). It is interesting to note that the latter was calculated using the same many body approach that successfully explained the excess of low-mass dileptons discussed in Section~\ref{subsec:sps} thus providing a consistent description of both the excess of low-mass dileptons and the excess of real photons in terms of thermal radiation from the hadronic phase. The QGP contribution was found to be relatively small in the entire \pt range.

An interesting and original analysis has recently been presented by PHENIX at RHIC \cite{phenix-photons08}. Capitalizing on the idea that every source of real photons should also emit virtual photons \cite{cobb78} and in order to avoid the huge physics background inherent to real photon measurements, PHENIX analyzed \ee pairs emitted in Au+Au collisions at \sqnR with low masses (\mee $<$ 300~MeV/c$^2$) and high transverse momentum (1$<$ \pt $<$ 5~GeV/c). In this restricted kinematic window, PHENIX observes a significant excess of \ee pairs beyond the expected yield 
 \begin{figure}[!h]
  \begin{center}
  \vspace{2mm}
     \includegraphics[width=105mm]{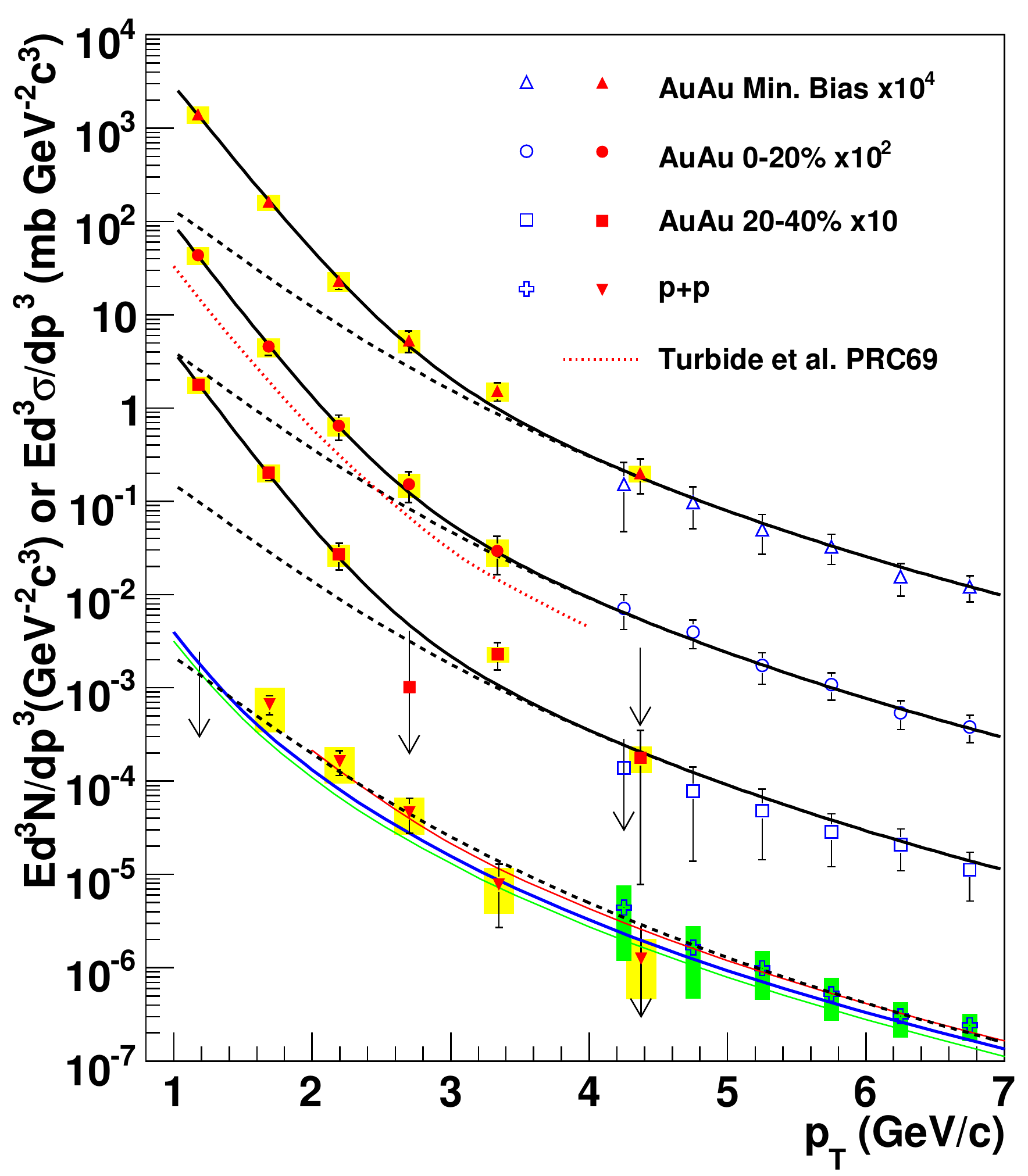}
  \end{center}
  \vspace{-3mm}
\caption{\pt distributions of direct photons in p+p collisions (invariant cross section) and in Au+Au collisions (invariant yield) for several centrality bins from PHENIX \cite{phenix-photons08}. The solid symbols are results from an analysis of  low-mass high \pt dileptons, whereas the open symbols are previous PHENIX results of direct photon measurements \cite{phenix-pp-photons,phenix-photons05}. See text for more details.}
  \label{fig:phenix-photons}
\vspace{3mm}
\end{figure}
from the hadronic cocktail of light mesons and open charm decays. The \ee invariant mass excess is transformed into a spectrum of real photons
under the assumption that the excess is entirely due to internal conversion of photons. Fig.~\ref{fig:phenix-photons} shows the resulting photon spectrum.
The figure includes also the direct photon data measured by PHENIX at high \pt in Au+Au collisions \cite{phenix-pp-photons,phenix-photons05}. In the \pt~range of overlap the two data sets, from dileptons and direct photons, are in good agreement within the experimental errors. The figure also shows the baseline spectrum measured in p+p collisions. The p+p data are well described by NLO pQCD direct photon calculations (see solid lines on the p+p data) and by a power-law fit (dashed line on the p+p data) \cite{vogelsang93}. The Au+Au data on the other hand are clearly above the T$_{AA}$ scaled p+p fit (see dashed line on the Au+Au data)  and require an additional source. The solid lines on the Au+Au data represent the power-law fit to the p+p data, scaled with T$_{AA}$, plus an exponential fit. For this exponential component the fit yields an inverse slope parameter which is within errors independent of centrality and found to be  T = 221 $\pm$ 19 $\pm$ 19 MeV in central collisions.
Interpreted as thermal radiation, this would represent the temperature of the system averaged over the space-time evolution of the collision and would correspond, according to hydrodynamical models to an initial temperature of $T_i$ = 300 (600) MeV for a thermalization time of 0.6 (0.15) fm/c \cite{d'enterria}.

\section{Summary and outlook}
\label{sec:summary}

In spite of the experimental difficulties, huge progress has been achieved over the past fifteen years in the measurement of electromagnetic observables. Like other aspects of the relativistic heavy ion program, the field is largely driven by experiment.

At SPS energies, all heavy-ion experiments have observed an enhancement of dileptons compared to the expected known sources and a coherent picture seems to emerge for the entire range of dilepton masses. In the LMR, the enhancement is associated with the thermal radiation (mainly $\pi^+\pi^- \rightarrow \rho \rightarrow \gamma^* \rightarrow l^+l^-$) emitted by the high density hadronic phase, close to the phase boundary, with strong in-medium modifications of the intermediate $\rho$. It seems that the onset of chiral symmetry restoration proceeds through broadening and subsequent melting of the resonances rather than by dropping masses. The enhancement in the IMR is not due to charm enhancement as originally suspected. It is also associated to thermal radiation but the temperature derived from the inverse slope of the exponential \mt distributions strongly suggests a partonic rather than hadronic origin.

At low energies of 1-2~AGeV, the DLS puzzle could be close to a solution with the recent HADES results on elementary  p+p and p+d collisions indicating no enhancement of low-mass \ee pairs in C+C collisions but rather a mere superposition of nucleon-nucleon collisions. If confirmed, this implies an onset of the low-mass dilepton enhancement somewhere in the energy range of 2-40~AGeV, almost the same energy range where the QCD critical point is being looked for in existing (RHIC, SPS) and planned (FAIR, NICA) facilities.

At the high energies of RHIC, first results from PHENIX are now available. The observed enhancement in the LMR appears different from the one observed at the SPS and may require an additional source. The apparent lack of enhancement in the IMR is intriguing. Further insight to elucidate these results will require measurements with better signal to background ratio and the ability to recognize displaced vertices, both expected with upgrade projects that are being implemented in the PHENIX detector. Combining results of low-mass dileptons with measurements of real photons, PHENIX identified an excess of direct photons (beyond the expectations based on measurements of p+p collisions scaled to the nuclear case with T$_{AA}$) with an exponential spectral shape. Interpreted as thermal radiation, this excess reflects an initial temperature of 300~MeV for a thermalization time of 0.6~fm/c.

A variety of precise spectroscopic studies to identify in-medium modifications of the LVM in elementary reactions where the conditions are very well defined, is underway. Effects of broadening and dropping mass are reported but there is no coherent picture so far and in particular not all results seem to be reconcilable with the experience gained in nuclear collisions at the SPS. More data are clearly needed.

Spectroscopic studies of the LVM in nuclear collisions are much more challenging, in particular the direct spectral shape analysis of the $\omega$ and $\phi$. An alternative method, the simultaneous measurement of the $\phi$ through its \kk and \ee decay modes, is pursued at SPS and RHIC to provide evidence for the expected in-medium modification of the $\phi$. At SPS, no effects are seen in In+In whereas the available results on Pb+Pb are inconclusive. Results from PHENIX at RHIC are also inconclusive due to the large uncertainties of the dilepton data. Higher quality results should be available soon.

\section*{Acknowledgments}
It is a pleasure to thank C. Gale, R. Rapp and R. Shahoyan for valuable discussions and information. This work is supported by the Israeli Science Foundation, the US-Israel Binational Science Foundation, the MINERVA Foundation and the Nella and Leon Benoziyo Center for High Energy Physics.



\begin{thebibliography}{99}
\bibitem{ceres-rapp} These calculations were performed by R. Rapp based on \cite{rapp-wambach}.
\end{thebibliography}
\end{document}